\documentclass[a4paper,fleqn,usenatbib]{mnras} 

\title[{\it Gaia} and the Hyades]{A {\it Gaia} study of the Hyades open cluster}

\usepackage[T1]{fontenc}
\usepackage{ae,aecompl}
\usepackage{graphicx}
\usepackage{amsmath}
\usepackage{amssymb}
\usepackage{txfonts}
\usepackage{color}

\author[S. Reino et al.]{
Stella Reino,$^{1,2}$\thanks{E-mail: reino@strw.leidenuniv.nl}
Jos de Bruijne,$^{1}$
Eleonora Zari,$^{2}$
Francesca d'Antona$^{3}$
and Paolo Ventura$^{3}$
\\
$^{1}$Science Support Office, Directorate of Science, European Space Research and Technology Centre (ESA/ESTEC), Keplerlaan 1, 2201 AZ, Noordwijk, The Netherlands\\
$^{2}$Leiden Observatory, Leiden University, Niels Bohrweg 2, 2333 CA Leiden, The Netherlands\\
$^{3}$INAF -- Osservatorio Astronomico di Roma, Via di Frascati 33, 00078 Monte Porzio Catone (Roma), Italy
}

\date{Accepted 2018 March 19. Received 2018 March 16; in original form 2018 March 2}

\pubyear{2018}

\hypersetup{draft}
\begin{document}
\label{firstpage}
\pagerange{\pageref{firstpage}--\pageref{lastpage}}
\maketitle

\begin{abstract}
We present a study of the membership of the Hyades open cluster, derive kinematically-modelled parallaxes of its members, and study the colour-absolute magnitude diagram of the cluster. We use {\it Gaia} DR1 {\it Tycho}-{\it Gaia} Astrometric Solution (TGAS) data complemented by {\it Hipparcos}-2 data for bright stars not contained in TGAS. We supplement the astrometric data with radial velocities collected from a dozen literature sources. By assuming that all cluster members move with the mean cluster velocity to within the velocity dispersion, we use the observed and the expected motions of the stars to determine individual cluster membership probabilities. We subsequently derive improved parallaxes through maximum-likelihood kinematic modelling of the cluster. This method has an iterative component to deal with 'outliers', caused for instance by double stars or escaping members. Our method extends an existing method and supports the mixed presence of stars with and without radial velocities. We find 251 candidate members, 200 of which have a literature radial velocity, and 70 of which are new candidate members with TGAS astrometry. The cluster is roughly spherical in its centre but significantly flattened at larger radii. The observed colour-absolute magnitude diagram shows a clear binary sequence. The kinematically-modelled parallaxes that we derive are a factor $\sim$1.7 /~ 2.9  more precise than the TGAS / {\it Hipparcos}-2 values and allow to derive an extremely sharp main sequence. This sequence shows evidence for fine-detailed structure which is elegantly explained by the full spectrum turbulence model of convection.
\end{abstract}


\section{Introduction}\label{sec:introduction}

As the nearest open cluster to the Sun, the Hyades are among the best-studied stellar groups  in the sky \citep[see, for instance,][for a review]{1998A&A...331...81P}. The proximity of the Hyades ($\sim$46~pc) is an advantage in many ways. For example, cluster members have negligible interstellar reddening and extinction. In addition, stars in the Hyades are relatively bright and, as a result of the large peculiar motion of the cluster, have large proper motions on the sky as well as large radial velocities along the line of sight. The proximity of the group does, however, also have its disadvantages. For example, the resulting angular extent on the sky leaves cluster members scattered over a wide field of view and interspersed with a great number of field stars.

The importance of the Hyades is illustrated by its use as a calibrator for various fundamental relations in astrophysics (e.g., the absolute magnitude-spectral type and the mass-luminosity relationships) and by its distance historically serving as a foundation for the cosmic distance ladder. The Hyades cluster, alongside other open clusters, is also ideal for studying stellar structure and evolution theories, as its members are thought to have formed at the same time from the same molecular cloud.

In this paper, we study the Hyades cluster using the {\it Tycho}-{\it Gaia} astrometric solution (TGAS) contained in the first data release \citep[{\it Gaia} DR1;][]{2016A&A...595A...2G} of the {\it Gaia} mission \citep{2016A&A...595A...1G}. We complement the TGAS data by {\it Hipparcos}-2 data \citep{2007ASSL..350.....V} for bright stars that are missing in TGAS and we combine this astrometric data set with  a newly compiled set of  radial velocities from various literature sources. The six-dimensional phase-space coordinates of the objects in the data set are used to determine kinematic membership probabilities. This provides a list of 251 candidate members, 70 of which are new compared to \citet{2017A&A...601A..19G}. Using this list of candidates as input, we apply  a newly-developed, extended version of  the iterative, maximum-likelihood kinematical modelling of the cluster, originally pioneered by \citet{2000A&A...356.1119L}, to determine improved parallaxes for individual stars while rejecting stars with discrepant motions, for instance caused by unrecognised binarity. At the turn of the millennium, such an approach applied to {\it Hipparcos}-1 data \citep{1997ESASP1200.....E} already provided a Hyades main sequence with unprecedented smoothness \citep{2001A&A...367..111D,2000ApJ...544L..65D}. Our study, owing to the ultra-precise kinematically-modelled parallaxes that are based on {\it Gaia} DR1 and {\it Hipparcos}-2 data for an extended membership list, provides further constraints on the modelling of convective energy transport in stars.

This paper is organised as follows. A description of our dataset is given in Sect.~\ref{sec:data}. Section~\ref{sec:membership} discusses our kinematic membership selection method and the derivation of membership probabilities. In Sect.~\ref{sec:membership_results}, we validate and discuss the results of our membership selection, including the spatial and velocity structure  of the cluster. In Sect.~\ref{sec:modelling}, we  outline the  kinematic modelling approach  and present the extension we have developed.  The results of applying the model to the TGAS / {\it Hipparcos}-2 data of the Hyades are presented and discussed in Sect.~\ref{sec:modelling_results}.  A summary and discussion of our work is provided in Sect.~\ref{sec:conclusions}.

A list of the 251 candidate members along with their membership probabilities and kinematically improved parallaxes is available online.

\begin{figure}
  \begin{center}
    \includegraphics[width=0.9\columnwidth]{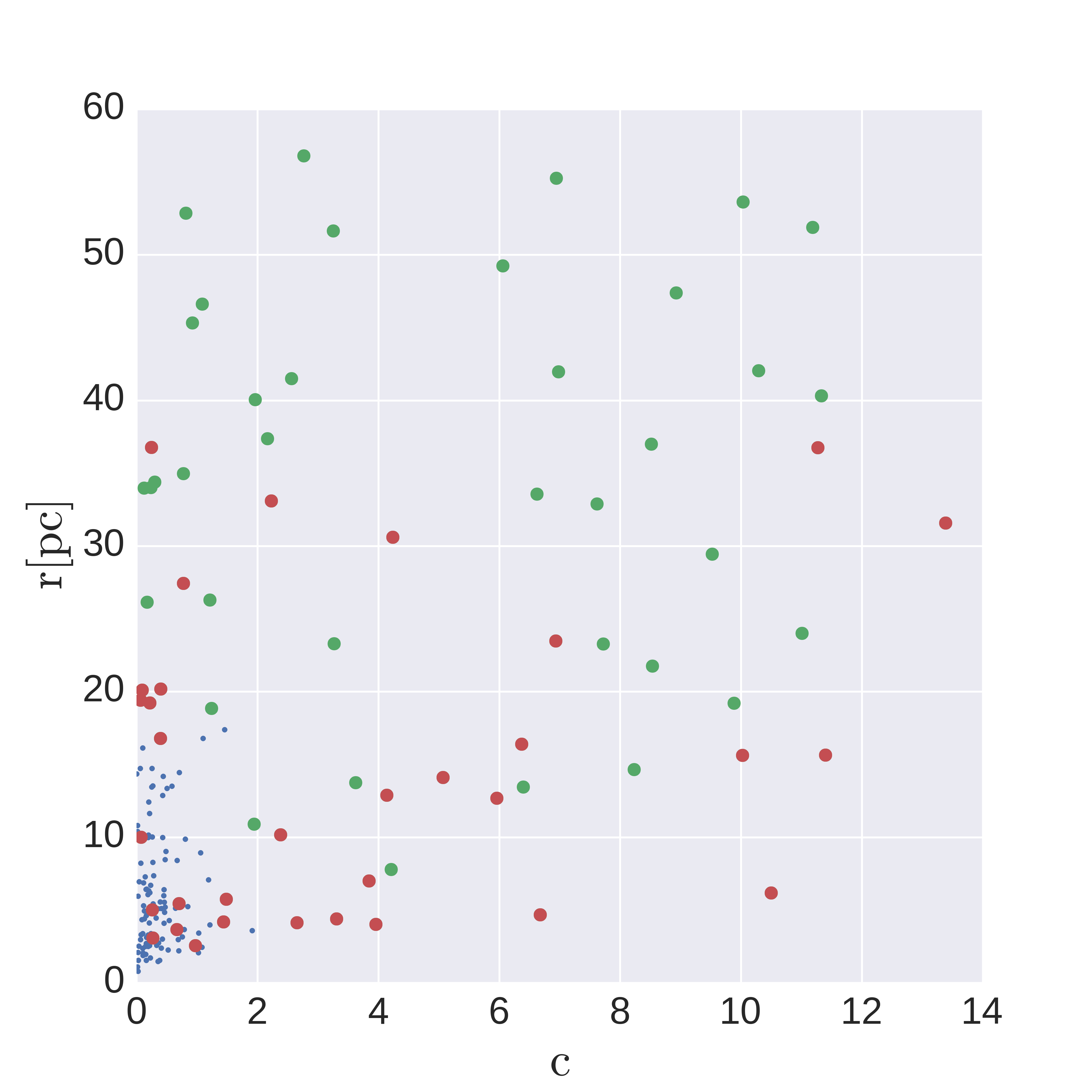}
    \caption{Membership statistic $c$ versus distance $r$ from the cluster centre for the 173 TGAS stars in our membership list (small blue dots). Large dots indicate the 70 TGAS members that were missed by \citet{2017A&A...601A..19G}; red points denote stars with known radial velocity while green points denote stars without known radial velocity.}
    \label{fig:70}
  \end{center}
\end{figure}

\begin{figure*}
  \begin{center}
    \includegraphics[width=0.78\textwidth]{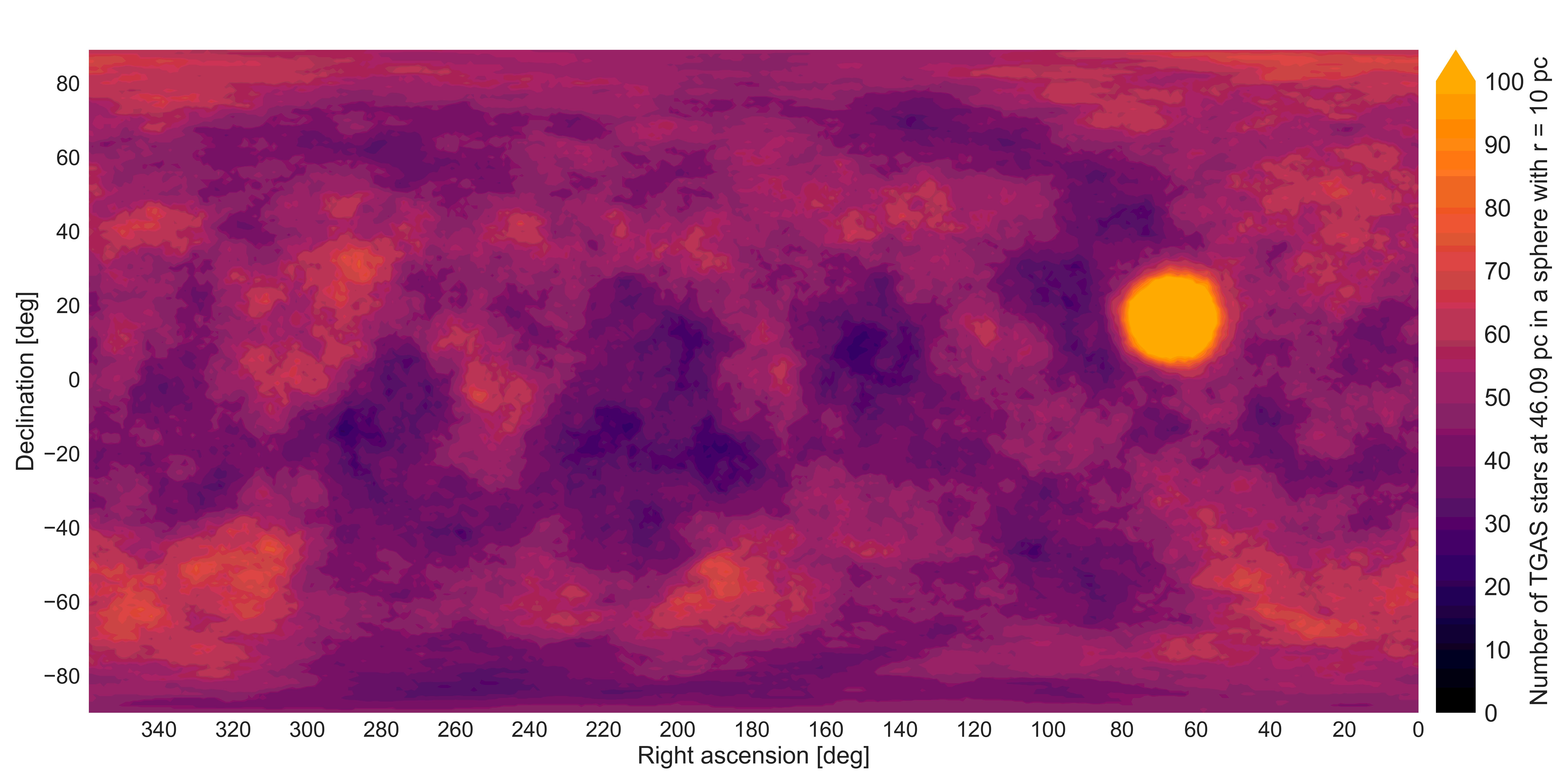}
    \caption{All-sky map showing, through colour coding, the number of TGAS stars in a 10-pc-radius sphere  centred on the Hyades centre of mass. The orange circle shows the overdensity representing the (core of the) Hyades cluster.}
    \label{fig:47h}
  \end{center}
\end{figure*}
\begin{figure*}
  \begin{center}
    \includegraphics[width=0.78\textwidth]{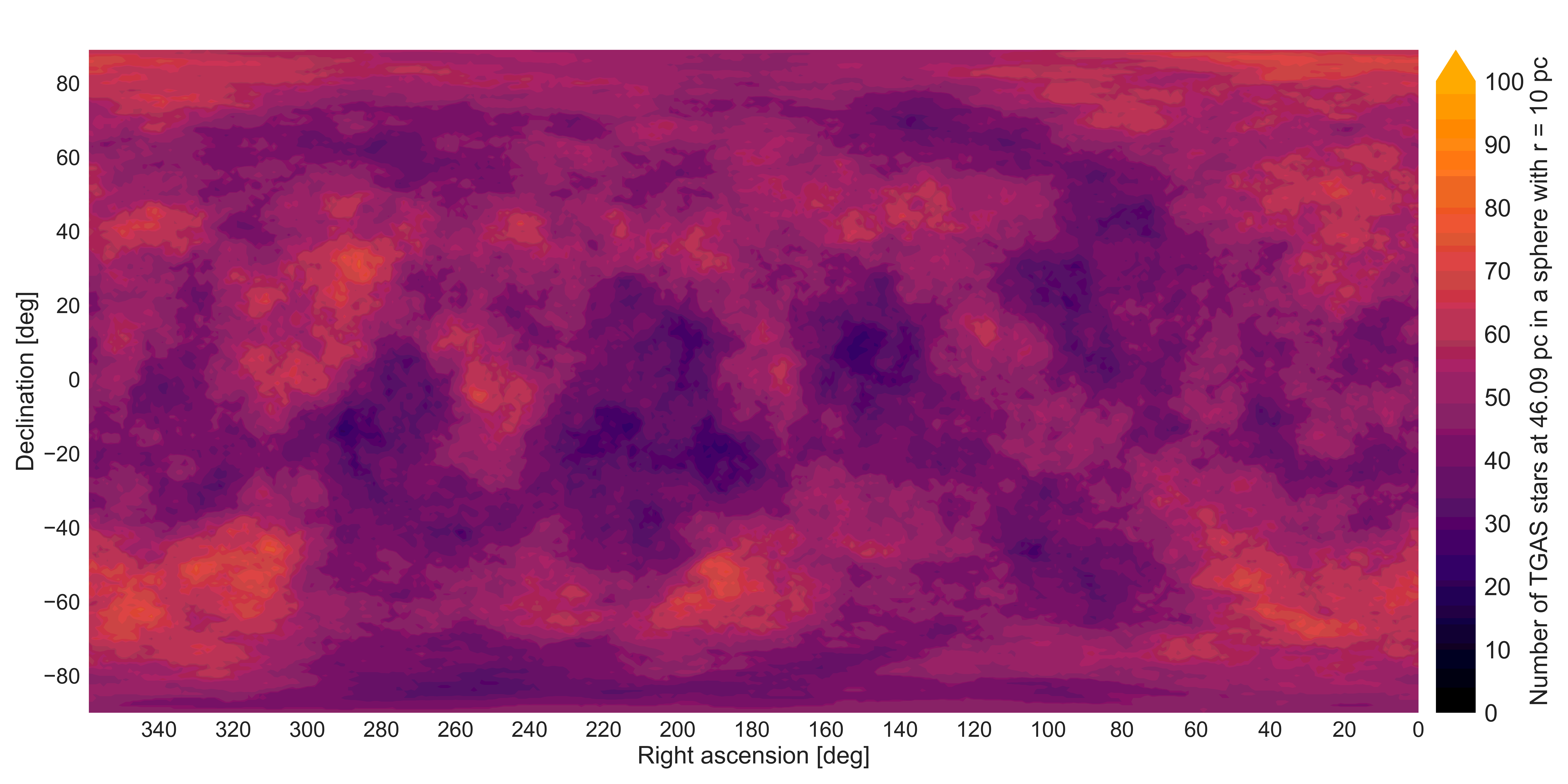}
    \caption{As Figure~\ref{fig:47h} but after the removal of the 173 TGAS members determined in this work. The galactic field inside the Hyades is  fully compatible with neighbouring areas, confirming  our membership selection is balanced.}
    \label{fig:47}
  \end{center}
\end{figure*}

\begin{figure}
  \begin{center}
    \includegraphics[width=0.9\columnwidth]{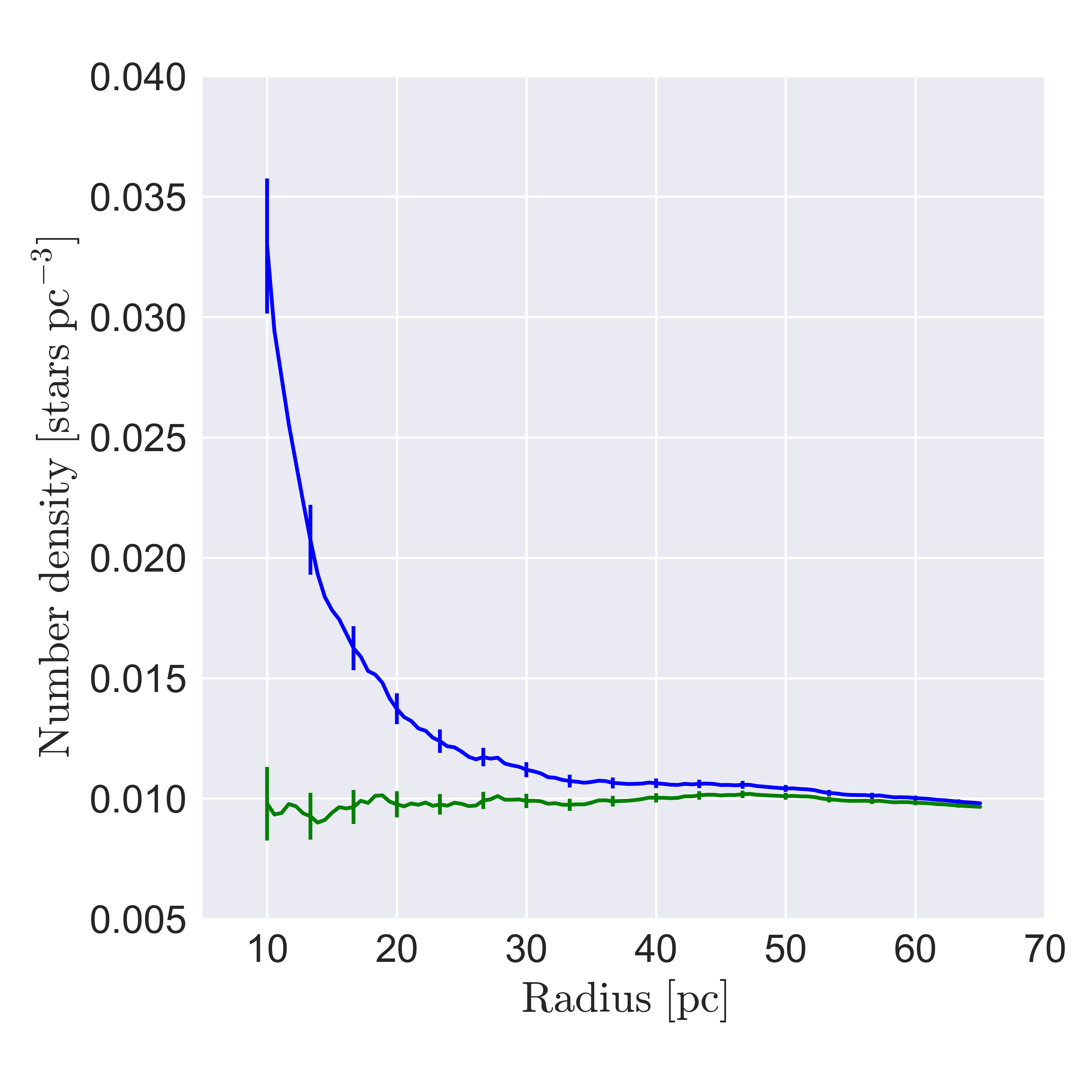}
    \caption{Volume density of the TGAS catalogue, as function of three-dimensional distance from the cluster centre, for the full TGAS catalogue (blue) and after subtraction of the 173 TGAS stars in our Hyades membership list (green).}
    \label{fig:volume_density}
  \end{center}
\end{figure}

\begin{figure}
  \begin{center}
    \includegraphics[width=\columnwidth]{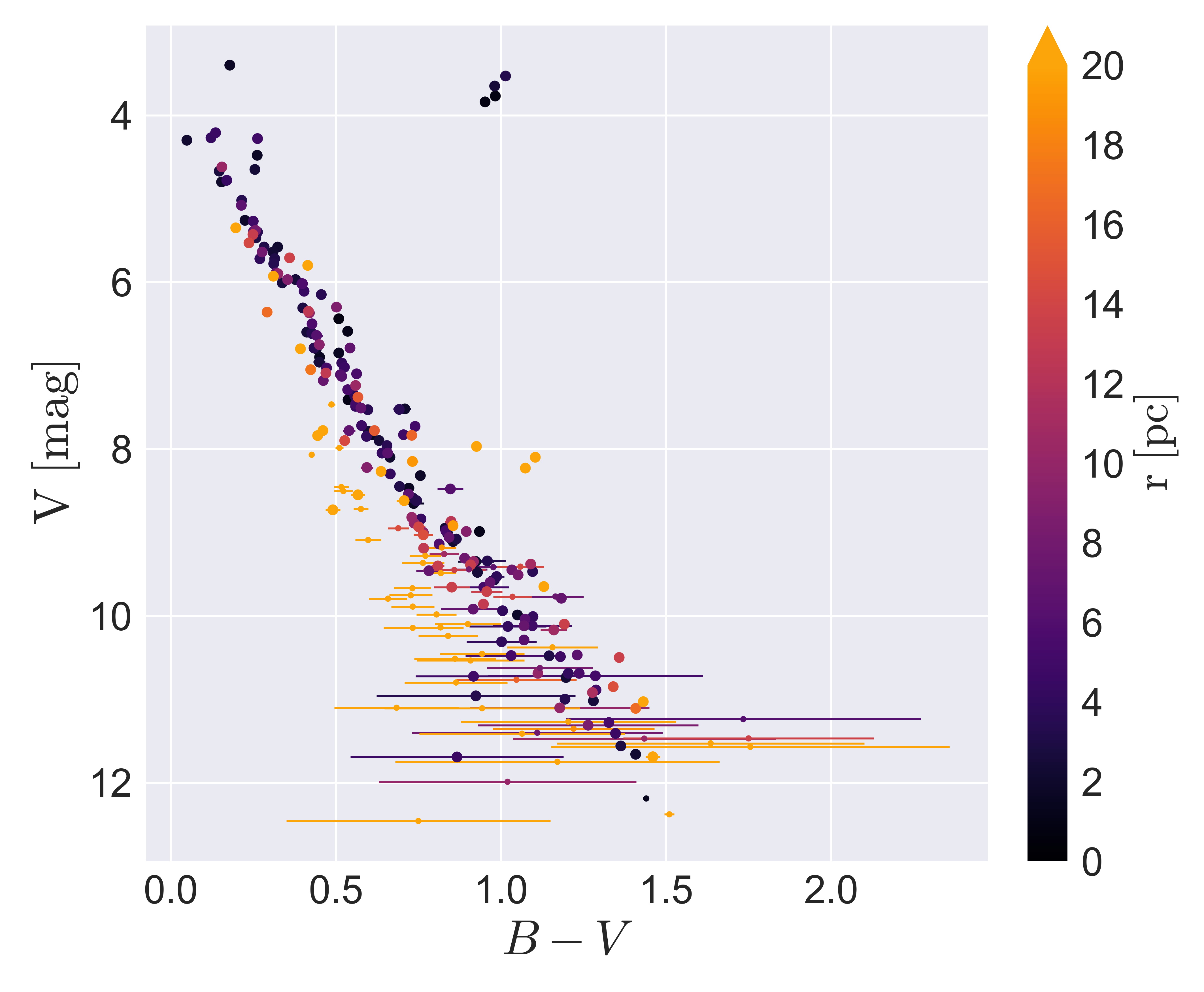}
    \caption{Colour-magnitude diagram based on Johnson photometry (Appendix~\ref{sec:Johnson}) for all 251 candidate members. The 200 objects with known radial velocity are indicated with a large markerpoint; the remaining 51 candidate members without known radial velocity are indicated with a small markerpoint. The colour of each point represents the distance $r$ of each star from the cluster centre.}
    \label{fig:CMD}
  \end{center}
\end{figure}

\begin{figure}
  \begin{center}
    \includegraphics[width=\columnwidth]{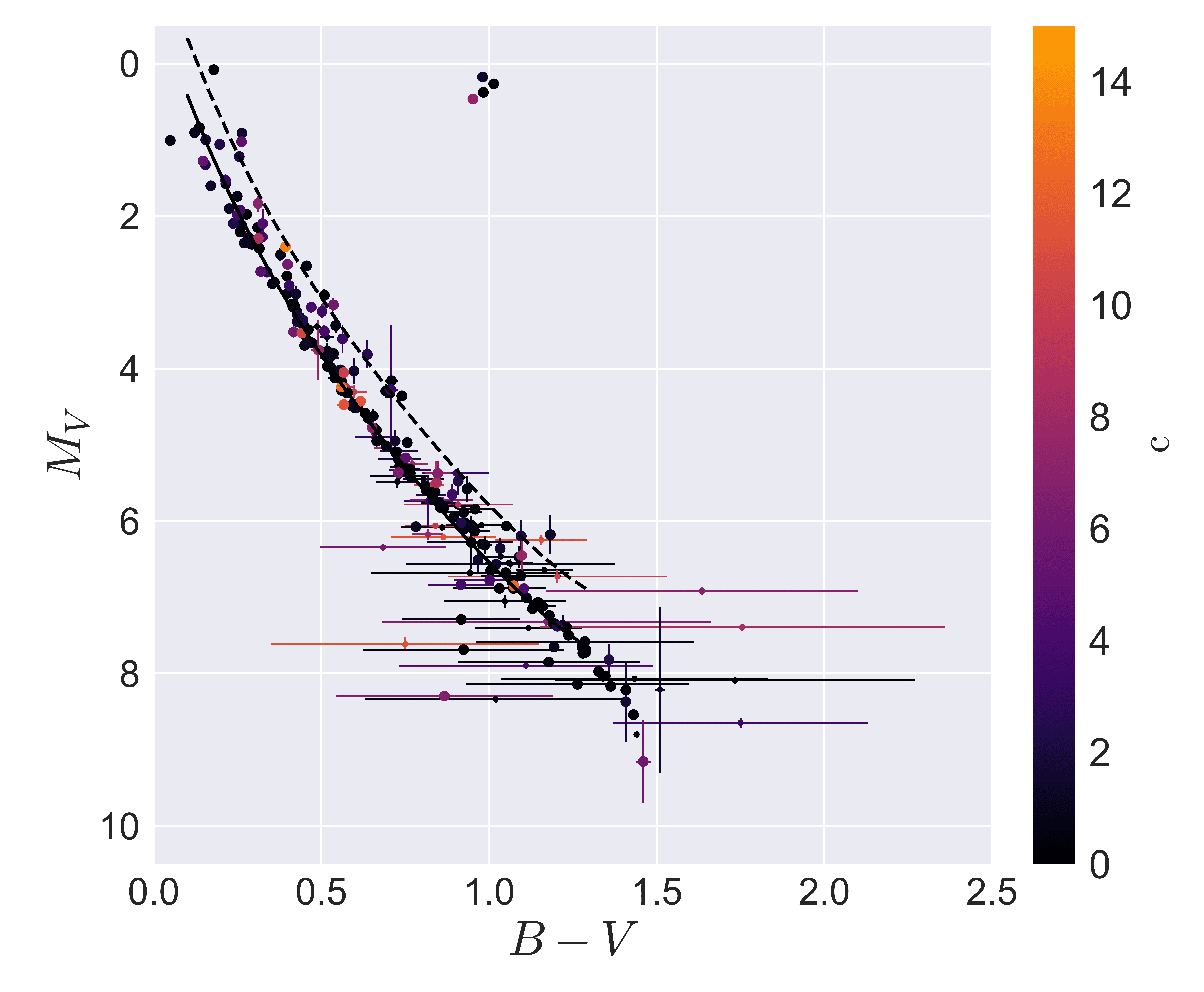}
    \caption{Colour-absolute magnitude diagram based on Johnson photometry (Appendix~\ref{sec:Johnson}) for all 251 member stars. The 200 members with known radial velocity are indicated with a large markerpoint; the remaining 51 candidate members without known radial velocity are indicated with a small markerpoint. The colour of each point represents the star's membership statistic ($c$ value; a small $c$ value means a high-fidelity member). The solid black line shows the empirical main sequence of the Hyades as derived by \citet{2012BASI...40..487S}; the dashed black line shows the associated equal-mass-binary sequence.}
    \label{fig:HR_bv}
  \end{center}
\end{figure}

\begin{figure}
  \begin{center}
    \includegraphics[width=\columnwidth]{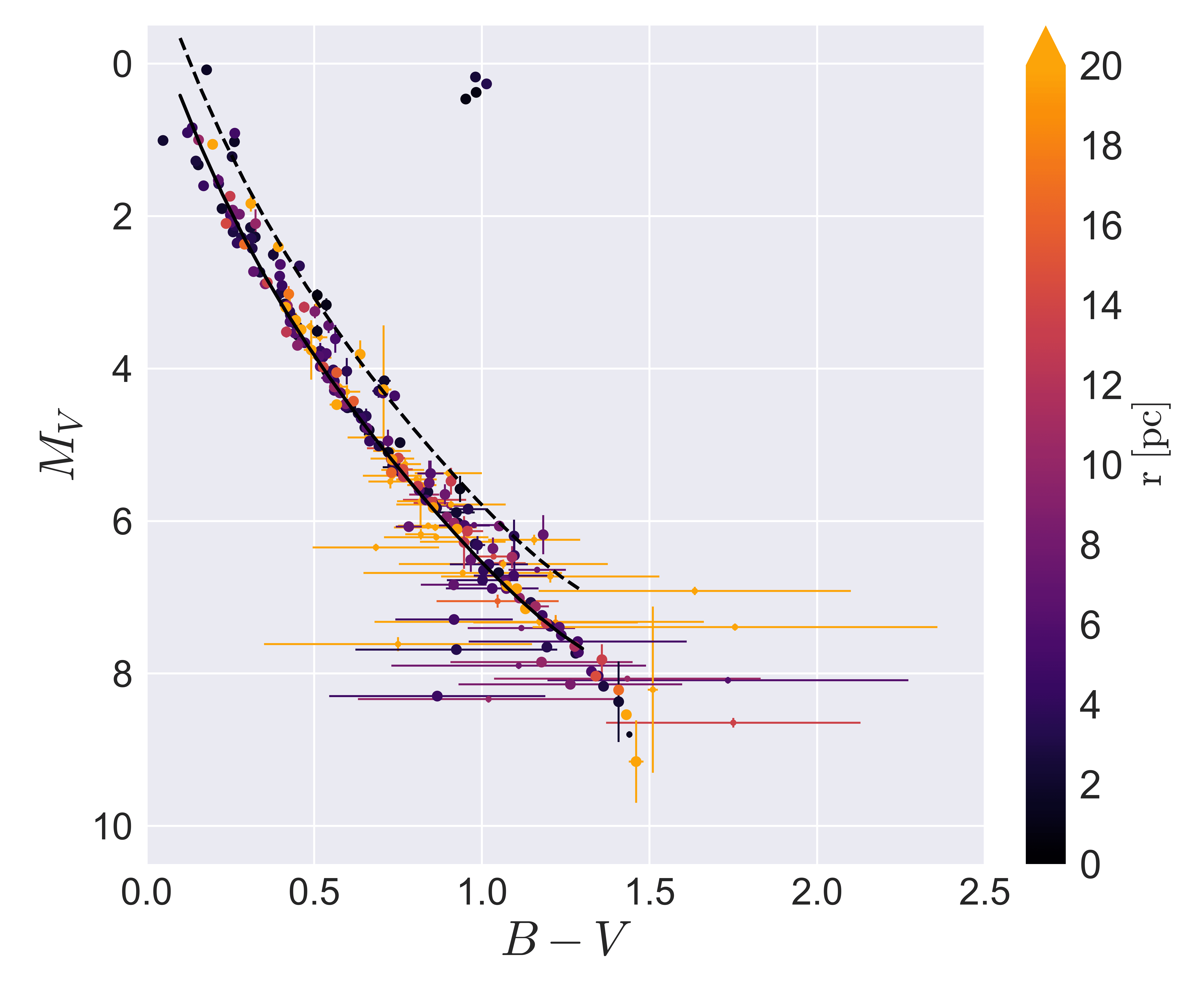}
    \caption{As Figure~\ref{fig:HR_bv}, but colour coded according to the distance $r$ of each star from the cluster centre.}
    \label{fig:HRr_bv}
  \end{center}
\end{figure}

\begin{figure*}
  \begin{center}
    \includegraphics[width=0.8\textwidth]{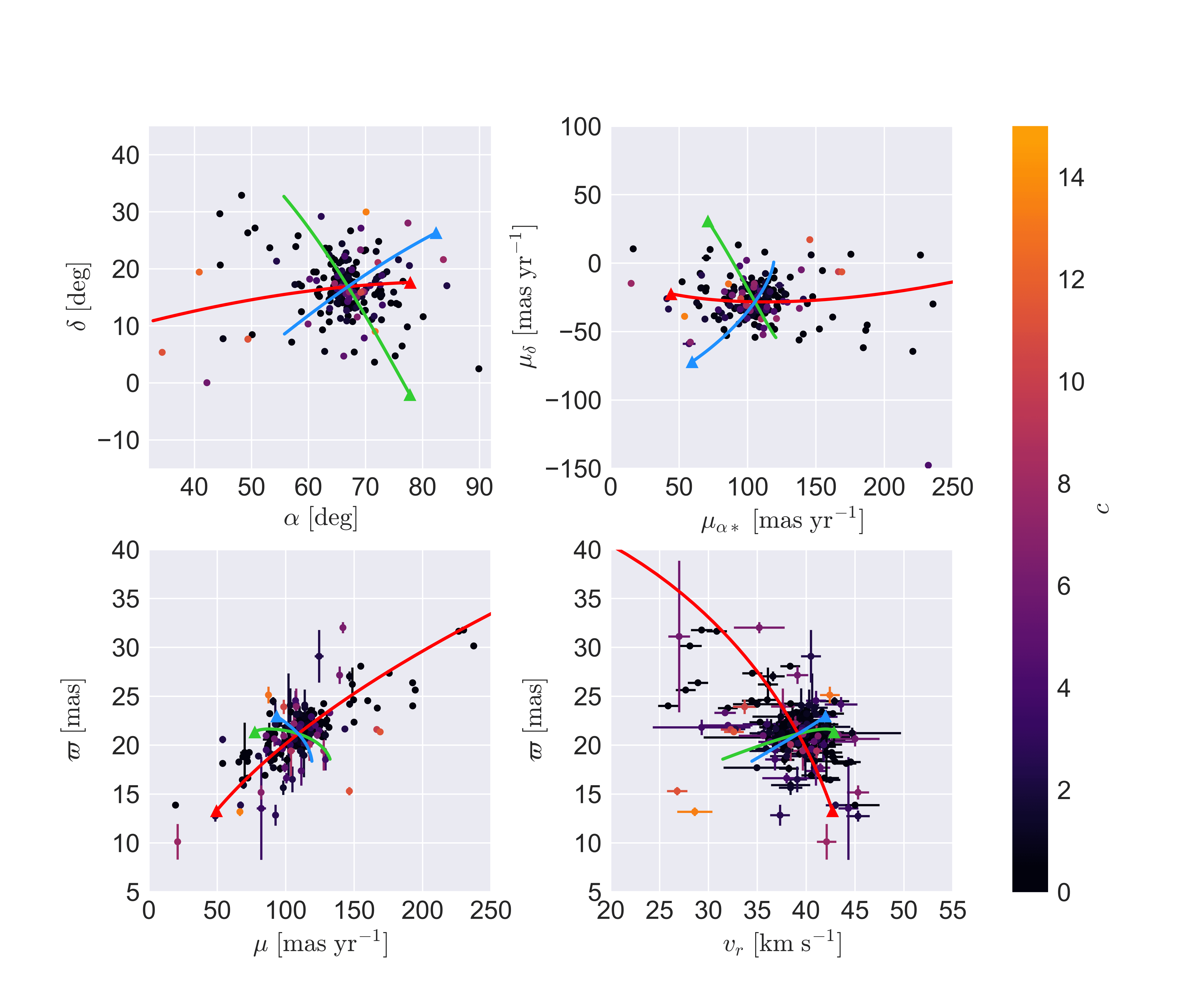}
    \caption{Distribution of the observables of the 200 candidate members with known radial velocity.
      \emph{Top left:} celestial distribution in right ascension and declination.
      \emph{Top right:} proper motion in declination versus proper motion in right ascension (vector-point diagram).
      \emph{Bottom left:} parallax versus total proper motion.
      \emph{Bottom right:} parallax versus radial velocity.
      Error bars are often too small to be visible. The colour of the points represents their $c$ value; a small $c$ value means a high-fidelity member. Whereas all 200 stars appear in the top left panel, four stars with large proper motion and large parallax fall outside the plotted ranges of the other panels. The lines reflect the projections of the major (red), intermediate (green), and minor (blue) axes of the cluster; the triangles can be used to cross-correlate the various panels.}
    \label{fig:sub}
  \end{center}
\end{figure*}

\begin{figure*}
  \begin{center}
    \includegraphics[width=0.8\textwidth]{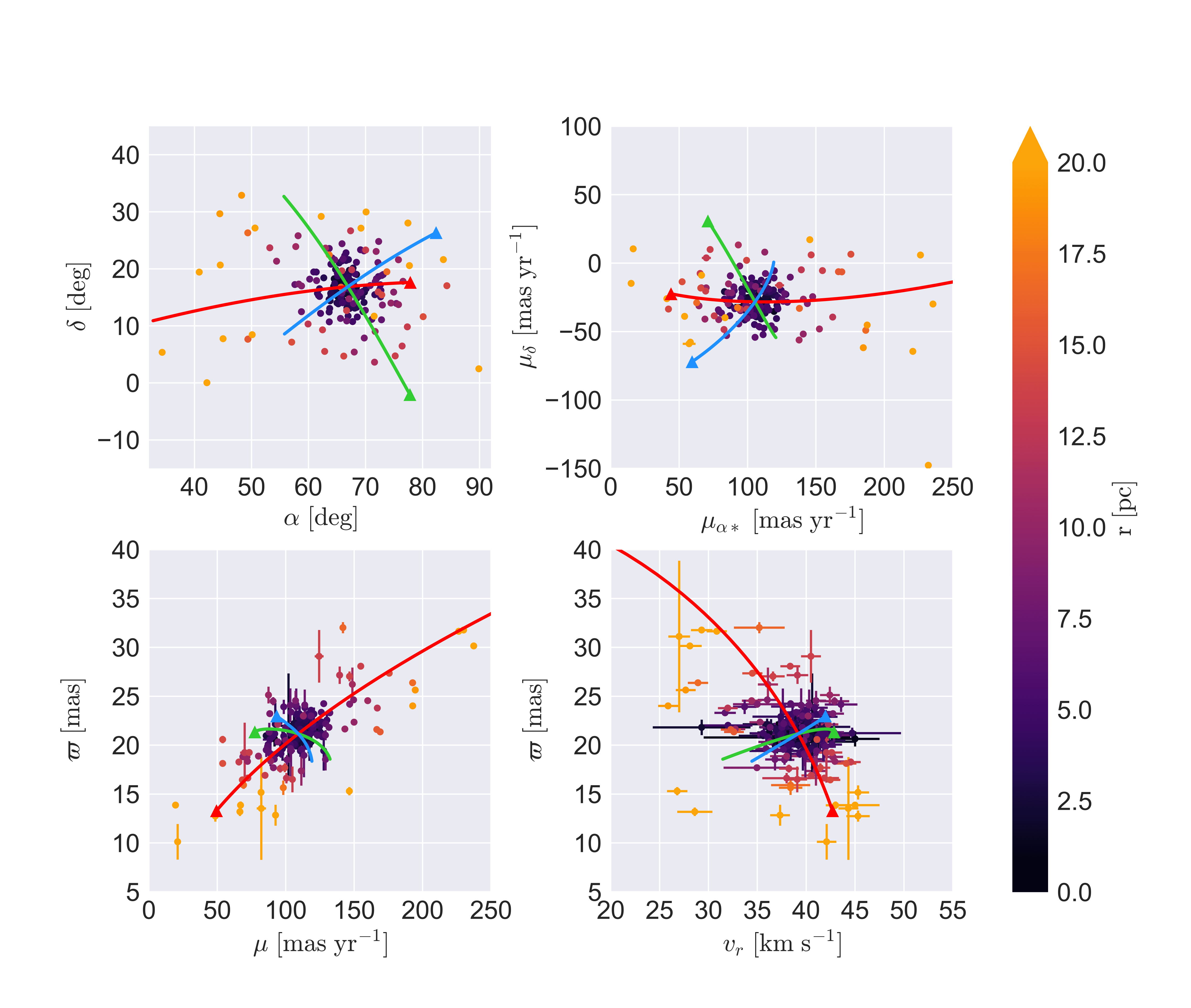}
    \caption{As Figure~\ref{fig:sub}, but colour coded according to the distance $r$ of each star from the cluster centre.}
    \label{fig:subr}
  \end{center}
\end{figure*}

\section{Data}\label{sec:data}

\subsection{Astrometry}\label{subsec:astrometry}

Our starting point is the {\it Gaia} DR1 TGAS data set \citep{2016A&A...595A...2G}. Following the seminal study by \citet{1998A&A...331...81P}, whose analysis of the Hyades cluster was based on the original {\it Hipparcos}-1 catalogue \citep{1997ESASP1200.....E}, we select all TGAS entries with $2^{\rm h}15^{\rm m} \leq \alpha \leq 6^{\rm h}5^{\rm m}$ and $-2^\circ \leq \delta \leq 35^\circ$. This leaves 67\,428 objects. Subsequently applying a conservative parallax cut retaining only nearby stars that have $\varpi > 10$~mas leaves 1913 objects. In the same area of sky, there are 383 {\it Hipparcos} stars which are not contained in TGAS. We add these stars, with the {\it Hipparcos}-2 astrometry from \citet{2007ASSL..350.....V}, to our sample. We reconstruct their covariance matrices using the recipe from \citet{2014A&A...571A..85M}. In total, this gives us a starting sample of 2296 objects with five-parameter astrometry.

We put the {\it Hipparcos}-2 proper motions on the {\it Gaia} reference frame by performing a correction based on Eq.~(8) in \citet{2016A&A...595A...4L}. For the stellar positions, we ignore the epoch difference between the {\it Hipparcos} astrometry (1991.25) and TGAS (2015.0). For Hyades members, this gives a systematic offset on the sky between {\it Hipparcos} and TGAS members of $23.75~{\rm year} \times 110~{\rm mas~yr}^{-1} \sim 2.6~{\rm arcsec}$, which is completely negligible in the direction cosines which are needed to establish space motions.

In this study, we derive distance estimates of Hyades members by inversion of their measured parallaxes. This is justified since the typical relative parallax error of objects in our sample is well below 10\% such that the inverse of the observed parallax is, in practice, an accurate and unbiased distance estimator \citep[e.g.,][]{2015PASP..127..994B,2016ApJ...832..137A,2016ApJ...833..119A}.

\subsection{Radial velocities}\label{subsec:radial_velocities}

For the 2296 objects with astrometry from Sect.~\ref{subsec:cluster_velocity}, we find literature radial velocities from the following sources: \citet{1998A&A...331...81P}, \citet{2012JApA...33...29G}, \citet{2013Obs...133..144G}, \citet{2009A&A...498..949M}, \citet{2007AJ....133.2524W}, \citet{2012yCat..35470013T}, \citet{2000A&A...354..881D}, HADES \citep{2017A&A...598A..27M}, the Extended {\it Hipparcos} Compilation \citep[XHIP;][]{2012AstL...38..331A}, RAVE DR5 \citep{2017AJ....153...75K}, APOGEE \citep[][SDSS DR13]{2016arXiv160802013S}, and GALAH DR1 \citep{2017MNRAS.465.3203M}. The radial velocity data is further described in Appendix~\ref{sec:RV}. In total, we find radial velocities for 908 stars in our sample, leaving 1388 stars without radial velocities. As detailed in Appendix~\ref{subsec:RV_final}, we conservatively inflate small ($<1$~km~s$^{-1}$) radial velocity standard errors in view of the inhomogeneity of our compilation such that radial velocity zero-point differences are washed out.

\section{Membership selection}\label{sec:membership}

Our approach to determining the membership of the cluster draws upon the method described by \citet{1998A&A...331...81P}. It follows three steps: 
(1) use the TGAS / {\it Hipparcos}-2 data of a preliminary set of members to determine the centre of the cluster and select the subset of the preliminary members that lie within a 10-pc radius from this cluster centre \citep[with 10~pc being the tidal radius;][]{1998A&A...331...81P},
(2) use this subset of stars to derive the mean velocity of the cluster, and
(3) use the difference between the observed and the expected motions of the stars to determine individual probabilities of membership.
Step (1) is described in Sect.~\ref{subsec:cluster_centre}. Sections~\ref{subsec:cluster_velocity} and \ref{subsec:membership} describe steps (2) and (3), respectively.  Sections~\ref{subsec:distance_velocity} and \ref{subsec:systematic_errors} discuss the cluster distance and velocity and the impact of the potential presence of systematic errors in the TGAS astrometry, respectively.

\subsection{Centre of the cluster}\label{subsec:cluster_centre}

The preliminary set of members is composed of the 197 Hyades members identified by \citet{1998A&A...331...81P} based on {\it Hipparcos}-1 data (these are stars which have '1' in column~x of their Table~2). In Perryman's study, these 197 candidate members had known radial velocities and their three-dimensional space motions were compatible with the Hyades cluster motion. We use, for each star, the astrometry from Sect.~\ref{subsec:astrometry}. In this step, we use the radial velocities and associated standard errors exclusively from Perryman's Table~2 (including zero-point corrections as described in Appendix~\ref{sec:RV}).

The centre of mass of the cluster is defined as $\vec{b}_{\rm c} = \sum_{i}(m_i \vec{b}_i) / \sum_{i}(m_i)$, where $\vec{b}_i$ are the position vectors of the $i = 1, \ldots, N = 197$ stars in Galactic Cartesian coordinates. We assume all $m_i = 1~{\rm M}_\odot$. This is justified by \citet{1998A&A...331...81P} who -- after experimenting with different weighting methods such as assigning binaries half the weight of single stars -- found that the centre of mass of the cluster is rather insensitive (i.e., showing differences at the level of a few tenths of a parsec, comparable to the error on the position of the cluster centre) to the weighting scheme used.  We did, nonetheless, perform tests with several mass-assignment schemes, as reported in Sect.~\ref{subsec:membership}. 

We find the centre of mass to be $\vec{b}_{\rm c} = (-44.16 \pm 0.74, 0.66 \pm 0.39, -17.76 \pm 0.41)$~pc. For comparison, using  only the members that are contained within 10~pc of the cluster centre, \citet{1998A&A...331...81P} find the cluster centre of mass to be at $\vec{b}^{\rm Perryman}_{\rm c} = (-43.08 \pm 0.25, 0.33 \pm 0.06, -17.09 \pm 0.11)$~pc. Of the 197 stars, 138 lie within a sphere of 10~pc radius from the centre (for 79 of these, we use TGAS astrometry while for 59 of these, we use {\it Hipparcos}-2 astrometry).  This set of core members is used to determine the velocity of the cluster in Sect.~\ref{subsec:cluster_velocity}.

The errors on our preliminary estimate of $\vec{b}_{\rm c}$, and this is also true for the final cluster-position estimate and the associated errors as presented in Sect.~\ref{subsec:membership}, are significantly larger than the errors quoted by \citet{1998A&A...331...81P}. We tried and failed to reproduce the Perryman et al.\ error estimates. For the 134 stars within 10~pc of their cluster centre (row 2 of their Table 3, labelled $r<10$~pc), we find, using {\it Hipparcos}-1 astrometry and assuming Solar-mass stars: $\vec{b}^{\rm Perryman,~reconstructed}_{\rm c,~r<10~pc} = (-42.97 \pm 0.30, 0.43 \pm 0.23, -17.08 \pm 0.22)$~pc versus $\vec{b}^{\rm Perryman}_{\rm c,~r<10~pc} = (-43.08 \pm 0.25, 0.33 \pm 0.06, -17.09 \pm 0.11)$~pc. Whereas the small position differences could be explained by Perryman's use of non-unit stellar masses, his smaller error bars remain a mystery, in particular since ``{\it for an estimation of the errors on the position of the centre of mass, associated standard errors were arbitrarily assigned [by Perryman] to be $0.1~M_\odot$ for single stars and $0.5~M_\odot$ for double stars}''. Following this approach would increase our error estimates and would make the discrepancy even larger.

\subsection{Velocity of the cluster}\label{subsec:cluster_velocity}

The 138 core members from Sect.~\ref{subsec:cluster_centre} are used to determine the mean cluster velocity along with the covariance matrix describing the velocity spread (sample variance) of the members.  The mean velocity vector in equatorial coordinates equals $\vec{v}_{\rm c} = (-6.30, 45.44, 5.32)~ {\rm km~s}^{-1}$ and the  velocity spread  equals:
\begin{equation}
\left( \begin{array}{rrr}
 1.69 & -0.02 & 0.43 \\
-0.02 &  1.95 & 0.28 \\
0.43 &  0.28 & 1.06 \end{array} \right),\label{eq:velocity_spread}
\end{equation}
where the diagonal elements are the standard  deviations  in each Cartesian component in units of ${\rm km~s}^{-1}$ and the off-diagonal elements are the correlation coefficients.
 For comparison, the \citet{1998A&A...331...81P} numbers are (their Eq.~17):
\begin{equation}
\left( \begin{array}{rrr}
 2.40 & -0.18 & 0.04 \\
-0.18 &  2.45 & 0.17 \\
0.04 &  0.17 & 1.26 \end{array} \right).
\end{equation}

The  above matrix Eq.~(\ref{eq:velocity_spread}) is related to, and has been derived from,  the sample covariance equation  which has elements $s_{jk}$ ($j=1,2,3$, $k=1,2,3$)  given by:
\begin{equation}
s_{jk} = \frac{1}{N-1} \sum_{i}^{N}(v_{ij}-\bar{v}_j)(v_{ik}-\bar{v}_k),\label{eq:first}
\end{equation}
where $v_{ij}$ are individual space velocities for all $i=1,\ldots,N=138$ core stars in Cartesian components $j=1,2,3$ and $\bar{v}_j$ denotes the mean velocity in component $j$.

The mean velocity transformed to Galactic coordinates is $\vec{v}_{\rm c} = (-41.92 \pm 0.16, -19.35 \pm 0.13, -1.11 \pm 0.11)~ {\rm km~s}^{-1}$, where the errors reflect the error on the mean. For comparison, the mean cluster motion determined by \citet{1998A&A...331...81P} for their final members situated within 10~pc of their cluster centre equals $\vec{v}^{\rm Perryman}_{\rm c} = (-41.70 \pm 0.16, -19.23 \pm 0.11, -1.08 \pm 0.11)~ {\rm km~s}^{-1}$  based on {\it Hipparcos}-1 data while \citet{2009A&A...497..209V} reports $\vec{v}_{\rm c} = (-41.1 \pm 0.9, -19.2 \pm 0.2, -1.4 \pm 0.4)~ {\rm km~s}^{-1}$ based on {\it Hipparcos}-2 data.

\subsection{Probability of membership}\label{subsec:membership}

We now use the mean velocity derived in Sect.~\ref{subsec:cluster_velocity} to determine the probability of membership for each of the 2296 sample stars from Sect.\ref{subsec:astrometry}. For each star, we first calculate the expected values of its (equatorial) transverse and radial velocities assuming it is comoving with the mean cluster velocity. We then compare the expected velocities with the observed velocities and, allowing for an uncertainty reflected by the sum of the sample covariance matrix around the mean cluster velocity (Eq.~\ref{eq:first}, representing the spread of cluster members) and the velocity covariance matrix of each star (representing the observed errors and correlations in the tangential and radial velocities), we calculate a dimensionless $\chi^2$ test statistic, $c$, for each star, defined as:
\begin{equation}
c \equiv \chi^2 = (\vec{v}^{\rm obs}-\vec{v}^{\rm exp})^{\rm T} {\vec C}^{-1} (\vec{v}^{\rm obs}-\vec{v}^{\rm exp}),\label{eq:chi^2}
\end{equation}
where the superscript T stands for transpose, $\vec{v}^{\rm obs}$ is the vector of observed tangential (and radial) velocities, ${\vec v}^{\rm exp}$ is the vector of expected tangential (and radial) velocities calculated using the mean cluster velocity, and $\vec{C}$ is the sum of the two covariance matrices as explained above. For stars with a literature radial velocity, the vectors $\vec{v}$ are three dimensional, whereas for stars without a literature radial velocity, they are two dimensional, exclusively reflecting the tangential velocity on the sky derived from the proper motion.  It is worth recalling that the tangential velocities depend on the proper motions and the parallaxes and that the errors on and correlations between these observables have been included through the covariance matrix $\vec{C}$ and its associated Jacobian matrices linked to the coordinate transformations we apply (cf.\ Eqs~13--15 in \citealt{1998A&A...331...81P}).

Since the $c$ statistic is expected to follow a $\chi^2$ distribution with 3 degrees of freedom (DOF) for stars with radial velocity data and 2 degrees of freedom for stars without radial velocity data, we calculate the probability of membership using the complementary cumulative distribution function (complementary CDF) at $c$, so $1-{\rm CDF}(c, {\rm DOF})$. We compute both the $\chi^2$ statistic $c$ and the corresponding membership probability for each of the 2296 stars in our sample. Depending on the scientific application, an appropriate membership threshold can be defined on a case by case basis. In this work, we consider a star to be a (candidate) member of the cluster if its membership probability, or $p$ value, exceeds $p_{\rm lim} = 0.0027$. That is, a star is considered a member if its $c$ value is less than 14.16 in case its radial velocity is known and less than 11.83 in case its radial velocity is not known \citep[see, e.g., Chapter 15.6 in][]{1992nrca.book.....P}. This results in a list of 251 (candidate) members, 200 of which have a known radial velocity (125 TGAS stars and 75 {\it Hipparcos}-2 stars) and 51 of which have an unknown radial velocity (48 TGAS stars and 3 {\it Hipparcos}-2~stars).

As mentioned in Sect.~\ref{subsec:cluster_centre}, we tested the sensitivity of our membership against the assumption that all stars have a mass of one Solar mass. As a reminder: this assumption impacts the location of the centre of mass of a preliminary set of members and, through the selection of the preliminary 'core members' that lie within a sphere of 10~pc radius from the centre, also the velocity of the cluster. Only the latter is used for our membership assignment. Using various mass-assignment schemes, the centre-of-mass position slightly changes, yet within the quoted errors. As a result, the number of preliminary 'core members', which are used only to define the cluster velocity, varies between 136 and 142 (compared to our 138). The impact of this variation on the final membership list is limited to the TGAS members without known radial velocity, which varies between 48 and 51 (compared to our 48). In short, our membership selection is insensitive to the assumed stellar masses, which only impact the identification of 1--2 stars at most.

\subsection{Final distance and velocity}\label{subsec:distance_velocity}

The distance to the centre of mass of the cluster depends not only on the subset of members that is used in its computation but also on the accuracy of the stellar masses, and hence on the recognition of double stars, as well as on the chosen method. With these caveats, the cluster centre of our 200 members that have a known radial velocity equals $\vec{b}_{\rm c} = (-43.94 \pm 0.66, 0.51 \pm 0.36, -17.37 \pm 0.30)$~pc in Galactic coordinates. The inverse of the mean parallax of these stars equals $46.09 \pm 0.73$~pc. For comparison, \citet{1998A&A...331...81P} reports a centre-of-mass distance of $46.34 \pm 0.27$~pc for 134 stars within 10~pc of the centre and $46.75 \pm 0.31$~pc for 180 stars within 20~pc of the centre based on {\it Hipparcos}-1 data, \citet{2009A&A...497..209V} reports $46.45 \pm 0.50$~pc based on {\it Hipparcos}-2 data, and \citet{2017A&A...601A..19G} report a weighted-mean distance for 103 probable member stars of $46.75 \pm 0.46$~pc based on TGAS data. We finally note that the possible systematic error of $\pm0.3$~mas in the TGAS parallaxes \citep[e.g.,][]{2016A&A...595A...2G} corresponds to $\pm0.7$~pc in distance.

There are 149 members with known radial velocity within 10~pc from the cluster centre. Their mean velocity, along with selected literature values, is listed in Table~\ref{tab:cluster_motion}.

\begin{table}
  \centering
  \caption{Mean cluster velocity $\vec{v}_{\rm c}$, in ${\rm km~s}^{-1}$, in equatorial coordinates for this study and various literature sources:
    (a) Sect.~\ref{subsec:distance_velocity}: TGAS / {\it Hipparcos}-2 data for 149 stars within 10~pc of the centre;
    (b) Sect.~\ref{subsec:modelling_results_velocity_distribution}: kinematic modelling of TGAS / {\it Hipparcos}-2 data;
    (c) \citet{1998A&A...331...81P}: {\it Hipparcos}-1 data for 134 stars within 10~pc of the centre;
    (d) \citet{2000A&A...356.1119L}: kinematic modelling of {\it Hipparcos}-1 data;
    (e) \citet{2007ASSL..350.....V}: {\it Hipparcos}-2 data;
    (f) \citet{2017A&A...601A..19G}: TGAS data.}
  \label{tab:cluster_motion}
  \begin{tabular}{clll}
    \hline
    Study & $v_x$ & $v_y$ & $v_z$ \\
    \hline
    (a) & $-6.20 \pm 0.11$ & $45.43 \pm 0.15$ & $5.38 \pm 0.08$ \\
    (b) & $-5.96 \pm 0.04$ & $45.60 \pm 0.07$ & $5.57 \pm 0.03$ \\
    (c) & $-6.28         $ & $45.19         $ & $5.31         $ \\
    (d) & $-5.90 \pm 0.13$ & $45.65 \pm 0.34$ & $5.56 \pm 0.10$ \\
    (e) & $-6.05 \pm 0.49$ & $44.57 \pm 1.15$ & $5.21 \pm 0.40$ \\
    (f) & $-6.03 \pm 0.08$ & $45.56 \pm 0.18$ & $5.57 \pm 0.06$ \\
    \hline
  \end{tabular}
\end{table}

\begin{figure}
  \begin{center}
    \includegraphics[width=0.8\columnwidth]{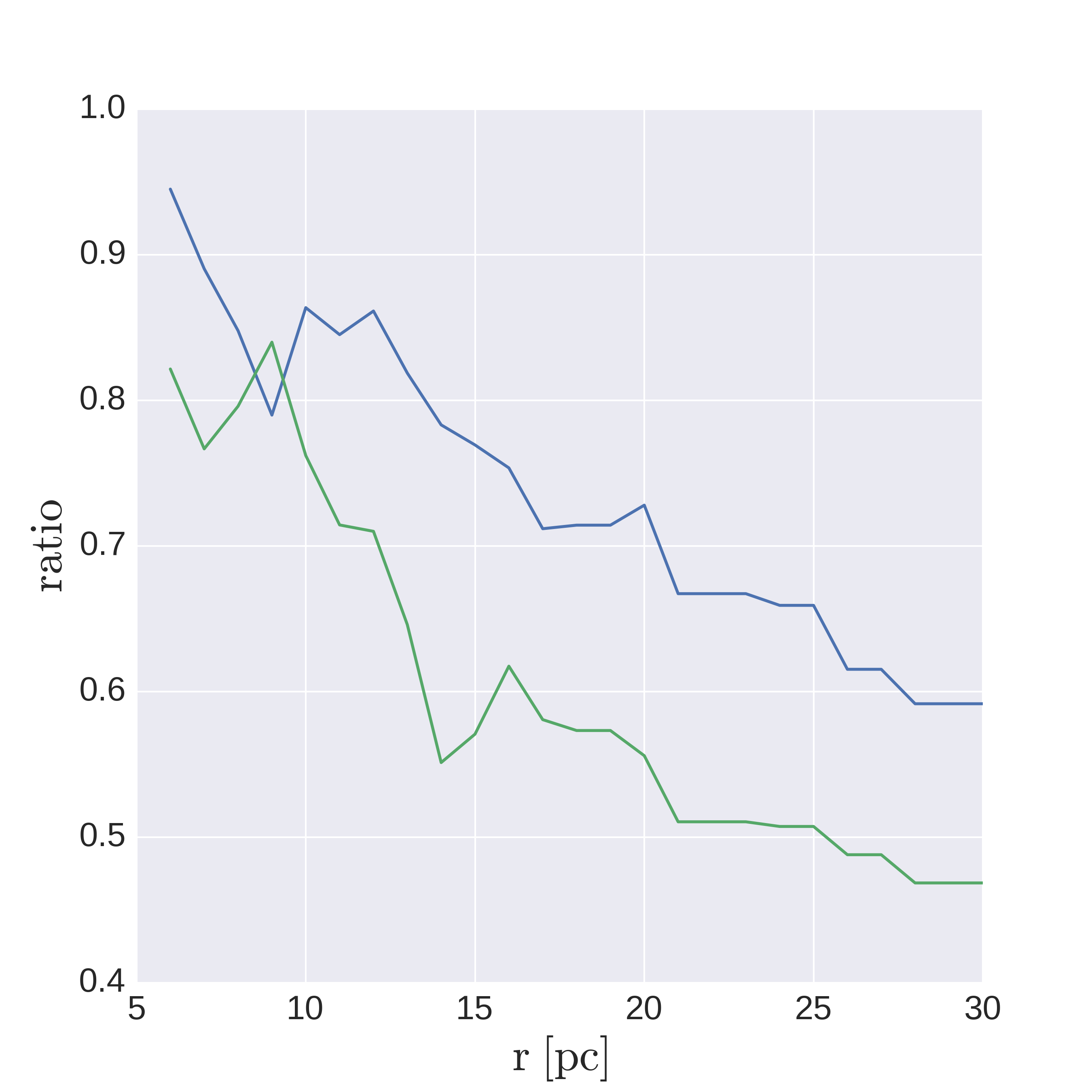}
    \caption{Shape of the cluster as function of the distance from the cluster centre. The blue curve denotes $b/a$ while the green curve denotes $c/a$.}
    \label{fig:ratios}
  \end{center}
\end{figure}

\subsection{Systematic astrometric errors}\label{subsec:systematic_errors}

In their validation assessment of the accuracy of the trigonometric TGAS parallaxes, \citet{2017A&A...599A..50A} compared them with the spectroscopic RAVE parallaxes from \citet{2014MNRAS.437..351B} for the $\sim200\,000$ stars in common between the two surveys. They identified a systematic difference of $\sim0.3$~mas in the parallaxes in a strip on the sky with ecliptic coordinates $\lambda \sim 180^{\circ}$ \citep[see Figure~28 in][]{2017A&A...599A..50A}. The Hyades centre lies at $\lambda \sim 67.5^{\circ}$ which -- although lying just outside the border of where (southern-hemisphere) RAVE data is available -- does not appear to be a badly affected region. Therefore, our conclusion is that the Hyades parallaxes are probably not particularly influenced by a  large  systematic parallax bias (but see Sect.~\ref{subsec:modelling_results_parallaxes}).  This is strengthened by the findings of \citet{2017A&A...601A..19G} who, using 2059 stars in common with the {\it Hipparcos}-2 Catalogue in an $18^{\circ}-$radius field centred on the Hyades, found a systematic parallax difference between the two data sets of only $0.14 \pm 0.03$~mas.

\section{Validation and discussion of membership results}\label{sec:membership_results}

\subsection{Comparison with earlier studies}

We compare our membership list with the results from \citet{1998A&A...331...81P} and \citet{2017A&A...601A..19G} in Sects.~\ref{subsubsec:P98} and \ref{subsubsec:Gaia_Collaboration}, respectively.

\subsubsection{Comparison with \citet{1998A&A...331...81P}}\label{subsubsec:P98}

Fifteen stars that were classified as Hyades members by \citet{1998A&A...331...81P} based on {\it Hipparcos}-1 data (column~x in their Table~2 equals '1') are not in our member list. Inspection of these cases shows no obvious reason why these stars are demoted using TGAS / {\it Hipparcos}-2 data, except for the new data having smaller uncertainties (than {\it Hipparcos}-1) and hence more discriminating power. Most of the demoted members are far from the cluster centre (e.g., 13 are beyond three core radii and 11 are beyond the tidal radius).

Compared to \citet{1998A&A...331...81P}, we find 18 new members with known radial velocity:
\begin{itemize}
\item HIP 24020 was classified by Perryman et al.\ as non-member (column~x equals '0' in their Table~2) with $c = 36.0$  based on {\it Hipparcos}-1 astrometry. We find $c = 3.5$ based on {\it Hipparcos}-2 astrometry. This difference is understood: HIP 24020 is a double star and the higher time-resolution treatment of the data in {\it Hipparcos}-2 \citep[see][]{2007ASSL..350.....V}, combined with the separation between the components, resulted in more accurate astrometry for this star.   
\item HIP 26159 was classified by Perryman et al.\ as unclassifiable (column~x equals '?' in their Table~2) with $c = 10.82$, meaning that its proper motion was consistent with membership but that a radial velocity was lacking as confirmation. With a radial velocity now available from \citet{2013Obs...133..144G}, we find $c = 7.8$ and hence confirm membership.
\item HIP 21760 does not appear in \citet{1998A&A...331...81P}, for unknown reasons.
\item The remaining 15 stars are new members by virtue of their new TGAS astrometry and literature radial velocities.
\end{itemize}

Of our 51 candidate members without a known radial velocity, only one is discussed in \citet{1998A&A...331...81P}; this is HIP 21092 which has '?' in column~x in their Table~2. The remaining 50 stars are new candidate members by virtue of their new TGAS astrometry.

\citet{1998A&A...331...81P} marked 21 stars as possible but unclassifiable member candidates (column~x equals '?' in their Table~2). Only two of these are members in our list (HIP 26159 and HIP 21092).

\subsubsection{Comparison with \citet{2017A&A...601A..19G}}\label{subsubsec:Gaia_Collaboration}

All 103 Hyades members identified by \citet{2017A&A...601A..19G} based on TGAS astrometry are in our member list. We identify 70 more TGAS members, mainly but not exclusively with lower probabilities and/or in the outer regions of the cluster (cf.\ Figure~\ref{fig:70}).

\subsection{Stellar density of the galactic field}

\begin{figure*}
  \begin{center}
    \includegraphics[width=0.9\textwidth]{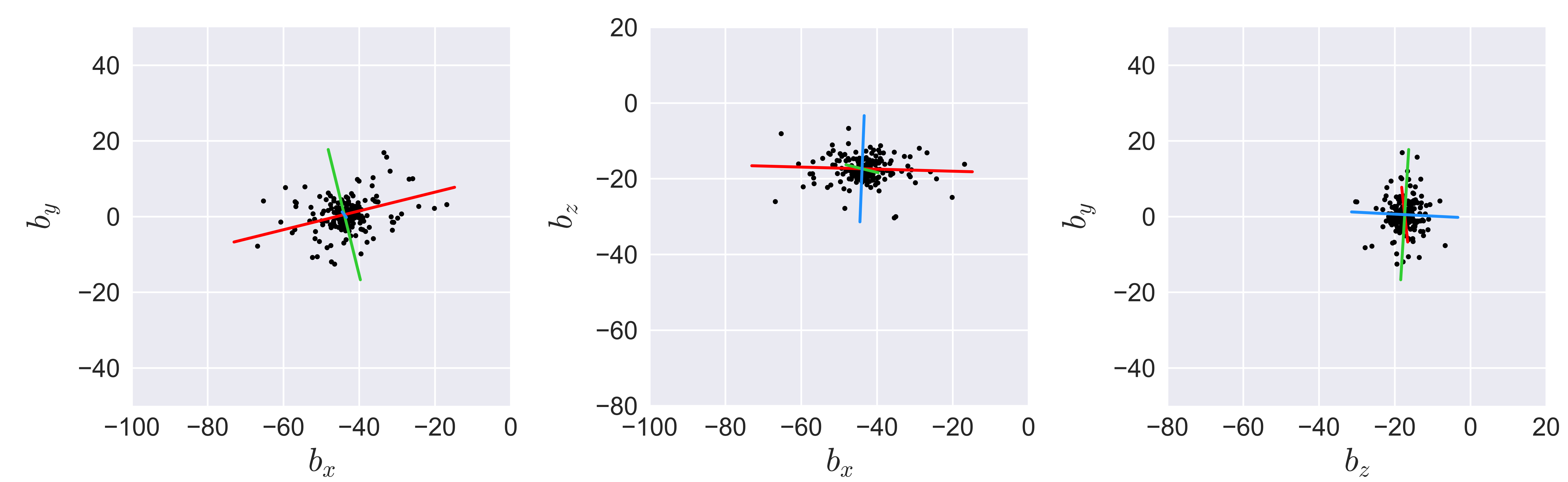}
    \caption{Projected positions (in pc) in Galactic coordinates of the candidate members with known radial velocity within 30~pc of the cluster centre. The major axis is shown in red, the intermediate axis in green, and the minor axis in blue. The Hyades are clearly flattened along the Galactic plane.}
    \label{fig:positions}
  \end{center}
\end{figure*}

As a consistency check of our membership selection, we establish the stellar density of the galactic field inside the Hyades and compare it to its surroundings.

First, as a visual check, we  sample the full sky on a 1-degree grid and, on each grid point, place a 10-pc-radius sphere into three-dimensional configuration space at a distance of 46~pc.  These numbers correspond to the distance and tidal radius of the Hyades \citep{1998A&A...331...81P}. We then count the number of TGAS stars in that sphere and colour code the sky (at that fixed, mean Hyades distance) accordingly. Figure~\ref{fig:47h} shows the result. The Hyades cluster is clearly seen as an overdense region at about $60^\circ$ in right ascension and $15^\circ$ in declination. Figure~\ref{fig:47} shows the same map after removing just our 173 TGAS Hyades members from the 2-milllion-star TGAS catalogue. The overdense Hyades region has disappeared and only the galactic-field density of $\sim$40 stars within a 10-pc-radius sphere remains.

Second, as a more quantitative check, we construct the volume density of the TGAS data as seen from the centre of the Hyades, as function of distance from the cluster centre, and repeat this exercise after removing our 173 TGAS Hyades members from the catalogue. The results, as shown in Figure~\ref{fig:volume_density}, confirm that the field-star distribution becomes flat after removal of our Hyades members, at a value of $\sim$0.1~objects~pc$^{-3}$, corresponding to the earlier-mentioned $\sim$40 stars within a 10-pc-radius.

 Both assessments confirm  that membership selection has been performed correctly: there is no overdensity left at the Hyades location, meaning the vast majority of the real members have been included in our list of members; at the same time, there is no underdensity created at the Hyades location, meaning our list of members (at least in the central 10~pc) contains few field stars. 

\subsection{Colour-absolute magnitude diagram}\label{subsec:cmd_obs}

Figure~\ref{fig:CMD} shows the colour-magnitude diagram of our 251 candidate members. Not surprisingly, it shows a large spread around the main sequence due to the resolved depth of the cluster along the line of sight. Replacing the observed magnitudes with their absolute counterparts, based on individual parallaxes, should provide a well-defined main sequence. Since our membership is based on kinematics only, the colour-absolute magnitude diagram of our candidate members can actually be used as consistency check. Figures~\ref{fig:HR_bv} and \ref{fig:HRr_bv} show the colour-absolute magnitude diagram of our 251 members, colour coded with membership statistic $c$ and distance $r$ to the cluster centre, respectively. Details on the Johnson photometry are provided in Appendix~\ref{sec:Johnson}. To guide the eye, an empirical main sequence \citep{2012BASI...40..487S} and its associated equal-mass-binary sequence have been added.

The members trace out a smooth main sequence over $\sim$9 magnitudes and actually also follow a distinct binary sequence. There are four (known) giants among the members \citep[e.g.,][]{2001A&A...367..111D}. The turn-off region looks slightly messy but this is a known feature explained by rotation, binarity, a blue straggler caused by magnetic mixing, etc. \citep[e.g.,][]{2001A&A...367..111D}. The colour coding in Figures~\ref{fig:HR_bv} and \ref{fig:HRr_bv} show that high-$c$ (lower-fidelity) members follow the main sequence equally well as low-$c$ (high-fidelity) members as well as that members without known radial velocity follow the main sequence equally well as members with known radial velocities. The large photometric $(B-V)$ errors at the faint end prevent making strong conclusions on objects appearing below the main sequence being unrelated field stars. All in all, our membership provides a plausible colour-absolute magnitude diagram.

\subsection{Spatial distribution of members}\label{subsec:spatial_distribution}

Figures~\ref{fig:sub} and \ref{fig:subr} show the distributions of various observables of the 200 members with known radial velocities, colour coded with membership statistic $c$ and distance $r$ to the cluster centre, respectively. In all cases, the cluster shows up with a clear, dense core \citep[with a core radius of 2.7~pc;][]{1998A&A...331...81P} and a significant spread of members up to a few tidal radii \citep[the tidal radius is 10~pc;][]{1998A&A...331...81P}. The features and correlations seen in the panels reflect (1) projection effects over the significant extent of the cluster on the sky, (2) the fact that all members share a common, three-dimensional space motion (such that, for instance, a large parallax implies a large proper motion), and (3) intrinsic structure of the cluster. The latter is clear from the projections of the three principal axes of the cluster shown in Figures~\ref{fig:sub} and \ref{fig:subr}, which were derived as follows.

The shape of the cluster can be determined by calculating the moment-of-inertia matrix $\vec{I}$, which has the components:
\begin{equation}\label{eq:MoI}
I_{jk} = \sum_{i=1}^{N} (m_i x_{ij} x_{ik}),
\end{equation}
where the index $i$ is over the $i = 1, \ldots, N$ stars. The three Cartesian coordinates $x_{ij}$ for star $i$ ($j=1,2,3$) are given in the Galactic coordinate system with the origin shifted to the cluster centre. As in Sect.~\ref{subsec:cluster_centre}, we assume all $m_i = 1~{\rm M}_\odot$  (essentially 'reducing' the moment-of-inertia matrix to a three-dimensional distribution matrix of the stars). The moment-of-inertia matrix $\vec{I}$ can be made a function of the distance $r$ from cluster centre by limiting the summation in Eq.~(\ref{eq:MoI}) to stars with $r_i \leq r$.

The principal axes of the cluster can be determined as the eigenvectors of the moment-of-inertia matrix. The lengths of the semi-axes $\rm a \geqslant b \geqslant c$ correspond to the square-roots of the eigenvalues of the matrix. Taking $r = 30$~pc as limit and including only those of the 251 candidate members that have a known radial velocity, we find the eigenvectors of the moment-of-inertia matrix to be:
\begin{equation}
\left( \begin{array}{rrr}
 0.9702 &  0.2390 &  0.0399\\
 0.2409 & -0.9692 & -0.0511\\
-0.0264 & -0.0592 &  0.9979 \end{array} \right).
\end{equation}
The corresponding eigenvalues are $(I_1, I_2, I_3) = (8952, 3134, 1966)~{\rm M}_\odot$~pc$^2$, such that the ratios of the semi-axes equal $b/a = 0.59$ and $c/a = 0.47$. Figure~\ref{fig:ratios} shows how the axes ratios vary with distance from the cluster centre.  We find that the cluster shape becomes more spherical towards the core, consistent with the findings of, e.g., \citet{1998A&A...331...81P} and \citet{2011A&A...531A..92R}.

Figure~\ref{fig:positions} shows the projected positions in Galactic coordinates of the 191 candidate members with known radial velocity that lie within $r \leq 30$~pc of the cluster centre along with the principal axes of the cluster. We confirm the findings of \citet{1998A&A...331...81P} and \citet{2011A&A...531A..92R} that, as a result of Galactic tidal forces, the major axis of the cluster is almost aligned with the Galactic $x$ axis while the cluster is flattened in the direction of the Galactic $z$ axis \citep[see also][]{1979A&A....78..312O}.

The presence of dozens of, preferentially low-mass, members in an extended 'halo' beyond the tidal radius ($\sim$10~pc), where stars are subject to velocity perturbations from the Galactic tidal field, was already noted by \citet{1975A&A....43..423P} and is commonly seen in N-body simulations (see Sections 7 and 8 in \citealt{1998A&A...331...81P}). Stars escape only through the Lagrangian opening points of the equipotential surface which, for the Hyades, are on the Galactic $x$-axis. This naturally explains the prolate shape of the cluster. The evaporating stars, which can 'stick around' for hundreds of millions of years, are expected to show systematically deviating motions, dependent on their distance to the cluster centre and on the time elapsed since escape. Modelling this process reliably is far beyond the scope of this work yet possible \citep[see, e.g.,][]{2017MNRAS.466.3937C,2017MNRAS.468.1453D} but requires careful considerations of various issues such as mass segregation, binarity, systematic velocity patterns, Galactic and cluster potentials, etc.

\subsection{Velocity distribution of members}\label{subsec:velocity_distribution}

Obtaining a reliable estimate of the velocity dispersion is non-trivial, not the least because velocity dispersion is likely to vary with radius from the centre as a result of mass segregation in the centre \citep[e.g.,][]{2001ASPC..228..506M}, escaping stars in the outer regions \citep[for a detailed discussion, see][their Section 7]{1998A&A...331...81P}, etc. Perryman et al.\ estimated the dispersion of the cluster, assuming it is homogeneous and isotropic, by realising that the $c$ values of the members should (approximately) follow a $\chi^2$ distribution (with 3 and 2 degrees of freedom for objects with and without known radial velocity, respectively). By adding an ever increasing value of the velocity dispersion ($\sigma_v$) to the covariance matrix $\vec{C}$ (Eq.~\ref{eq:chi^2}) and comparing the resulting distribution of $c$ values with a $\chi^2$ distribution, they derived $\sigma_v \sim 0.3~ {\rm km~s^{-1}}$ as best-fit value. \citet{1998A&A...331...81P} limited themselves to a high-precision subset of members containing 38 stars for which there was no indication of multiplicity, whose radial velocities were determined by \citet{1988AJ.....96..172G}, and whose standard errors on the {\it Hipparcos} parallax and proper motions were less than 2~mas and 2~mas~$\rm yr^{-1}$, respectively. \citet{1988AJ.....96..172G} noted that 0.2--0.4~km~s$^{-1}$ is also the expected value for the one-dimensional velocity dispersion for a Hyades-like cluster from theoretical as well as $N$-body considerations.

\begin{figure}
  \begin{center}
    \includegraphics[width=0.9\columnwidth]{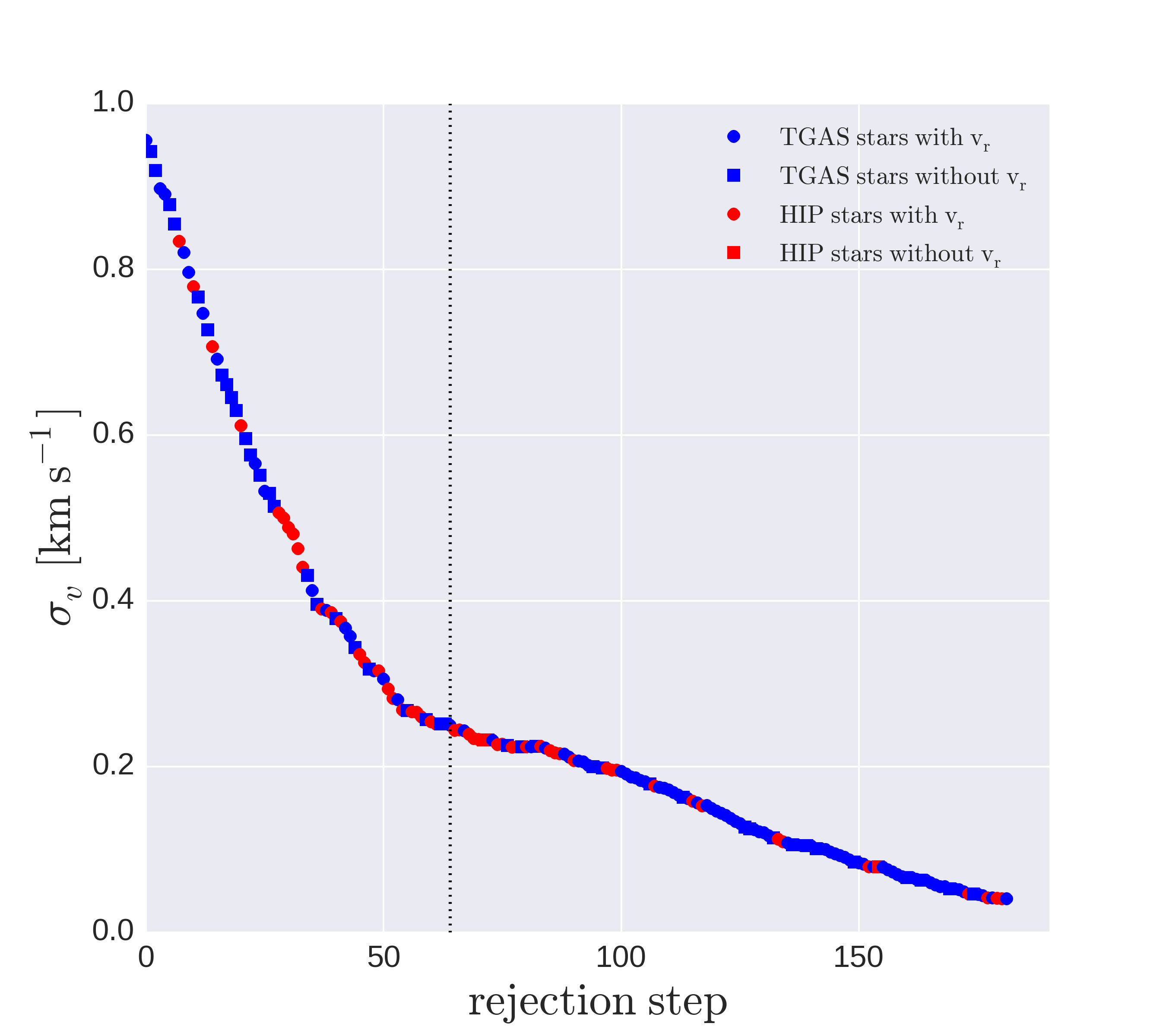}
    \caption{Maximum-likelihood velocity dispersion as a function of rejection step. Blue symbols denote TGAS members; red symbols denote {\it Hipparcos}-2 members. Circles denote objects with known radial velocity; squares denote objects without known radial velocity. The dotted vertical line shows the end point of our iterations for our default stopping criterion, associated with a significance level $p_{\rm lim} = 0.0027$.}
    \label{fig:sigmav}
  \end{center}
\end{figure}

\begin{figure}
  \begin{center}
    \includegraphics[width=0.9\columnwidth]{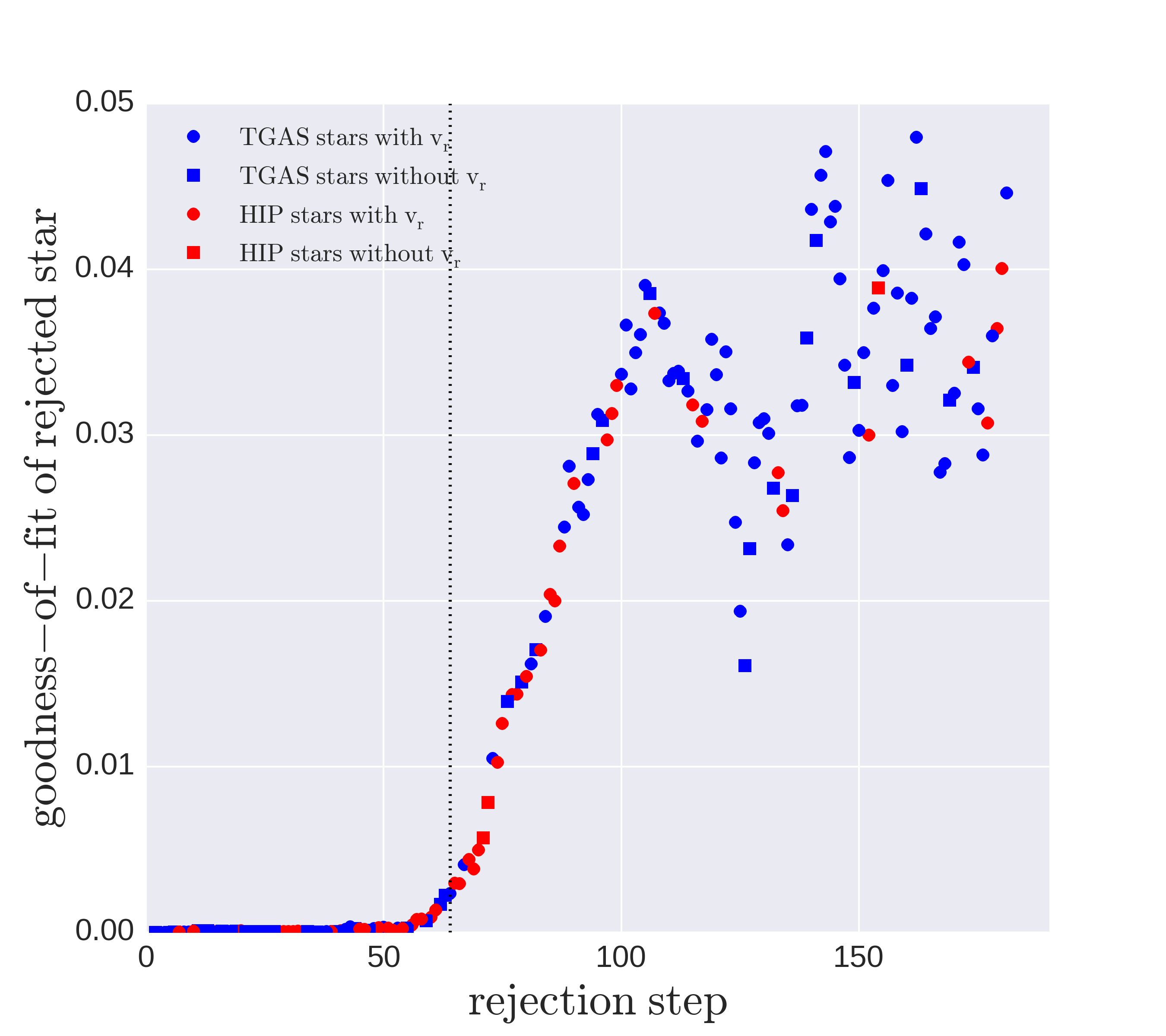}
    \caption{Goodness-of-fit statistic ($p$ value) of the rejected star (the worst outlier at each iteration step) as a function of the rejection step. Blue symbols denote TGAS members; red symbols denote {\it Hipparcos}-2 members. Circles denote objects with known radial velocity; squares denote objects without known radial velocity. The dotted vertical line shows the end point of our iterations for our default stopping criterion, associated with a significance level $p_{\rm lim} = 0.0027$.}
    \label{fig:g}
  \end{center}
\end{figure}

When we repeat this exercise with our 200 members with known radial velocity, we find $\sigma_v \sim 0.6~ {\rm km~s}^{-1}$. This, clearly, is larger than the Perryman et al.\ value  (yet in good agreement with the 0.58~km~s$^{-1}$ derived by \citealt{2017A&A...601A..19G}). In fact, our estimate should be considered as an upper limit (and we actually derive a lower limit $\sigma_v \gtrsim 0.25~ {\rm km~s}^{-1}$ in Sect.~\ref{subsec:modelling_results_velocity_dispersion}) for various reasons:
\begin{itemize}
\item We have not excluded (known) multiple stars;
\item We have not excluded the subset of objects with the largest astrometric standard errors (which are likely the objects which have {\it Tycho-2} positions for the 1991.25 epoch in TGAS);
\item We have an inhomogeneous set of literature radial velocities in our sample with, in particular, inflated standard errors to wash out the effect of different radial velocity zero-points (see Appendix~\ref{sec:RV}).
\end{itemize}
In addition, there is the complicating factor that \citet{2000A&A...356.1119L} found that, for a simulated Hyades cluster, only a scaled version of $c$, namely $c_{\rm scaled} = \beta~ c$ (with $\beta \sim 0.66$ for the Hyades), follows a $\chi^2$ distribution. For this value of $\beta$, we find $\sigma_v \sim 0.5~ {\rm km~s}^{-1}$. When letting $\beta$ a free parameter, we find $\sigma_v \sim 0.13~ {\rm km~s}^{-1}$ with $\beta = 0.25$ when using a Kolmogorov-Smirnov test and $\sigma_v \sim 0.27~ {\rm km~s}^{-1}$ with $\beta = 0.43$ using an Anderson-Darling test \citep[see, e.g.,][]{2012msma.book.....F}.
Finally, there is the complicating factor that the $c$ values are affected by the estimates of the standard errors, both on the astrometry and on the radial velocities (the standard errors are part of $\vec{C}$ in Eq.~\ref{eq:chi^2}). In theory, therefore, the $c$ distribution of the members could be used to determine the reliability of the TGAS (and radial velocity) standard errors \citep[note in particular that the TGAS standard errors have been claimed to be systematically wrongly estimated; for a review, see Sect.~~2 in][]{2017arXiv170901216B}. In practice, however, any attempts in this direction lead to inconclusive results.

\begin{figure*}
  \begin{center}
    \includegraphics[width=\textwidth]{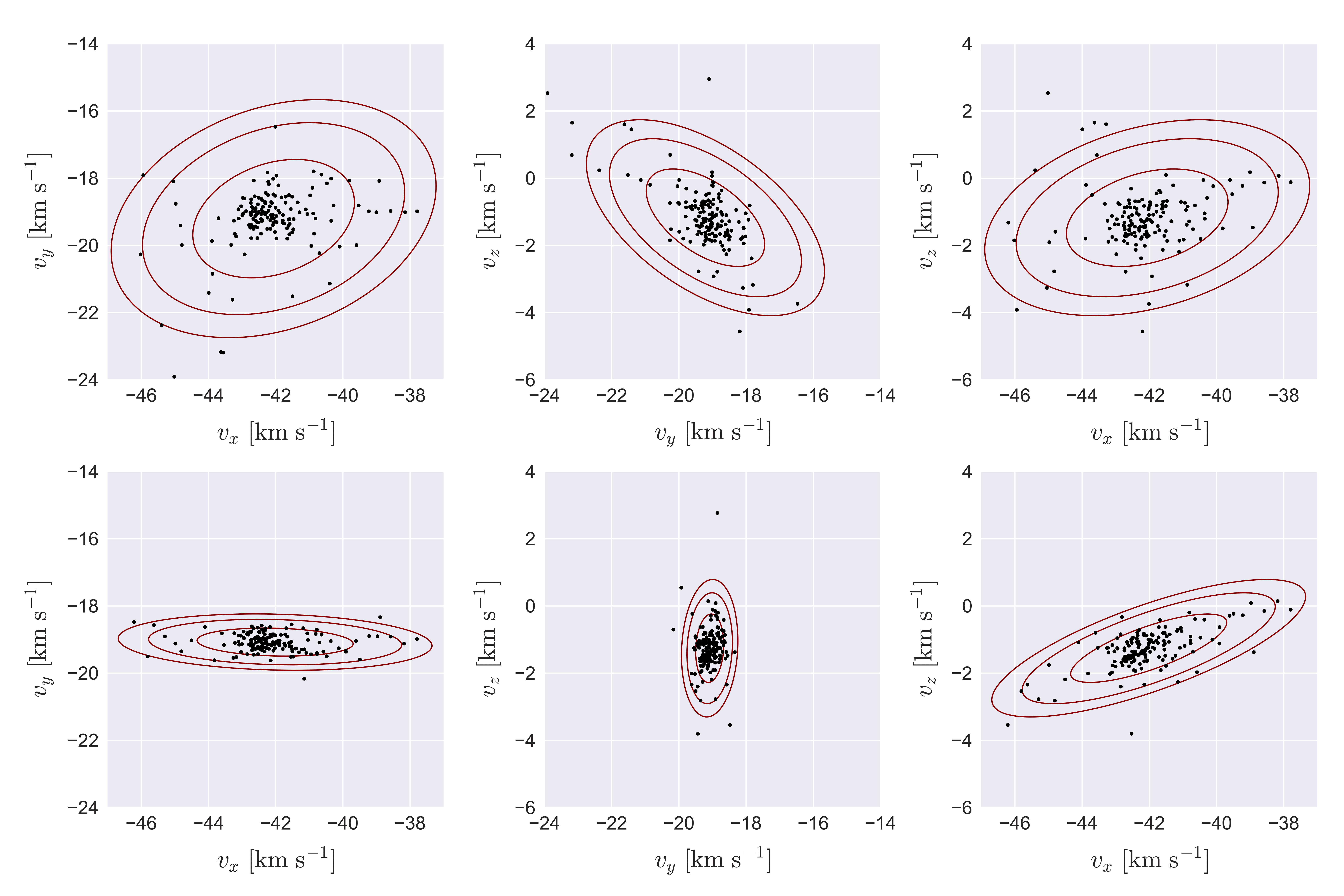}
    \caption{Three-dimensional velocity distribution, in Galactic coordinates, of the Hyades, using trigonometric (top panels) and kinematically-modelled parallaxes (bottom panels) for the 187 objects that survived the kinematic modelling (Sect.~\ref{sec:modelling_results}). The contours show 68.27, 95.45, and 99.73\% confidence levels. The $v_x$ component does not change drastically since it depends primarily on radial velocity, which is independent from parallax. On the contrary, the $v_y$ component improves significantly since it depends primarily on proper motion such that improved parallaxes reduce the spread in $v_y$ velocities.}
    \label{fig:velocities}
  \end{center}
\end{figure*}

\begin{figure}
  \begin{center}
    \includegraphics[width=\columnwidth]{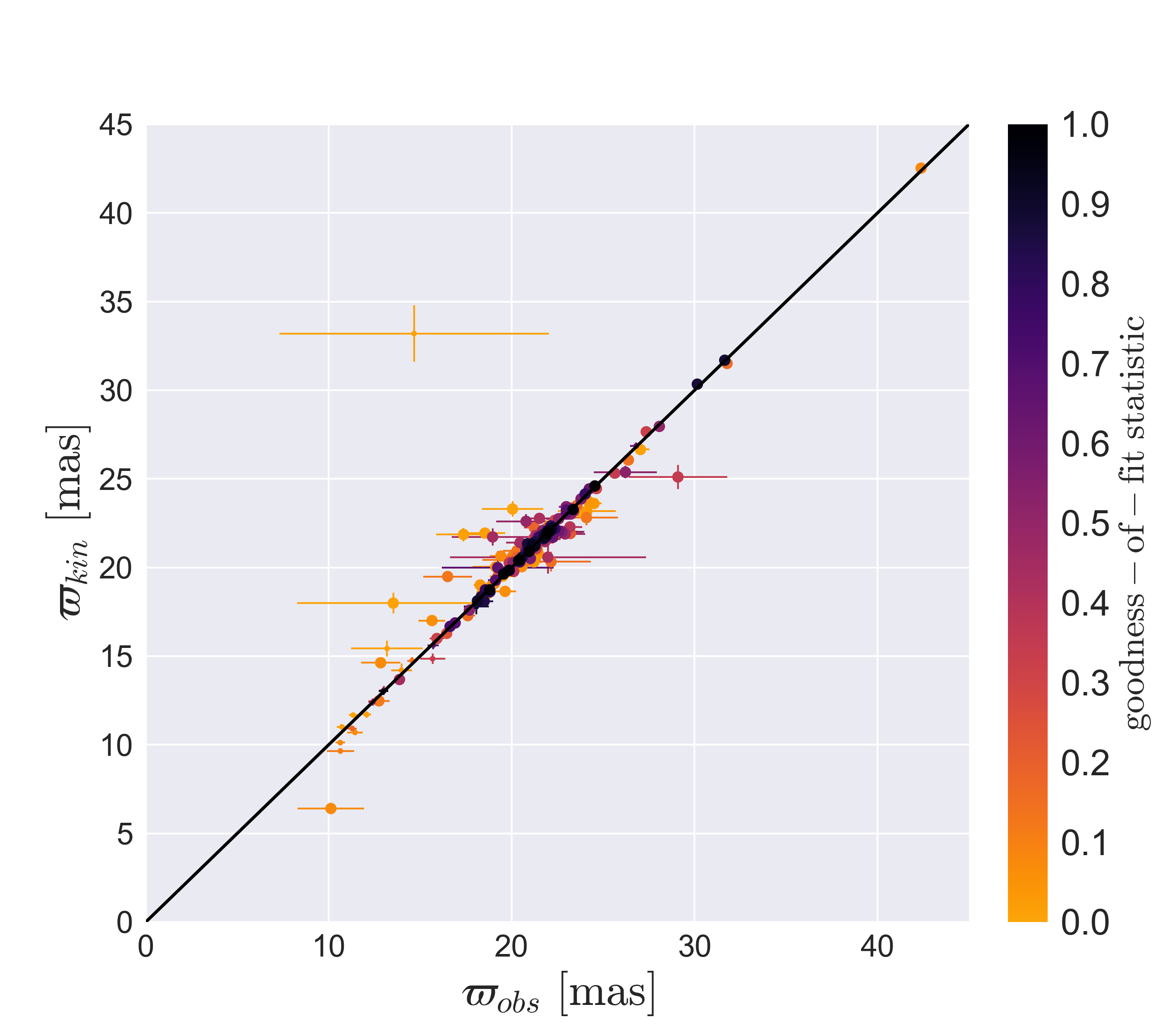}
    \caption{Observed, trigonometric parallaxes (from either TGAS or {\it Hipparcos}-2) versus the kinematically-modelled parallaxes (for 187 objects). The colour of the markerpoints reflects the goodness-of-fit (or $p$ value) of the stars. The black solid line shows the 1:1 relation. Stars with known radial velocity are indicated with a large markerpoint.  Outliers are discussed in the text; objects with large error bars are mostly double stars.}
    \label{fig:parallax1_1}
  \end{center}
\end{figure}

\section{Kinematic modelling}\label{sec:modelling}

We now use an iterative maximum-likelihood method to derive improved parallaxes for individual stars from their proper motions (and radial velocities) by kinematically modelling the cluster. This method is outlined in Sect.~\ref{subsec:modelling_overview}. Section~\ref{subsec:modelling_initial_conditions} and \ref{subsec:modelling_stopping_criterion} describe tests to investigate the impact of the initial conditions and stopping criterion used in the optimisation and the iterations. The results of the method after application to the Hyades members are described in Sect.~\ref{sec:modelling_results}.
      
\subsection{Overview}\label{subsec:modelling_overview}

 Our method builds upon yet extends  the maximum-likelihood method developed by \citet{2000A&A...356.1119L}. This method is meant to be applied to cluster members and these are assumed to share the same, three-dimensional space motion apart from a small, random dispersion term. The maximum-likelihood  approach essentially is a rejuvenated 'moving cluster' method \citep[e.g.,][ itself comparable to the direct, 'reduced proper motion' approach used in, e.g., \citealt{2007ASSL..350.....V} and \citealt{2017A&A...601A..19G}]{1952BAN....11..385V,1998gaas.book.....B} and simultaneously determines the individual, kinematically-modelled parallaxes of the stars ($\varpi_{\rm kin}$), the mean cluster velocity (${\vec v}_{\rm c}$), and the velocity dispersion ($\sigma_v$) from the observational data while using an iterative outlier exclusion procedure to remove deviating objects, for instance caused by (unrecognised) binarity.

Whereas \citet{2000A&A...356.1119L}, being interested in deriving astrometric radial velocities  \citep[see also][]{1999A&A...348.1040D}, formulated their model in terms of proper motions as main observables, we generalised this to include  measured, spectroscopic  radial velocities:
\begin{enumerate}
\item We added radial velocity, whenever available, as fourth observable, besides trigonometric parallax and proper motion;
\item We made a transition from the $\chi^2$ statistic used in \citet{2000A&A...356.1119L}, and denoted $g$, to a $p$ value or $1 - {\rm CDF}(g, {\rm DOF})$ as a goodness-of-fit statistic (the $p$ value is the area under the $\chi^2$ probability density function to the right of the observed test statistic and signifies the probability that the star has a $\chi^2$ statistic exceeding $g$ under the assumption that it is a member star). This is required since $g$ values are approximately distributed as a $\chi^2$-distribution with two degrees of freedom when the radial velocity is unknown and three degrees of freedom when the radial velocity is known; 
\item We used a mixed three- and four-dimensional likelihood function so that both stars with and without known radial velocity can be treated simultaneously.
\end{enumerate}
Details on these  extensions  of the method are given in Appendix~\ref{sec:ML}.

The maximum-likelihood method requires that the kinematic model is a statistically correct description of the data. In particular, we assume, following \citet{2000A&A...356.1119L}, that the (Hyades) cluster has no net expansion (or contraction) and that it does not rotate. As shown in Appendix~\ref{sec:expansion_rotation}, these assumptions are not violated by the TGAS / {\it Hipparcos} data since, for the subset of 149 members within 10~pc of the cluster centre, the net expansion equals $0.08 \pm 0.12$~km~s$^{-1}$ while the net rotation equals $0.46 \pm 0.28$~km~s$^{-1}$.

The method is best applied only to the actual members of the cluster as identified in Sect.~\ref{sec:membership}. This set still contains 'outliers', meaning members which have (slightly) discrepant astrometry (and/or radial velocities) as a result of unrecognised multiplicity, them escaping from the cluster, etc. Such outliers can be found, after maximising the likelihood function, by computing the $p$ value (associated with a particular $g$ value) for each star in the solution \citep[Eq.~(19) in][]{2000A&A...356.1119L}. The largest outlier is removed from the sample and a new maximum likelihood solution is determined, until all $g$ values are acceptably small ($g_i \leq g_{\rm lim}$ or $p_i \geq p_{\rm lim}$). The stopping criterion is further discussed in Sect.~\ref{subsec:modelling_stopping_criterion}.

We developed a new implementation of the method, with the  extensions  as described above, in {\it Python}\footnote{\url{https://github.com/eleonorazari/KinematicModelling}}. The method of choice for performing the multi-dimensional optimisation is the Newton Conjugate Gradient method from the {\it Python} package {\it SciPy} \citep{SciPy}. This method is generally fast since it makes use of the gradient and the Hessian of the objective function. The Newton Conjugate Gradient method was successfully tested against the much slower Nelder-Mead simplex method. The sensitivity of the Newton Conjugate Gradient method to the initial guesses is assessed in Sect.~\ref{subsec:modelling_initial_conditions}. The best-fit parameters come with associated covariance data (i.e., standard errors in particular) derived directly from the Hessian matrix.

\begin{figure}
  \begin{center}
    \includegraphics[width=\columnwidth]{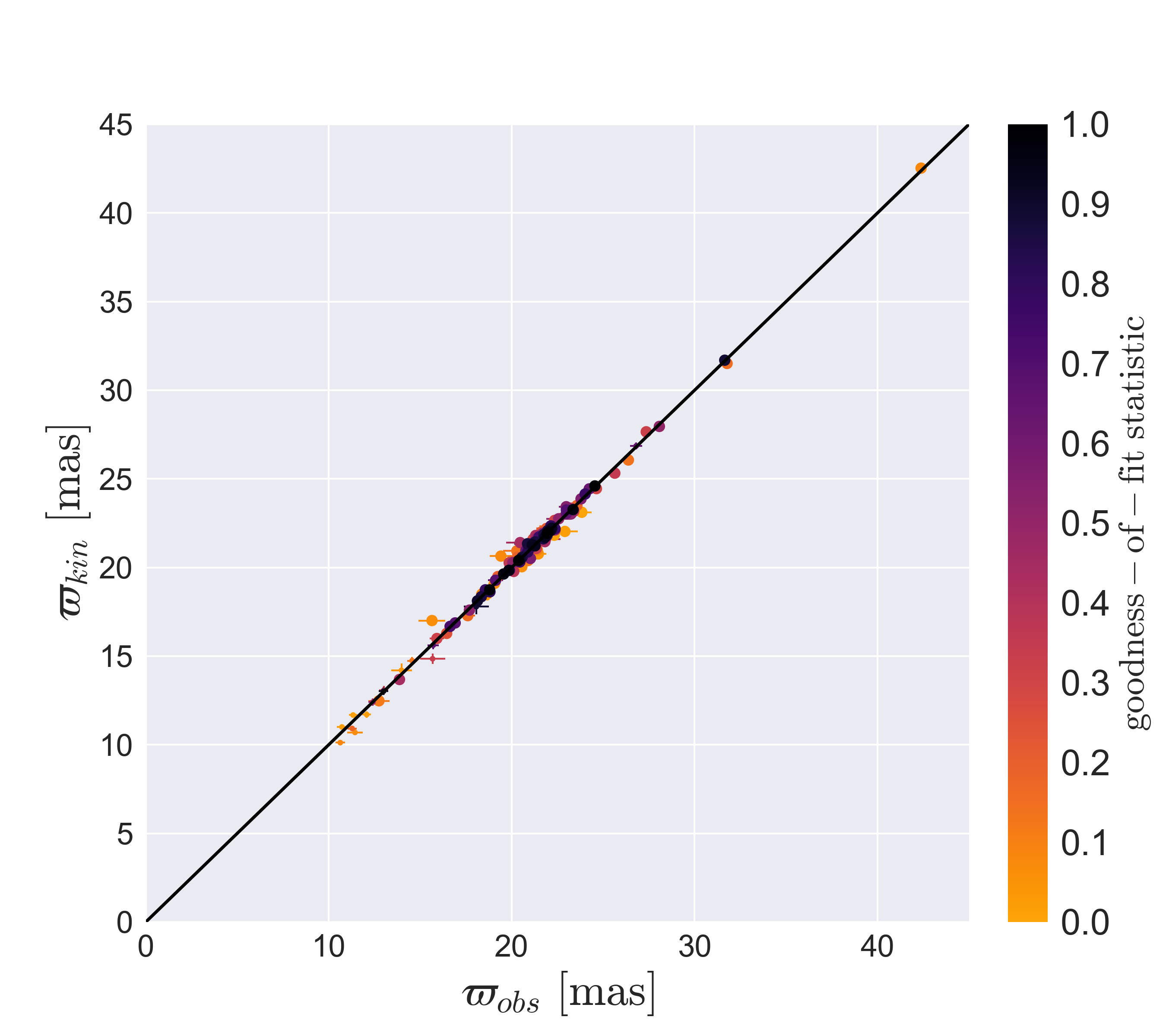}
    \caption{As Figure~\ref{fig:parallax1_1}, but showing TGAS stars only.  Cf.\ Figure~11 in \citet{2017A&A...601A..19G}.}
    \label{fig:parallax1_1_TGAS}
  \end{center}
\end{figure}

\subsection{Initial conditions}\label{subsec:modelling_initial_conditions}

\begin{figure}
  \begin{center}
    \includegraphics[width=\columnwidth]{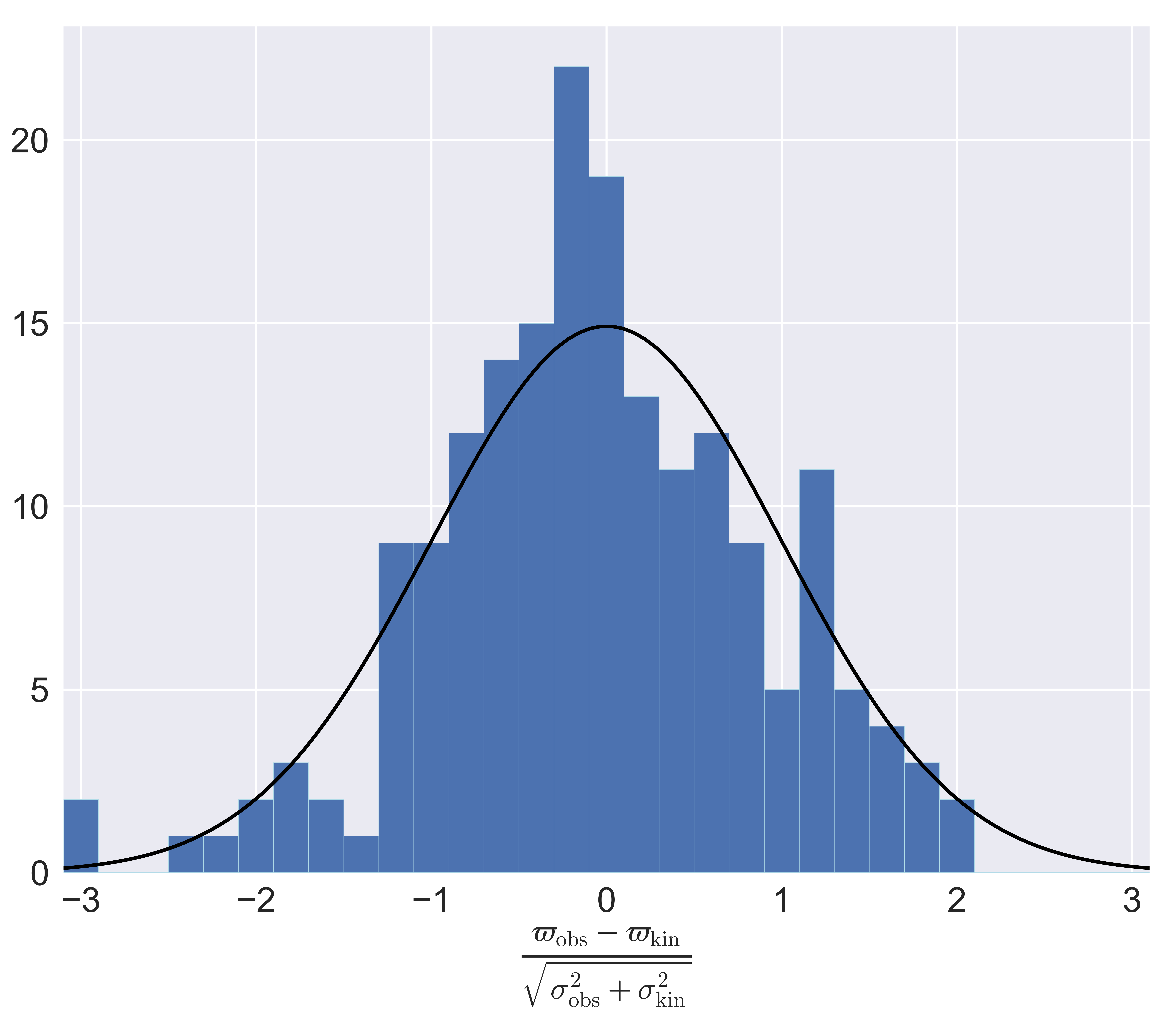}
    \caption{Histogram of the error-normalised difference between the observed and kinematically-modelled parallaxes. The mean and median of the distribution are $-0.06$ and $-0.09$, respectively. The curve represents a unit-variance, zero-centred Gaussian and is not a best fit to the data.}
    \label{fig:histogram}
  \end{center}
\end{figure}

\begin{figure*}
  \begin{center}
    \includegraphics[width=0.8\textwidth]{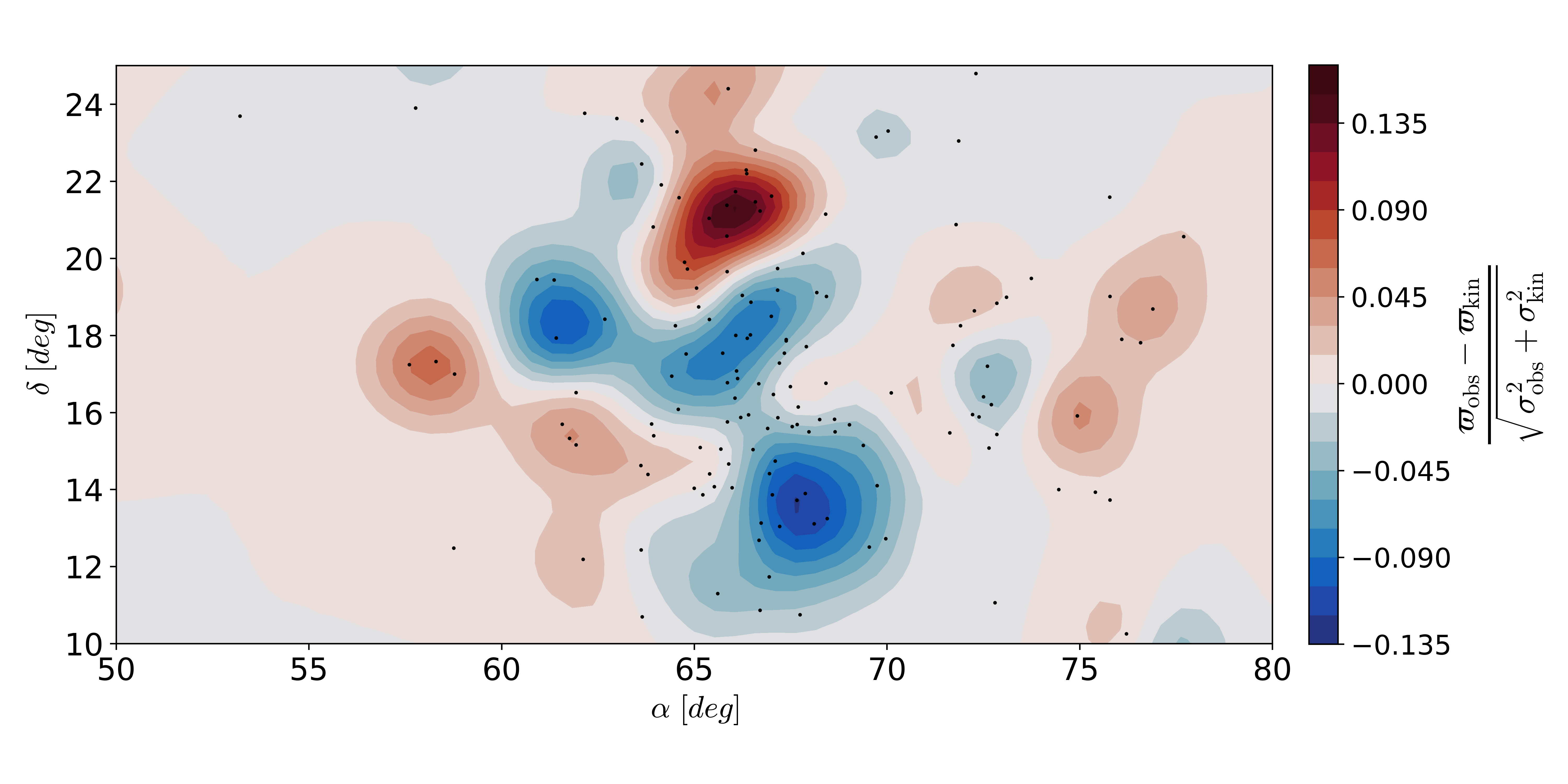}
    \caption{Sky map, for the central region of the Hyades, of the error-normalised difference between the observed and kinematically-modelled parallaxes, after smoothing the signal of each star with a two-dimensional Gaussian with $\sigma = 1^\circ$. Blue / red areas indicate regions in which the TGAS / {\it Hipparcos}-2 parallaxes are systematically smaller / larger than the kinematically-modelled parallaxes. As a result of the smoothing, which has been applied to visually bring out patterns, the contour levels themselves are not directly interpretable.}
    \label{fig:correlations}
  \end{center}
\end{figure*}

To verify the absence of a dependency of the kinematic modelling results on the initial conditions of the optimisation code, we performed 1000 runs with the initial guess for the mean cluster velocity components $v_x$, $v_y$, and $v_z$ drawn randomly from a Gaussian distribution centred on  the mean cluster velocity  with $\sigma = 1\ \rm km\ s^{-1}$; this uncertainty is conservative since it is a factor $\sim$5--10 larger than the standard errors of the mean cluster velocity (components) derived in Sect.~\ref{subsec:cluster_velocity}. We start in all cases with the 251 candidate members identified in this work and iterations consistently stop with 187 members left (see Sect.~\ref{subsec:modelling_stopping_criterion}). Whereas the final kinematically-modelled parallax varies from run to run for a given star, this variation (expressed as a standard deviation computed over the 1000 runs) is more than a few dozen times smaller than the standard error of the kinematically-modelled parallax itself. The only exception is HIP 21092, where this factor is 4.6 (see Sect.~\ref{subsec:modelling_results_parallaxes} for details on this peculiar star). In short, the method is robust to the choice of initial conditions.

\subsection{Stopping criterion}\label{subsec:modelling_stopping_criterion}

There is no perfect stopping criterion in the iterative process of removing outliers. If one stops too early, real outliers will be left and the best-fit velocity dispersion will remain too high. On the contrary, one can keep on iterating and removing outliers until just two stars with very similar three-dimensional motions are left, severely underestimating the velocity dispersion. After some experimenting, we settled for a stopping criterion associated with a significance level $p_{\rm lim} = 0.0027$. Starting the iterations with the 251 candidate members identified in Sect.~\ref{sec:membership}, the algorithm removes 64 outliers, leaving 187 stars. As already mentioned earlier, the rejected outliers are not necessarily non-members. Outliers are simply stars which have observables that do not perfectly fit the common space motion of the cluster. They might, however, very well be members with perturbed astrometry due to unrecognised multiplicity, escaping members with slightly deviating motions, members with wrongly estimated astrometric standard errors, etc.

Figures~\ref{fig:sigmav} and \ref{fig:g} show what happens to the velocity dispersion and the goodness-of-fit statistic ($p$ value) of the rejected star (the worst outlier) during the iterations and also show what would happen if one would keep on rejecting outliers:
\begin{itemize}
\item Figure~\ref{fig:sigmav} shows that the best-fit velocity dispersion monotonically decreases with rejection step, which is not surprising. The slope of the decline changes with rejection step, first around step $\sim$40 and second around step $\sim$55. The physical interpretation of these slope changes is unclear: we have searched for correlations of the rejected stars (e.g., are they primarily in the outer regions, are they primarily without known radial velocity, are they primarily bright, etc.) but found no clues.
\item Figure~\ref{fig:g} shows how the goodness-of-fit of the rejected star, i.e., the worst outlier, varies with rejection step. Whereas the goodness-of-fit remains small up to around step $\sim$60, suggesting the rejection affects uncertain / deviating candidate members, it starts to rise rapidly thereafter, suggesting the rejected stars evolve into well-matching members. The oscillation and noisy behaviour starting around step $\sim$100 is explained by the interplay with the velocity dispersion: the decreasing velocity dispersion results in decreased $p$ values since fewer stars fit the mean cluster motion given the allowed error.
\end{itemize}

\begin{figure}
  \begin{center}
    \includegraphics[width=\columnwidth]{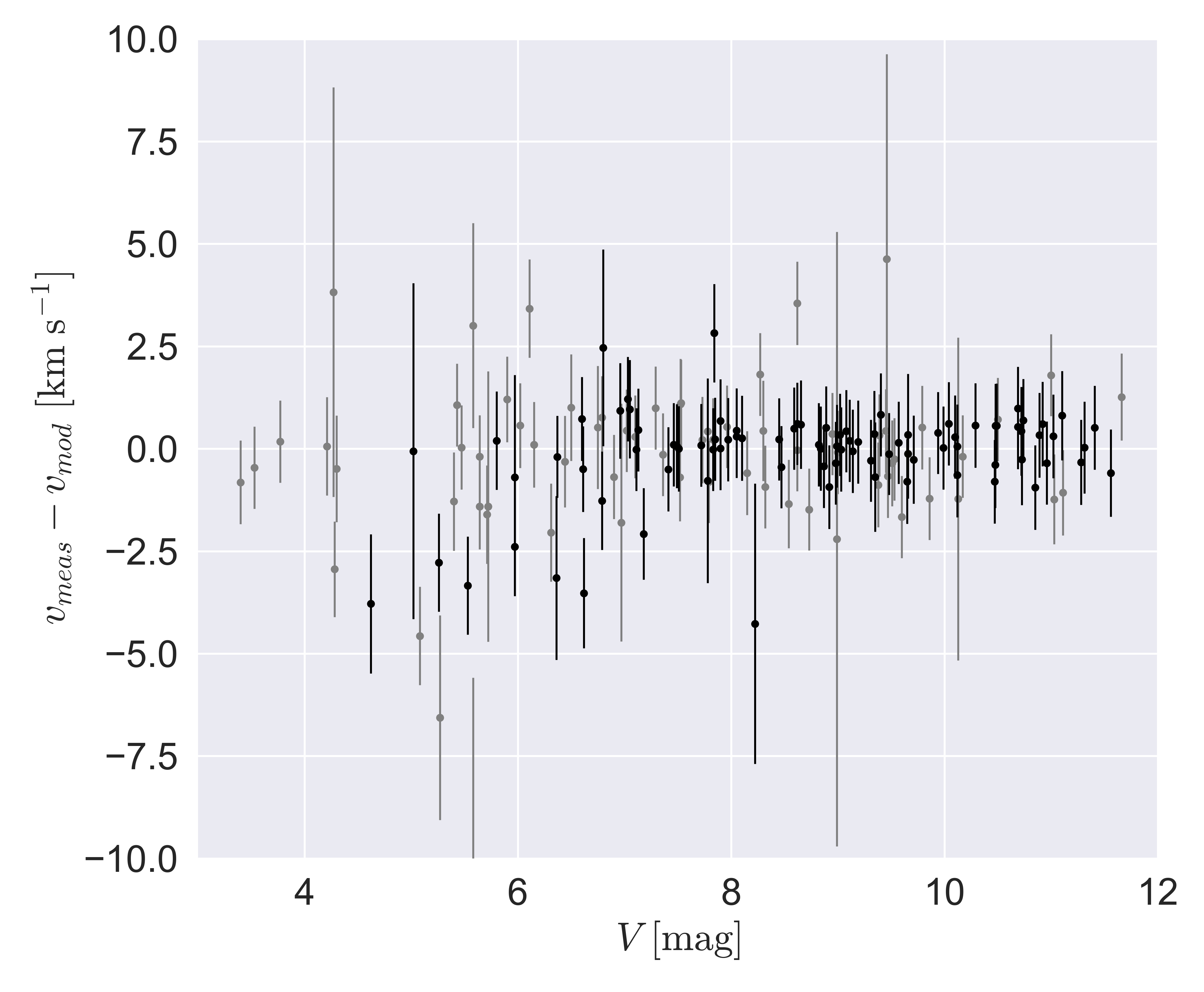}
    \caption{Difference of kinematically-modelled ('astrometric') and spectroscopic radial velocities for 161 candidate members as fuction of $V$ magnitude (which is 'equivalent' to $(B-V)$ colour index for a cluster main sequence). The vertical error bars reflect the errors on the spectroscopic radial velocities only. Known or suspect double stars are grey (Sect.~\ref{sec:double_stars}). HIP 20995, a known double star with literature radial velocities ranging from 20.7 to 35.6~km~s$^{-1}$, falls outside the plotted range, below $-10$~km~s$^{-1}$. The median difference of all data is 0.03~km~s$^{-1}$ while the median absolute deviation is 0.53~km~s$^{-1}$. The velocity dispersion of the cluster derived from proper-motion residuals equals $\sim 0.32$~km~s$^{-1}$ (Sect.~\ref{subsec:modelling_results_velocity_dispersion}).}
    \label{fig:rv_v}
  \end{center}
\end{figure}

\section{Results}\label{sec:modelling_results}

We apply the method outlined in Sect.~\ref{subsec:modelling_overview} to the 251 candidate Hyades members identified in Sect.~\ref{sec:membership}. As initial guesses for the mean cluster velocity and velocity dispersion, we use $\vec{v}_{\rm c} = (-6.30, 45.44, 5.32)~ {\rm km~s}^{-1}$ in equatorial coordinates (derived in Sect.~\ref{subsec:cluster_velocity}) and $\sigma_v = 0.3~ {\rm km~s}^{-1}$ \citep[from][]{1998A&A...331...81P}, respectively. We select $p_{\rm lim} = 0.0027$ (Sect.~\ref{subsec:modelling_stopping_criterion}) associated with a confidence level of 0.9973. With these settings, the method rejects 64 objects, leaving 187 members with associated kinematically-modelled parallaxes and standard errors (160 of these have a known radial velocity while 27 do not).

\subsection{Velocity distribution}\label{subsec:modelling_results_velocity_distribution}

The best-fit mean velocity of the cluster equals $\vec{v}_{\rm c} = (-5.96 \pm 0.04, 45.60 \pm 0.07, 5.57 \pm 0.03)~ {\rm km~s}^{-1}$ in equatorial coordinates, translating to $\vec{v}_{\rm c} = (-42.20, -19.07, -1.32)~ {\rm km~s}^{-1}$ in Galactic coordinates  (see Sect.~\ref{subsec:distance_velocity} for a comparison with other studies). The three-dimensional velocity distribution of the members, using either trigonometric or kinematically-modelled parallaxes, is shown in Figure~\ref{fig:velocities}. The kinematically-modelled parallaxes show a more concentrated velocity distribution, indicative of higher-precision parallaxes (Sect.~\ref{subsec:modelling_results_parallaxes}).

\subsection{Velocity dispersion}\label{subsec:modelling_results_velocity_dispersion}

The best-fit velocity dispersion equals $\sigma_v = 0.25 \pm 0.01~ {\rm km~s}^{-1}$, where $0.01~ {\rm km~s}^{-1}$ refers to the formal error. A side-effect of the maximum-likelihood method, extensively discussed by \citet[][their Sects 4.1, 4.3, and 5.1]{2000A&A...356.1119L}, is that the best-fit dispersion is typically significantly underestimated. Together with the upper limit $\sigma_v \lesssim 0.6~ {\rm km~s}^{-1}$ derived in Sect.~\ref{subsec:velocity_distribution}, this limits $0.25 \lesssim \sigma_v \lesssim 0.6~ {\rm km~s}^{-1}$. When estimating the velocity dispersion from the proper-motion residuals perpendicular to the centroid velocity projected on the sky \cite[see][their Sect.~4.3 and Appendix A.4, in particular Eqs~A.23 and A.24]{2000A&A...356.1119L}, we find $\sigma_v = 0.32 \pm 0.01~ {\rm km~s}^{-1}$, which is fully in line with the \citet{1998A&A...331...81P} value of $\sim0.3~ {\rm km~s}^{-1}$  and the \citet{2000A&A...356.1119L} value of $0.31 \pm 0.02~ {\rm km~s}^{-1}$.

\subsection{Parallax comparison}\label{subsec:modelling_results_parallaxes}

The comparison of the observed and the kinematically-modelled parallaxes is shown in Figures~\ref{fig:parallax1_1} for all 187 stars and \ref{fig:parallax1_1_TGAS} for the 133 TGAS stars only. With the exception of a few outliers, the data follow the 1:1 relation well. The largest outlier, in the top left corner of Figure~\ref{fig:parallax1_1}, is HIP 21092,  a faint  M2V dwarf. This object has $\varpi_{\rm obs} = 14.67 \pm 7.33$~mas (from {\it Hipparcos}-2) and $\varpi_{\rm kin} = 33.20 \pm 1.60$~mas; the {\it Hipparcos}-1 parallax is $19.64 \pm 9.61$~mas. HIP 21092 is a {\it Hipparcos} component binary with a primary component with $Hp = 12.834$~mag and a secondary component with $Hp = 14.039$~mag with a separation of 0.476~arcsec. The outlier in the lower left corner is HIP 26159 (discussed in Sect.~\ref{subsubsec:P98}).

Figure~\ref{fig:histogram} shows the histogram of the error-normalised difference between the observed and kinematically-modelled parallaxes. The mean and median of the distribution are slightly negative ($-0.06$ and $-0.09$, respectively), reflected also in a negative mean and median value of $\varpi_{\rm obs}-\varpi_{\rm kin}$ of $-0.17$ and $-0.04$~mas, respectively (TGAS stars: $-0.02$ and $-0.04$~mas; {\it Hipparcos}-2 stars: $-0.53$\footnote{We find $-0.19$~mas when excluding HIP 21092, which has $\varpi_{\rm obs}-\varpi_{\rm kin} = -18.53$~mas.} and $-0.03$~mas; for reference, 0.1~mas corresponds to 0.2~pc at the distance of the Hyades). This cannot necessarily be interpreted as evidence for a bias in the TGAS (and/or {\it Hipparcos}-2) parallaxes since the kinematically-modelled parallaxes can be biased by a wrong cluster space motion \citep[e.g.,][their Sect.~6.1]{2001A&A...367..111D}. The sensitivity to this effect is fairly large: with $\varpi = 4.74~ \mu~ v_{\rm tan}^{-1}$, this gives $\Delta \varpi = \Delta v_{\rm tan}~ \varpi~ v_{\rm tan}^{-1}$. Since, for the centre of the Hyades, $\varpi~ v_{\rm tan}^{-1} \approx 22~ 24^{-1} \approx 1$~mas~s~km$^{-1}$, we have $\Delta \varpi \ {\rm [mas]} \approx \Delta v_{\rm tan} \ {\rm [km~ s^{-1}]}$. This means that a bias in the kinematically-modelled parallaxes of 0.1~mas is already induced by a space-motion error of 0.1~km~s$^{-1}$. There is currently no way to exclude the presence of such an error in the best-fit cluster velocity $\vec{v}_{\rm c}$. The similarity of the widths of the observed distribution and the unit-variance Gaussian in Figure~\ref{fig:histogram} suggests that the standard errors of both the observed and the kinematically-modelled parallaxes are fair estimates.

\subsection{Systematic errors and correlations}\label{subsec:modelling_results_systematics}

Figure~\ref{fig:correlations} shows a sky map, for the central region of the Hyades, of the error-normalised difference between the observed and kinematically-modelled parallaxes, after smoothing the signal of each star with a two-dimensional Gaussian with $\sigma = 1^\circ$. This selected smoothing length is somewhat arbitrary, and roughly represents the maximum length scale on which significant spatial correlations in the {\it Gaia} DR1 astrometry are expected \citep[][their Appendix D.3]{2016A&A...595A...4L}. We have verified that the appearance of Figure~\ref{fig:correlations} does not vary drastically when doubling or halving the smoothing length; in particular the pronounced north-south (red-blue) asymmetry is preserved. Plots such as Figure~\ref{fig:correlations}, for {\it Hipparcos}-1 data of the Hyades, have been presented earlier by \citet[][their Figure~9]{1999ApJ...523..328N} and \citet[][their Figure~4]{2001A&A...367..111D} and hence confirmed the (known) presence of correlated astrometric parameters on angular scales of $\sim$1--$3^\circ$ in the {\it Hipparcos}-1  data. Owing to the intermediate character of {\it Gaia} DR1, the presence in TGAS of both systematic errors that are strongly correlated over angular scales of tens of degrees and spatial correlations in the astrometric parameters for stars that are separated by angular separations up to a few degrees is known \cite[][their Appendix~D.3; see also \citealt{2017A&A...599A..50A}]{2016A&A...595A...4L}. This is confirmed in Figure~\ref{fig:correlations}. A similar plot was published recently for TGAS parallaxes, based on a comparison with asteroseismic parallaxes in the Kepler field \citep[][their Figure~9]{2017ApJ...844..166Z}, suggesting a systematic TGAS parallax error of 0.059~mas / 0.011~mas exists on scales of $0.3^\circ /~ 8^\circ$ \citep[using Cepheids, ][report spatially correlated TGAS parallaxes at the level of $19 \pm 34~ \mu$as on angular scales smaller than $10^\circ$]{2017A&A...599A..67C}.

\subsection{Astrometric radial velocities}\label{subsec:modelling_results_radial_velocities}

\begin{figure}
  \begin{center}
    \includegraphics[width=\columnwidth]{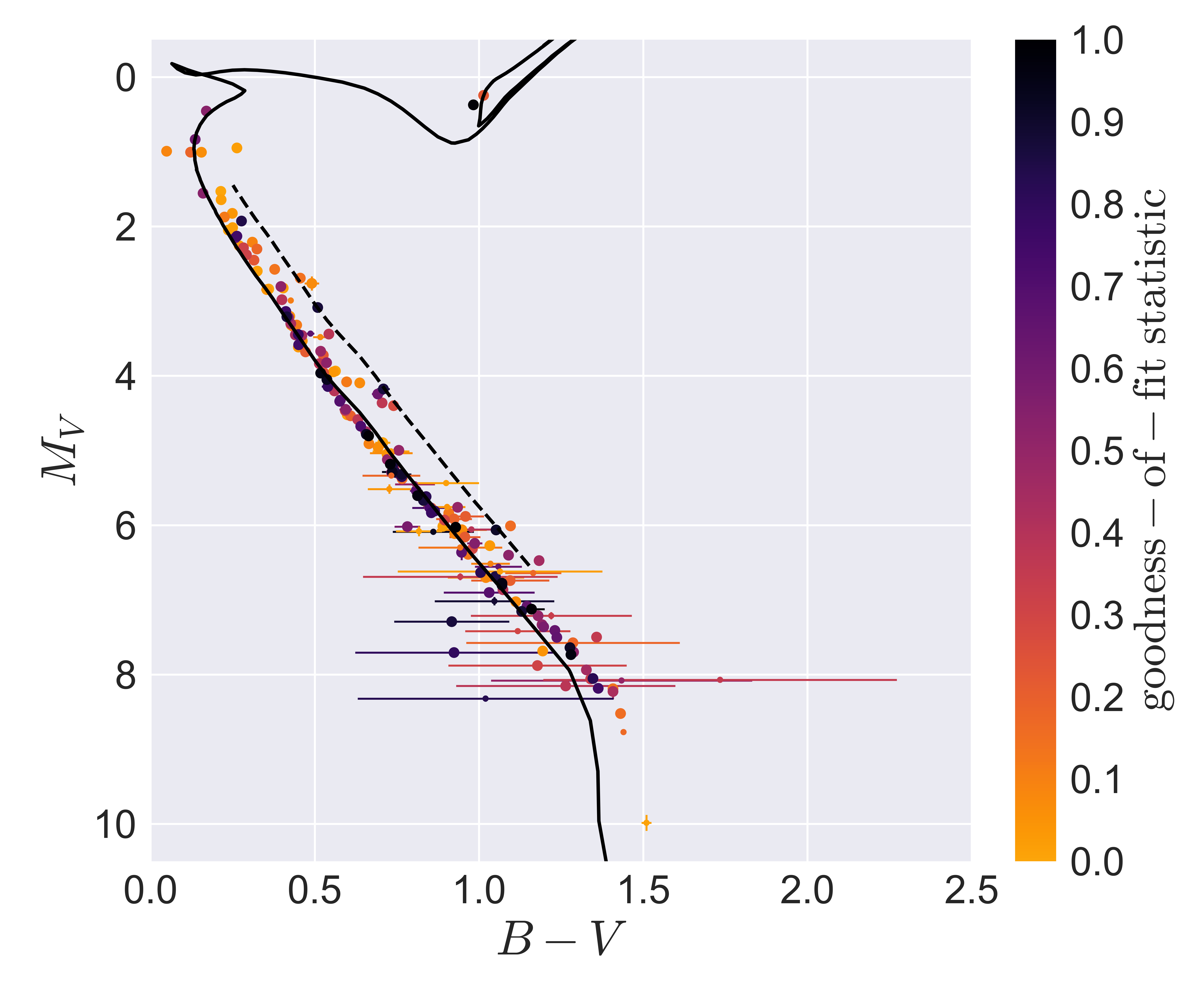}
\caption{As Figure~\ref{fig:HR_bv}, but using the goodness-of-fit statistic $p$ and using kinematically-modelled parallaxes. Large $p$ values correspond to higher-fidelity members, i.e., stars that closely follow the mean cluster velocity. The solid black line shows the Padova 675 Myr isochrone for $Z = 0.024$ while the dashed black line shows (part of) the associated binary sequence; both are meant to guide the eye and are not best fits. HIP 20894 has been split into its two components following Appendix~B.3 in \citet{2001A&A...367..111D}. The lowest point is HIP 21092, an M2V star (Sect.~\ref{subsec:modelling_results_parallaxes}).}
\label{fig:members_cl}
  \end{center}
\end{figure}

The maximum-likelihood modelling not only provides kinematically-modelled parallaxes but also kinematically-modelled radial velocities. Such 'astrometric radial velocities' are independent of, e.g., convective motions in stellar atmospheres or gravitational redshifts of spectral lines and therefore do not depend on spectral type, template choice, atmosphere model, etc., and hence can be used to study these effects \cite[for a review, see][]{1999A&A...348.1040D}. Figure~\ref{fig:rv_v} shows, for 161 candidate members, how our astrometric radial velocities differ from the spectroscopically determined radial velocities that we collected (see~Sect.~\ref{sec:RV}). Inspired by \citet{2007ASSL..350.....V}, who presented the differences as function of $(B-V)$ colour index, we present the differences as function of $V$ magnitude. Overall, there is a good correspondence, with the median difference only being $0.03$~km~s$^{-1}$. This difference can be due to a (slightly) wrong space motion at which the kinematic modelling converged, a global radial-velocity zero-point offset in the spectroscopic data, or a combination of both. Whereas significant outliers are most likely unrecognised double stars, the systematically larger dispersion at the bright end likely primarily reflects the increased velocity dispersion in the centre of the cluster, which contains most massive, bright stars as a result of mass segregation in the cluster \citep[e.g.,][]{1998A&A...331...81P}. Putting further constraints on these elements requires a homogeneous set of radial velocities with full control of the associated zero-point, including possible variations with $V$ magnitude (Sect.~\ref{subsec:RV_P98}), in other words: requires {\it Gaia} DR2 (Sect.~\ref{sec:conclusions}).

\begin{figure}
  \begin{center}
    \includegraphics[width=\columnwidth]{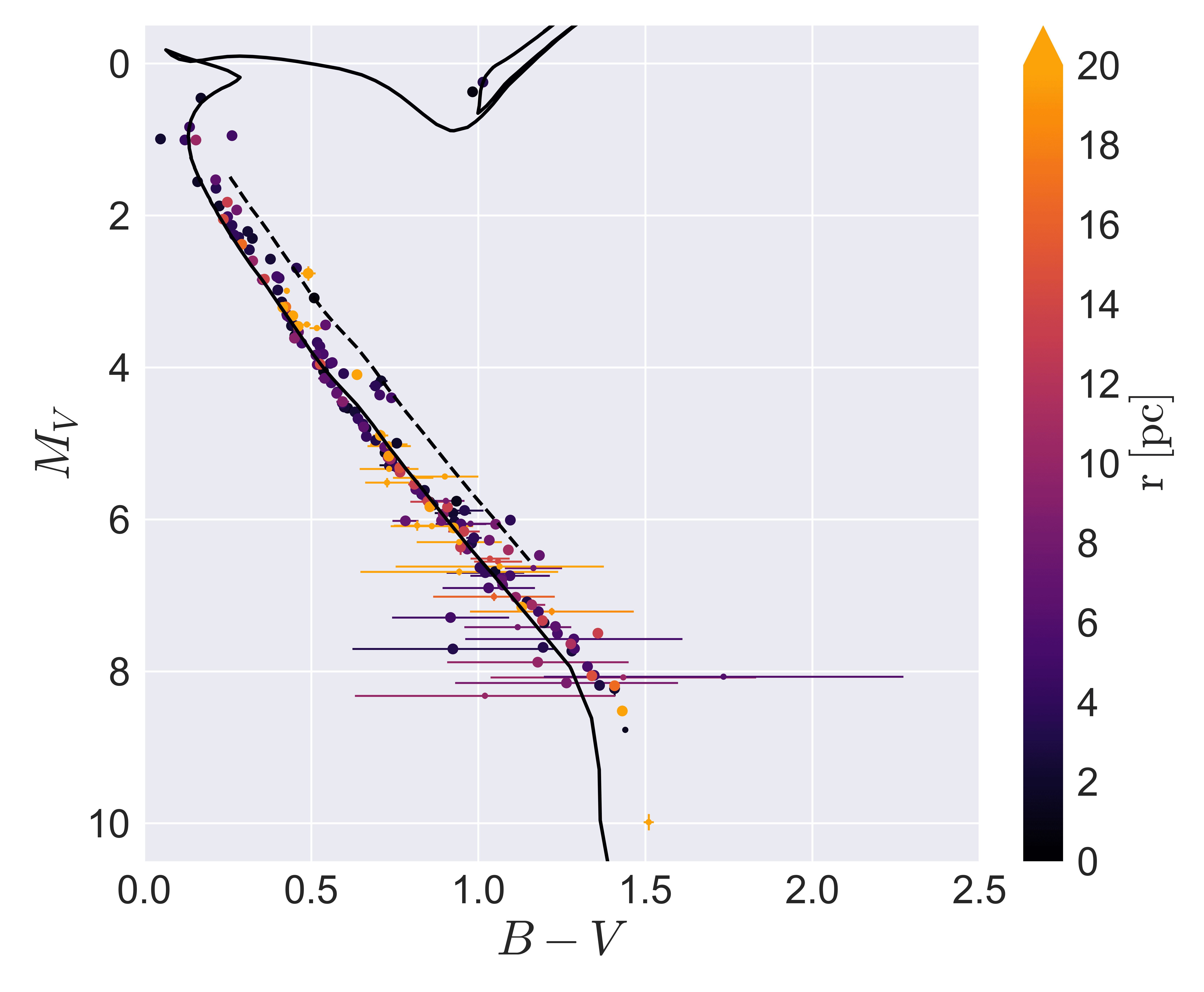}
    \caption{As Figure~\ref{fig:members_cl}, but colour coded according to the distance $r$ of each star from the cluster centre.} 
\label{fig:members_r}
  \end{center}
\end{figure}

\subsection{Colour-absolute magnitude diagram}\label{subsec:modelling_results_CMD}

\begin{figure*}
  \begin{center}
    \includegraphics[width=\textwidth]{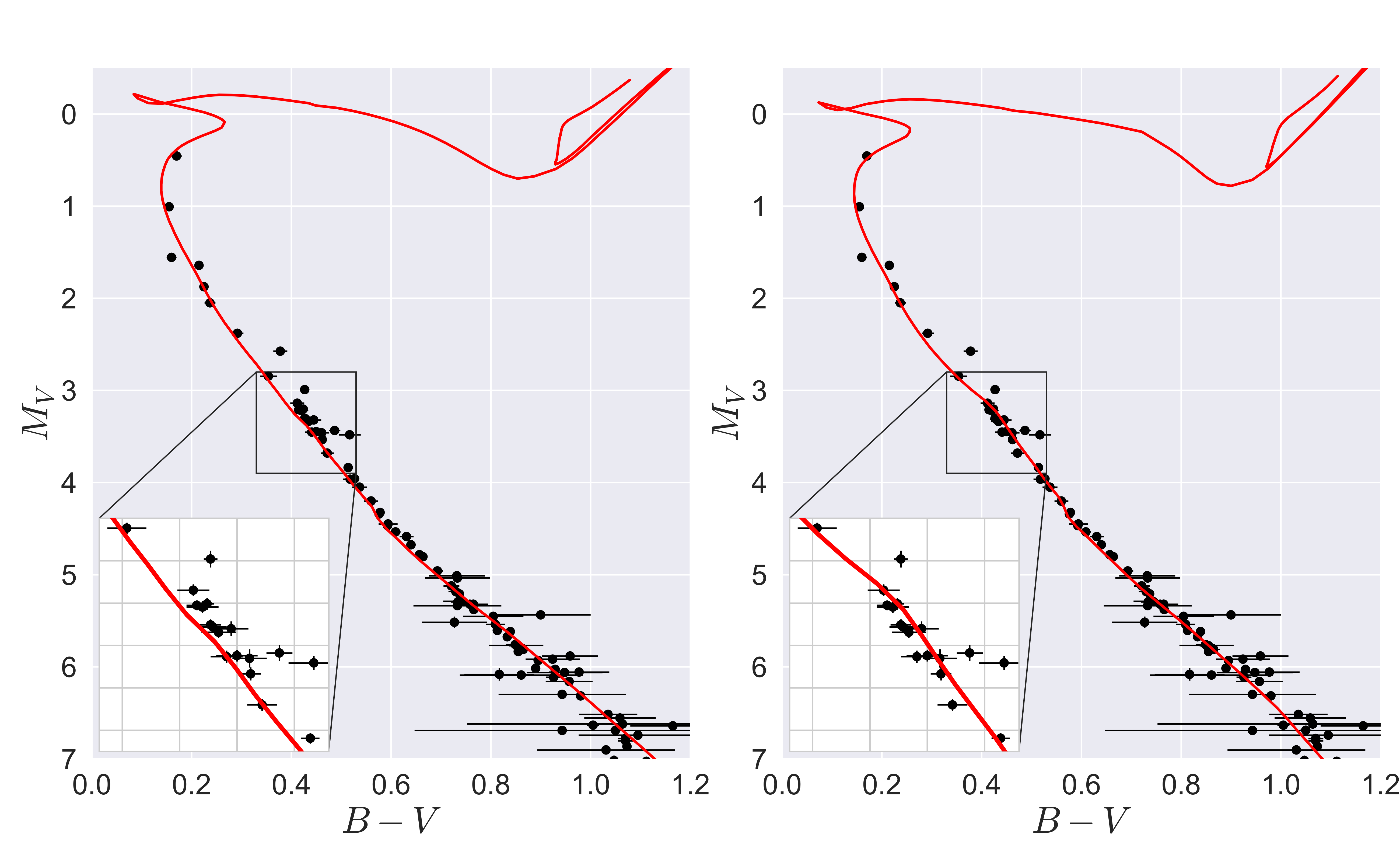}
    \caption{Bright-end colour-absolute magnitude diagrams based on kinematically-modelled parallaxes, after removing known double stars (Sect.~\ref{sec:double_stars}). As in Figure~\ref{fig:members_cl}, the double star HIP 20894 has been added with its two components following Appendix~B.3 in \citet{2001A&A...367..111D}. The left panel shows the mixing-length theory (MLT) isochrone; the right panel shows the full spectrum of turbulence (FST) model isochrone (see text and Sect.~\ref{sec:isochrones} for details). The latter isochrone seems to describe the 'upturn' in the main sequence around $(B-V) \sim 0.4$~mag better (highlighted in the insets; recall that a bias of a few tenths of a mas, i.e., a few hundredths of a mag, in the set of kinematically-modelled parallaxes cannot be excluded as a result of a wrong cluster space motion; see Sect.~\ref{subsec:modelling_results_parallaxes}). The four stars above the main sequence between $M_V \sim 2.5$ and 3.5~mag are, from bright/blue to faint/red:
HIP 20614: rapdily rotating F4V star ($v_{\rm rot} = 150$~km~s$^{-1}$; \citealt{2009A&A...493.1099S});
HIP 15624: F2V star at $\sim$5 tidal radii, so possibly escaping with slightly deviating motion and hence parallax;
TYC 118-854-1: F5V star at $\sim$3 tidal radii, so possibly escaping;
TYC 1272-67-1: F5 star at $\sim$5 tidal radii, so possibly escaping.}
    \label{fig:FST}
  \end{center}
\end{figure*}

The colour-absolute magnitude diagram based on the kinematically-modelled parallaxes of 187 stars is shown in Figures~\ref{fig:members_cl} and \ref{fig:members_r}, colour coded with goodness-of-fit statistic $p$ and distance $r$ to the cluster centre, respectively. To guide the eye, a Padova isochrone \citep{2012MNRAS.427..127B,2014MNRAS.444.2525C,2015MNRAS.452.1068C,2014MNRAS.445.1538T} for 675~Myr and $Z = 0.024$ plus its associated binary sequence have been added. Both figures show -- besides a noticeable isochrone mismatch for red, low-mass stars -- a well-defined and narrow main sequence (cf.\ Figures~\ref{fig:HR_bv} and \ref{fig:HRr_bv} which are based on the trigonometric parallaxes) which provides another proof of the high quality of the kinematically-modelled parallaxes.

Figure~\ref{fig:FST} shows the bright end of the colour-absolute magnitude diagram based on kinematically-modelled parallaxes. We have removed, using criteria defined in Sect.~\ref{sec:double_stars}, known or suspected double stars not only since (unresolved) double stars can have deviating photometry but also since their {\it Hipparcos} / {\it Gaia} astrometry, in particular their proper motions, can be affected, leading to erroneous kinematically-modelled parallaxes. As a sample, indeed, the 64 members that were removed during the kinematic-modelling iterations as a result of having the largest kinematic (astrometric) deviations have a higher double-star percentage (33/64) than the 187 surviving stars (72/187).

The data in Figure~\ref{fig:FST} suggest the presence of a 'turn-off' around $(B-V) \sim 0.4$~mag (a blueward gap seems conspicuously present as well but one should recall that double stars and members rejected during the iterations are not plotted here; see Figure~\ref{fig:HR_bv}). Such features have been discussed before and have been argued to be related to the properties of convective atmospheres \citep{1970A&A.....8..283B,1981ARA&A..19..295B,1982ApJ...255..191B,1995A&A...297L..25B,1995AJ....110..228B}. The existence of B{\"o}hm-Vitense gaps in the Hyades main sequence has previously been discussed by \citet{2000ApJ...544L..65D}, who used kinematically-modelled parallaxes based on {\it Hipparcos}-1 data. \citet{2002ApJ...564L..93D} realised that the B{\"o}hm-Vitense gap in the Hyades around $(B-V) \sim 0.4 $~mag is naturally explained as an effective-temperature effect by the full spectrum of turbulence model \citep{1996ApJ...473..550C}. This model yields a very sharp transition, around $T_{\rm eff} \sim 6800$~K, between stars that are convective only in the surface layers and stars that show a well-developed convection also in the interior. Figure~\ref{fig:FST} reinforces this conclusion based on the new data presented in this paper by showing two isochrones, one based on the canonical mixing length theory (MLT, left panel) and one based on the full spectrum of turbulence (FST, right panel) model of convection. Details on both isochrones are provided in Appendix~\ref{sec:isochrones}. Whereas both isochrones fit the main sequence, the FST isochrone better follows the turn-off in the data around $(B-V) \sim 0.4 $~mag. This confirms the interpretation of \citet{2002ApJ...564L..93D} of this feature in terms of an effective-temperature effect.

It is finally important to realise that both isochrones have been constructed from models that have been specifically computed for this comparison with as main aim showing differences induced by changing the convection modelling. As such, the isochrones are meant as examples of reasonable fits and not as unique or best-fit descriptions of the data; in particular, they do not provide consolidated constraints on Helium content, age, etc. Clearly, fitting the Hyades colour-absolute magnitude diagram based on the high-precision kinematically-modelled parallaxes derived in this work does allow a detailed exploration of models, which, however, is beyond the scope of this paper.

\section{Discussion and conclusions}\label{sec:conclusions}

We study the Hyades open cluster using {\it Gaia} DR1 astrometry complemented by {\it Hipparcos}-2 astrometry for bright stars and by literature radial velocities from 12 sources. This combination of data provides three-dimensional velocities for a large subset of our initial sample of 2296 stars and permits making a reliable kinematic selection of cluster members. We determine membership probabilities using a $\chi^2$ test statistic to quantify the difference between observed and expected three-dimensional velocities of stars. We accept 251 stars as candidate members of the Hyades by choosing a membership probability threshold of 0.27\% (confidence level of 99.73\%). Our list demotes 15 members from the canonical {\it Hipparcos}-1-based member list with 197 objects constructed by \citet{1998A&A...331...81P} and adds 18 new members with radial velocity and 51 new members without known radial velocity. The stellar density of the Galactic field inside the (central region of the) Hyades does not differ from neighbouring areas on the sky.

We find, in line with previous studies, that the major axis of the cluster is nearly aligned with the Galactic $x$ axis and that the cluster is significantly flattened parallel to the Galactic plane. While the major axis of the cluster is about two times larger than the minor axis when considering stars within 30~pc of the cluster centre, the cluster shape becomes nearly spherical in the central regions. We actually find members beyond three tidal radii ($\sim$30~pc). Clearly, these distant,  likely  escaping members are interesting objects for future studies of the evaporation of the cluster.

We employ an iterative, maximum-likelihood kinematic modelling method \citep[originally developed by][and amended by us to include radial velocities when available]{2000A&A...356.1119L} to derive the mean velocity, velocity dispersion, and kinematically-modelled parallaxes for cluster members. The iterative element removes those stars which have the least compatible three-dimensional space motions, for instance due to unrecognised multiplicity. The kinematically-modelled parallaxes for the 187 objects that survive the iterations are $\sim$1.7 /~ 2.9 times more precise than the TGAS / {\it Hipparcos}-2 trigonometric parallaxes although they could be biased by a few tenths of a mas as a result of a wrong cluster space motion. The internal velocity dispersion of the surviving stars is $\sigma_v = 0.32 \pm 0.01\ \rm km\ s^{-1}$, fullly compatible with existing literature values \citep[e.g., 0.3~km~s$^{-1}$ derived in][]{1998A&A...331...81P}.

The colour-absolute magnitude diagram based on the kinematically-modelled parallaxes shows a well-defined main sequence with a 'turn-off' feature around $(B-V)\ \sim 0.4$~mag. We construct  two, custom-made  isochrones, one based on the mixing length theory and the other on the full spectrum of turbulence model of convection, and our data  seem to  favour the latter model, in line with earlier results \citep{2002ApJ...564L..93D}. The formidable challenge of fitting isochrones to the Hyades colour-absolute magnitude diagram based on the high-precision kinematically-modelled parallaxes derived in this work is left for the future.

In terms of future prospects, {\it Gaia} DR2 will open many avenues. Although DR2 will have the same bright-star problem as DR1, hence still necessitating the addition of {\it Hipparcos}-2 data for the brightest members, DR2 will not only have higher-accuracy parallaxes and proper motions but will also allow extending the cluster membership by $\sim$8 magnitudes, up to and including early L-type brown dwarfs \citep{2014arXiv1404.3896D}. In addition, DR2 will contain (homogeneous) radial velocities for a good fraction of the bright stars, down to $\sim$12~mag, as well as homogeneous, milli-magnitude-accuracy BP and RP photometry. In short, the Hyades story is not over yet.

\section*{Acknowledgements}

This work has made use of data from the European Space Agency (ESA) mission {\it Gaia} (\url{https://www.cosmos.esa.int/gaia}), processed by the {\it Gaia} Data Processing and Analysis Consortium (DPAC, \url{https://www.cosmos.esa.int/web/gaia/dpac/consortium}). Funding for the DPAC has been provided by national institutions, in particular the institutions participating in the {\it Gaia} Multilateral Agreement. We would like to thank Carlo Manara, Timo Prusti, Anthony Brown, and Jordan Voirin for stimulating discussions and Roger Griffin for valuable insight into radial-velocity zero-points. This research is based on data obtained by ESA's {\it Hipparcos} satellite and has made use of the ADS (NASA), SIMBAD / VizieR (CDS), and TOPCAT \citep[][\url{http://www.starlink.ac.uk/topcat/}]{2005ASPC..347...29T} tools.

\bibliographystyle{mnras} 
\bibliography{hyades} 

\begin{thebibliography}{}
\makeatletter
\relax
\def\mn@urlcharsother{\let\do\@makeother \do\$\do\&\do\#\do\^\do\_\do\%\do\~}
\def\mn@doi{\begingroup\mn@urlcharsother \@ifnextchar [ {\mn@doi@}
  {\mn@doi@[]}}
\def\mn@doi@[#1]#2{\def\@tempa{#1}\ifx\@tempa\@empty \href
  {http://dx.doi.org/#2} {doi:#2}\else \href {http://dx.doi.org/#2} {#1}\fi
  \endgroup}
\def\mn@eprint#1#2{\mn@eprint@#1:#2::\@nil}
\def\mn@eprint@arXiv#1{\href {http://arxiv.org/abs/#1} {{\tt arXiv:#1}}}
\def\mn@eprint@dblp#1{\href {http://dblp.uni-trier.de/rec/bibtex/#1.xml}
  {dblp:#1}}
\def\mn@eprint@#1:#2:#3:#4\@nil{\def\@tempa {#1}\def\@tempb {#2}\def\@tempc
  {#3}\ifx \@tempc \@empty \let \@tempc \@tempb \let \@tempb \@tempa \fi \ifx
  \@tempb \@empty \def\@tempb {arXiv}\fi \@ifundefined
  {mn@eprint@\@tempb}{\@tempb:\@tempc}{\expandafter \expandafter \csname
  mn@eprint@\@tempb\endcsname \expandafter{\@tempc}}}

\bibitem[\protect\citeauthoryear{{Anderson} \& {Francis}}{{Anderson} \&
  {Francis}}{2012}]{2012AstL...38..331A}
{Anderson} E.,  {Francis} C.,  2012, \mn@doi [Astronomy Letters]
  {10.1134/S1063773712050015}, \href
  {http://adsabs.harvard.edu/abs/2012AstL...38..331A} {38, 331}

\bibitem[\protect\citeauthoryear{{Arenou} et~al.,}{{Arenou}
  et~al.}{2017}]{2017A&A...599A..50A}
{Arenou} F.,  et~al., 2017, \mn@doi [\aap] {10.1051/0004-6361/201629895}, \href
  {http://adsabs.harvard.edu/abs/2017A%26A...599A..50A} {599, A50}

\bibitem[\protect\citeauthoryear{{Astraatmadja} \&
  {Bailer-Jones}}{{Astraatmadja} \&
  {Bailer-Jones}}{2016a}]{2016ApJ...832..137A}
{Astraatmadja} T.~L.,  {Bailer-Jones} C.~A.~L.,  2016a, \mn@doi [\apj]
  {10.3847/0004-637X/832/2/137}, \href
  {http://adsabs.harvard.edu/abs/2016ApJ...832..137A} {832, 137}

\bibitem[\protect\citeauthoryear{{Astraatmadja} \&
  {Bailer-Jones}}{{Astraatmadja} \&
  {Bailer-Jones}}{2016b}]{2016ApJ...833..119A}
{Astraatmadja} T.~L.,  {Bailer-Jones} C.~A.~L.,  2016b, \mn@doi [\apj]
  {10.3847/1538-4357/833/1/119}, \href
  {http://adsabs.harvard.edu/abs/2016ApJ...833..119A} {833, 119}

\bibitem[\protect\citeauthoryear{{Bailer-Jones}}{{Bailer-Jones}}{2015}]{2015PASP..127..994B}
{Bailer-Jones} C.~A.~L.,  2015, \mn@doi [\pasp] {10.1086/683116}, \href
  {http://adsabs.harvard.edu/abs/2015PASP..127..994B} {127, 994}

\bibitem[\protect\citeauthoryear{{Binney} \& {Merrifield}}{{Binney} \&
  {Merrifield}}{1998}]{1998gaas.book.....B}
{Binney} J.,  {Merrifield} M.,  1998, {Galactic Astronomy}

\bibitem[\protect\citeauthoryear{{Binney} et~al.,}{{Binney}
  et~al.}{2014}]{2014MNRAS.437..351B}
{Binney} J.,  et~al., 2014, \mn@doi [\mnras] {10.1093/mnras/stt1896}, \href
  {http://adsabs.harvard.edu/abs/2014MNRAS.437..351B} {437, 351}

\bibitem[\protect\citeauthoryear{{Bobylev}, {Bajkova}  \&
  {Gontcharov}}{{Bobylev} et~al.}{2006}]{2006A&AT...25..143B}
{Bobylev} V.~V.,  {Bajkova} A.~T.,   {Gontcharov} G.~A.,  2006, \mn@doi
  [Astronomical and Astrophysical Transactions] {10.1080/10556790600893104},
  \href {http://adsabs.harvard.edu/abs/2006A%26AT...25..143B} {25, 143}

\bibitem[\protect\citeauthoryear{{B{\"o}hm-Vitense}}{{B{\"o}hm-Vitense}}{1958}]{1958ZA.....46..108B}
{B{\"o}hm-Vitense} E.,  1958, \zap, \href
  {http://adsabs.harvard.edu/abs/1958ZA.....46..108B} {46, 108}

\bibitem[\protect\citeauthoryear{{B{\"o}hm-Vitense}}{{B{\"o}hm-Vitense}}{1970}]{1970A&A.....8..283B}
{B{\"o}hm-Vitense} E.,  1970, \aap, \href
  {http://adsabs.harvard.edu/abs/1970A%26A.....8..283B} {8, 283}

\bibitem[\protect\citeauthoryear{{B{\"o}hm-Vitense}}{{B{\"o}hm-Vitense}}{1981}]{1981ARA&A..19..295B}
{B{\"o}hm-Vitense} E.,  1981, \mn@doi [\araa]
  {10.1146/annurev.aa.19.090181.001455}, \href
  {http://adsabs.harvard.edu/abs/1981ARA%26A..19..295B} {19, 295}

\bibitem[\protect\citeauthoryear{{B{\"o}hm-Vitense}}{{B{\"o}hm-Vitense}}{1982}]{1982ApJ...255..191B}
{B{\"o}hm-Vitense} E.,  1982, \mn@doi [\apj] {10.1086/159817}, \href
  {http://adsabs.harvard.edu/abs/1982ApJ...255..191B} {255, 191}

\bibitem[\protect\citeauthoryear{{B{\"o}hm-Vitense}}{{B{\"o}hm-Vitense}}{1995a}]{1995AJ....110..228B}
{B{\"o}hm-Vitense} E.,  1995a, \mn@doi [\aj] {10.1086/117511}, \href
  {http://adsabs.harvard.edu/abs/1995AJ....110..228B} {110, 228}

\bibitem[\protect\citeauthoryear{{B{\"o}hm-Vitense}}{{B{\"o}hm-Vitense}}{1995b}]{1995A&A...297L..25B}
{B{\"o}hm-Vitense} E.,  1995b, \aap, \href
  {http://adsabs.harvard.edu/abs/1995A%26A...297L..25B} {297, L25}

\bibitem[\protect\citeauthoryear{{Bressan}, {Marigo}, {Girardi}, {Salasnich},
  {Dal Cero}, {Rubele}  \& {Nanni}}{{Bressan}
  et~al.}{2012}]{2012MNRAS.427..127B}
{Bressan} A.,  {Marigo} P.,  {Girardi} L.,  {Salasnich} B.,  {Dal Cero} C.,
  {Rubele} S.,   {Nanni} A.,  2012, \mn@doi [\mnras]
  {10.1111/j.1365-2966.2012.21948.x}, \href
  {http://adsabs.harvard.edu/abs/2012MNRAS.427..127B} {427, 127}

\bibitem[\protect\citeauthoryear{{Brown}}{{Brown}}{2017}]{2017arXiv170901216B}
{Brown} A.~G.~A.,  2017, preprint, \href
  {http://adsabs.harvard.edu/abs/2017arXiv170901216B} {} (\mn@eprint {arXiv}
  {1709.01216})

\bibitem[\protect\citeauthoryear{{Canuto} \& {Mazzitelli}}{{Canuto} \&
  {Mazzitelli}}{1991}]{1991ApJ...370..295C}
{Canuto} V.~M.,  {Mazzitelli} I.,  1991, \mn@doi [\apj] {10.1086/169815}, \href
  {http://adsabs.harvard.edu/abs/1991ApJ...370..295C} {370, 295}

\bibitem[\protect\citeauthoryear{{Canuto}, {Goldman}  \& {Mazzitelli}}{{Canuto}
  et~al.}{1996}]{1996ApJ...473..550C}
{Canuto} V.~M.,  {Goldman} I.,   {Mazzitelli} I.,  1996, \mn@doi [\apj]
  {10.1086/178166}, \href {http://adsabs.harvard.edu/abs/1996ApJ...473..550C}
  {473, 550}

\bibitem[\protect\citeauthoryear{{Casertano}, {Riess}, {Bucciarelli}  \&
  {Lattanzi}}{{Casertano} et~al.}{2017}]{2017A&A...599A..67C}
{Casertano} S.,  {Riess} A.~G.,  {Bucciarelli} B.,   {Lattanzi} M.~G.,  2017,
  \mn@doi [\aap] {10.1051/0004-6361/201629733}, \href
  {http://adsabs.harvard.edu/abs/2017A%26A...599A..67C} {599, A67}

\bibitem[\protect\citeauthoryear{{Chen}, {Girardi}, {Bressan}, {Marigo},
  {Barbieri}  \& {Kong}}{{Chen} et~al.}{2014}]{2014MNRAS.444.2525C}
{Chen} Y.,  {Girardi} L.,  {Bressan} A.,  {Marigo} P.,  {Barbieri} M.,   {Kong}
  X.,  2014, \mn@doi [\mnras] {10.1093/mnras/stu1605}, \href
  {http://adsabs.harvard.edu/abs/2014MNRAS.444.2525C} {444, 2525}

\bibitem[\protect\citeauthoryear{{Chen}, {Bressan}, {Girardi}, {Marigo}, {Kong}
   \& {Lanza}}{{Chen} et~al.}{2015}]{2015MNRAS.452.1068C}
{Chen} Y.,  {Bressan} A.,  {Girardi} L.,  {Marigo} P.,  {Kong} X.,   {Lanza}
  A.,  2015, \mn@doi [\mnras] {10.1093/mnras/stv1281}, \href
  {http://adsabs.harvard.edu/abs/2015MNRAS.452.1068C} {452, 1068}

\bibitem[\protect\citeauthoryear{{Claydon}, {Gieles}  \& {Zocchi}}{{Claydon}
  et~al.}{2017}]{2017MNRAS.466.3937C}
{Claydon} I.,  {Gieles} M.,   {Zocchi} A.,  2017, \mn@doi [\mnras]
  {10.1093/mnras/stw3309}, \href
  {http://adsabs.harvard.edu/abs/2017MNRAS.466.3937C} {466, 3937}

\bibitem[\protect\citeauthoryear{{D'Antona}, {Montalb{\'a}n}, {Kupka}  \&
  {Heiter}}{{D'Antona} et~al.}{2002}]{2002ApJ...564L..93D}
{D'Antona} F.,  {Montalb{\'a}n} J.,  {Kupka} F.,   {Heiter} U.,  2002, \mn@doi
  [\apjl] {10.1086/338911}, \href
  {http://adsabs.harvard.edu/abs/2002ApJ...564L..93D} {564, L93}

\bibitem[\protect\citeauthoryear{{Daniel}, {Heggie}  \& {Varri}}{{Daniel}
  et~al.}{2017}]{2017MNRAS.468.1453D}
{Daniel} K.~J.,  {Heggie} D.~C.,   {Varri} A.~L.,  2017, \mn@doi [\mnras]
  {10.1093/mnras/stx571}, \href
  {http://adsabs.harvard.edu/abs/2017MNRAS.468.1453D} {468, 1453}

\bibitem[\protect\citeauthoryear{{Debernardi}, {Mermilliod}, {Carquillat}  \&
  {Ginestet}}{{Debernardi} et~al.}{2000}]{2000A&A...354..881D}
{Debernardi} Y.,  {Mermilliod} J.-C.,  {Carquillat} J.-M.,   {Ginestet} N.,
  2000, \aap, \href {http://adsabs.harvard.edu/abs/2000A%26A...354..881D} {354,
  881}

\bibitem[\protect\citeauthoryear{{Dravins}, {Lindegren}  \& {Madsen}}{{Dravins}
  et~al.}{1999}]{1999A&A...348.1040D}
{Dravins} D.,  {Lindegren} L.,   {Madsen} S.,  1999, \aap, \href
  {http://adsabs.harvard.edu/abs/1999A%26A...348.1040D} {348, 1040}

\bibitem[\protect\citeauthoryear{{ESA}}{{ESA}}{1997}]{1997ESASP1200.....E}
{ESA} ed. 1997, {The HIPPARCOS and TYCHO catalogues. Astrometric and
  photometric star catalogues derived from the ESA HIPPARCOS Space Astrometry
  Mission}  ESA Special Publication Vol. 1200

\bibitem[\protect\citeauthoryear{{Fabricius}, {H{\o}g}, {Makarov}, {Mason},
  {Wycoff}  \& {Urban}}{{Fabricius} et~al.}{2002}]{2002A&A...384..180F}
{Fabricius} C.,  {H{\o}g} E.,  {Makarov} V.~V.,  {Mason} B.~D.,  {Wycoff}
  G.~L.,   {Urban} S.~E.,  2002, \mn@doi [\aap] {10.1051/0004-6361:20011822},
  \href {http://adsabs.harvard.edu/abs/2002A%26A...384..180F} {384, 180}

\bibitem[\protect\citeauthoryear{{Feigelson} \& {Babu}}{{Feigelson} \&
  {Babu}}{2012}]{2012msma.book.....F}
{Feigelson} E.~D.,  {Babu} G.~J.,  2012, {Modern Statistical Methods for
  Astronomy}

\bibitem[\protect\citeauthoryear{{Ferguson}, {Alexander}, {Allard}, {Barman},
  {Bodnarik}, {Hauschildt}, {Heffner-Wong}  \& {Tamanai}}{{Ferguson}
  et~al.}{2005}]{2005ApJ...623..585F}
{Ferguson} J.~W.,  {Alexander} D.~R.,  {Allard} F.,  {Barman} T.,  {Bodnarik}
  J.~G.,  {Hauschildt} P.~H.,  {Heffner-Wong} A.,   {Tamanai} A.,  2005,
  \mn@doi [\apj] {10.1086/428642}, \href
  {http://adsabs.harvard.edu/abs/2005ApJ...623..585F} {623, 585}

\bibitem[\protect\citeauthoryear{{Gaia Collaboration} et~al.,}{{Gaia
  Collaboration} et~al.}{2016a}]{2016A&A...595A...1G}
{Gaia Collaboration} et~al., 2016a, \mn@doi [\aap]
  {10.1051/0004-6361/201629272}, \href
  {http://adsabs.harvard.edu/abs/2016A%26A...595A...1G} {595, A1}

\bibitem[\protect\citeauthoryear{{Gaia Collaboration} et~al.,}{{Gaia
  Collaboration} et~al.}{2016b}]{2016A&A...595A...2G}
{Gaia Collaboration} et~al., 2016b, \mn@doi [\aap]
  {10.1051/0004-6361/201629512}, \href
  {http://adsabs.harvard.edu/abs/2016A%26A...595A...2G} {595, A2}

\bibitem[\protect\citeauthoryear{{Gaia Collaboration} et~al.,}{{Gaia
  Collaboration} et~al.}{2017}]{2017A&A...601A..19G}
{Gaia Collaboration} et~al., 2017, \mn@doi [\aap]
  {10.1051/0004-6361/201730552}, \href
  {http://adsabs.harvard.edu/abs/2017A%26A...601A..19G} {601, A19}

\bibitem[\protect\citeauthoryear{{Griffin}}{{Griffin}}{2012}]{2012JApA...33...29G}
{Griffin} R.~F.,  2012, \mn@doi [Journal of Astrophysics and Astronomy]
  {10.1007/s12036-012-9137-5}, \href
  {http://adsabs.harvard.edu/abs/2012JApA...33...29G} {33, 29}

\bibitem[\protect\citeauthoryear{{Griffin}}{{Griffin}}{2013}]{2013Obs...133..144G}
{Griffin} R.~F.,  2013, The Observatory, \href
  {http://adsabs.harvard.edu/abs/2013Obs...133..144G} {133, 144}

\bibitem[\protect\citeauthoryear{{Griffin}, {Griffin}, {Gunn}  \&
  {Zimmerman}}{{Griffin} et~al.}{1988}]{1988AJ.....96..172G}
{Griffin} R.~F.,  {Griffin} R.~E.~M.,  {Gunn} J.~E.,   {Zimmerman} B.~A.,
  1988, \mn@doi [\aj] {10.1086/114800}, \href
  {http://cdsads.u-strasbg.fr/abs/1988AJ.....96..172G} {96, 172}

\bibitem[\protect\citeauthoryear{{Hauschildt}, {Allard}  \&
  {Baron}}{{Hauschildt} et~al.}{1999a}]{1999ApJ...512..377H}
{Hauschildt} P.~H.,  {Allard} F.,   {Baron} E.,  1999a, \mn@doi [\apj]
  {10.1086/306745}, \href {http://adsabs.harvard.edu/abs/1999ApJ...512..377H}
  {512, 377}

\bibitem[\protect\citeauthoryear{{Hauschildt}, {Allard}, {Ferguson}, {Baron}
  \& {Alexander}}{{Hauschildt} et~al.}{1999b}]{1999ApJ...525..871H}
{Hauschildt} P.~H.,  {Allard} F.,  {Ferguson} J.,  {Baron} E.,   {Alexander}
  D.~R.,  1999b, \mn@doi [\apj] {10.1086/307954}, \href
  {http://adsabs.harvard.edu/abs/1999ApJ...525..871H} {525, 871}

\bibitem[\protect\citeauthoryear{{Iglesias} \& {Rogers}}{{Iglesias} \&
  {Rogers}}{1996}]{1996ApJ...464..943I}
{Iglesias} C.~A.,  {Rogers} F.~J.,  1996, \mn@doi [\apj] {10.1086/177381},
  \href {http://adsabs.harvard.edu/abs/1996ApJ...464..943I} {464, 943}

\bibitem[\protect\citeauthoryear{Jones, Oliphant, Peterson  et~al.}{Jones
  et~al.}{2017}]{SciPy}
Jones E.,  Oliphant T.,  Peterson P.,   et~al., 2001--2017, {SciPy}: Open
  source scientific tools for {Python}, \url {http://www.scipy.org/}

\bibitem[\protect\citeauthoryear{{Kharchenko}}{{Kharchenko}}{2001}]{2001KFNT...17..409K}
{Kharchenko} N.~V.,  2001, Kinematika i Fizika Nebesnykh Tel, \href
  {http://adsabs.harvard.edu/abs/2001KFNT...17..409K} {17, 409}

\bibitem[\protect\citeauthoryear{{Kharchenko}, {Scholz}, {Piskunov},
  {R{\"o}ser}  \& {Schilbach}}{{Kharchenko} et~al.}{2007}]{2007AN....328..889K}
{Kharchenko} N.~V.,  {Scholz} R.-D.,  {Piskunov} A.~E.,  {R{\"o}ser} S.,
  {Schilbach} E.,  2007, \mn@doi [Astronomische Nachrichten]
  {10.1002/asna.200710776}, \href
  {http://adsabs.harvard.edu/abs/2007AN....328..889K} {328, 889}

\bibitem[\protect\citeauthoryear{{Kunder} et~al.,}{{Kunder}
  et~al.}{2017}]{2017AJ....153...75K}
{Kunder} A.,  et~al., 2017, \mn@doi [\aj] {10.3847/1538-3881/153/2/75}, \href
  {http://adsabs.harvard.edu/abs/2017AJ....153...75K} {153, 75}

\bibitem[\protect\citeauthoryear{{Lindegren}, {Madsen}  \&
  {Dravins}}{{Lindegren} et~al.}{2000}]{2000A&A...356.1119L}
{Lindegren} L.,  {Madsen} S.,   {Dravins} D.,  2000, \aap, \href
  {http://cdsads.u-strasbg.fr/abs/2000A%26A...356.1119L} {356, 1119}

\bibitem[\protect\citeauthoryear{{Lindegren} et~al.,}{{Lindegren}
  et~al.}{2016}]{2016A&A...595A...4L}
{Lindegren} L.,  et~al., 2016, \mn@doi [\aap] {10.1051/0004-6361/201628714},
  \href {http://adsabs.harvard.edu/abs/2016A%26A...595A...4L} {595, A4}

\bibitem[\protect\citeauthoryear{{Lodders}}{{Lodders}}{2003}]{2003ApJ...591.1220L}
{Lodders} K.,  2003, \mn@doi [\apj] {10.1086/375492}, \href
  {http://adsabs.harvard.edu/abs/2003ApJ...591.1220L} {591, 1220}

\bibitem[\protect\citeauthoryear{{Maderak}, {Deliyannis}, {King}  \&
  {Cummings}}{{Maderak} et~al.}{2013}]{2013AJ....146..143M}
{Maderak} R.~M.,  {Deliyannis} C.~P.,  {King} J.~R.,   {Cummings} J.~D.,  2013,
  \mn@doi [\aj] {10.1088/0004-6256/146/6/143}, \href
  {http://adsabs.harvard.edu/abs/2013AJ....146..143M} {146, 143}

\bibitem[\protect\citeauthoryear{{Madsen}, {Lindegren}  \& {Dravins}}{{Madsen}
  et~al.}{2001}]{2001ASPC..228..506M}
{Madsen} S.,  {Lindegren} L.,   {Dravins} D.,  2001, in {Deiters} S.,  {Fuchs}
  B.,  {Just} A.,  {Spurzem} R.,   {Wielen} R.,  eds,  Astronomical Society of
  the Pacific Conference Series Vol. 228, Dynamics of Star Clusters and the
  Milky Way. p.~506

\bibitem[\protect\citeauthoryear{{Maldonado} et~al.,}{{Maldonado}
  et~al.}{2017}]{2017A&A...598A..27M}
{Maldonado} J.,  et~al., 2017, \mn@doi [\aap] {10.1051/0004-6361/201629223},
  \href {http://adsabs.harvard.edu/abs/2017A%26A...598A..27M} {598, A27}

\bibitem[\protect\citeauthoryear{{Martell} et~al.,}{{Martell}
  et~al.}{2017}]{2017MNRAS.465.3203M}
{Martell} S.~L.,  et~al., 2017, \mn@doi [\mnras] {10.1093/mnras/stw2835}, \href
  {http://adsabs.harvard.edu/abs/2017MNRAS.465.3203M} {465, 3203}

\bibitem[\protect\citeauthoryear{{Mendoza}}{{Mendoza}}{1967}]{1967BOTT....4..149M}
{Mendoza} E.~E.,  1967, Boletin de los Observatorios Tonantzintla y Tacubaya,
  \href {http://adsabs.harvard.edu/abs/1967BOTT....4..149M} {4, 149}

\bibitem[\protect\citeauthoryear{{Mermilliod}, {Mayor}  \& {Udry}}{{Mermilliod}
  et~al.}{2009}]{2009A&A...498..949M}
{Mermilliod} J.-C.,  {Mayor} M.,   {Udry} S.,  2009, \mn@doi [\aap]
  {10.1051/0004-6361/200810244}, \href
  {http://cdsads.u-strasbg.fr/abs/2009A%26A...498..949M} {498, 949}

\bibitem[\protect\citeauthoryear{{Michalik}, {Lindegren}, {Hobbs}  \&
  {Lammers}}{{Michalik} et~al.}{2014}]{2014A&A...571A..85M}
{Michalik} D.,  {Lindegren} L.,  {Hobbs} D.,   {Lammers} U.,  2014, \mn@doi
  [\aap] {10.1051/0004-6361/201424606}, \href
  {http://adsabs.harvard.edu/abs/2014A%26A...571A..85M} {571, A85}

\bibitem[\protect\citeauthoryear{{Narayanan} \& {Gould}}{{Narayanan} \&
  {Gould}}{1999}]{1999ApJ...523..328N}
{Narayanan} V.~K.,  {Gould} A.,  1999, \mn@doi [\apj] {10.1086/307716}, \href
  {http://adsabs.harvard.edu/abs/1999ApJ...523..328N} {523, 328}

\bibitem[\protect\citeauthoryear{{Oort}}{{Oort}}{1979}]{1979A&A....78..312O}
{Oort} J.~H.,  1979, \aap, \href
  {http://adsabs.harvard.edu/abs/1979A%26A....78..312O} {78, 312}

\bibitem[\protect\citeauthoryear{{Patel}, {Pandey}, {Savanov}, {Prasad}  \&
  {Srivastava}}{{Patel} et~al.}{2013}]{2013MNRAS.430.2154P}
{Patel} M.~K.,  {Pandey} J.~C.,  {Savanov} I.~S.,  {Prasad} V.,   {Srivastava}
  D.~C.,  2013, \mn@doi [\mnras] {10.1093/mnras/stt036}, \href
  {http://adsabs.harvard.edu/abs/2013MNRAS.430.2154P} {430, 2154}

\bibitem[\protect\citeauthoryear{{Pels}, {Oort}  \& {Pels-Kluyver}}{{Pels}
  et~al.}{1975}]{1975A&A....43..423P}
{Pels} G.,  {Oort} J.~H.,   {Pels-Kluyver} H.~A.,  1975, \aap, \href
  {http://adsabs.harvard.edu/abs/1975A%26A....43..423P} {43, 423}

\bibitem[\protect\citeauthoryear{{Perryman} et~al.,}{{Perryman}
  et~al.}{1998}]{1998A&A...331...81P}
{Perryman} M.~A.~C.,  et~al., 1998, \aap, \href
  {http://adsabs.harvard.edu/abs/1998A%26A...331...81P} {331, 81}

\bibitem[\protect\citeauthoryear{{Potekhin}, {Baiko}, {Haensel}  \&
  {Yakovlev}}{{Potekhin} et~al.}{1999}]{1999A&A...346..345P}
{Potekhin} A.~Y.,  {Baiko} D.~A.,  {Haensel} P.,   {Yakovlev} D.~G.,  1999,
  \aap, \href {http://cdsads.u-strasbg.fr/abs/1999A%26A...346..345P} {346, 345}

\bibitem[\protect\citeauthoryear{{Pourbaix} et~al.,}{{Pourbaix}
  et~al.}{2004}]{2004A&A...424..727P}
{Pourbaix} D.,  et~al., 2004, \mn@doi [\aap] {10.1051/0004-6361:20041213},
  \href {http://adsabs.harvard.edu/abs/2004A%26A...424..727P} {424, 727}

\bibitem[\protect\citeauthoryear{{Press}, {Teukolsky}, {Vetterling}  \&
  {Flannery}}{{Press} et~al.}{1992}]{1992nrca.book.....P}
{Press} W.~H.,  {Teukolsky} S.~A.,  {Vetterling} W.~T.,   {Flannery} B.~P.,
  1992, {Numerical recipes in C. The art of scientific computing}

\bibitem[\protect\citeauthoryear{{R{\"o}ser}, {Schilbach}, {Piskunov},
  {Kharchenko}  \& {Scholz}}{{R{\"o}ser} et~al.}{2011}]{2011A&A...531A..92R}
{R{\"o}ser} S.,  {Schilbach} E.,  {Piskunov} A.~E.,  {Kharchenko} N.~V.,
  {Scholz} R.-D.,  2011, \mn@doi [\aap] {10.1051/0004-6361/201116948}, \href
  {http://cdsads.u-strasbg.fr/abs/2011A%26A...531A..92R} {531, A92}

\bibitem[\protect\citeauthoryear{{SDSS Collaboration} et~al.,}{{SDSS
  Collaboration} et~al.}{2016}]{2016arXiv160802013S}
{SDSS Collaboration} et~al., 2016, preprint, \href
  {http://adsabs.harvard.edu/abs/2016arXiv160802013S} {} (\mn@eprint {arXiv}
  {1608.02013})

\bibitem[\protect\citeauthoryear{{Schr{\"o}der}, {Reiners}  \&
  {Schmitt}}{{Schr{\"o}der} et~al.}{2009}]{2009A&A...493.1099S}
{Schr{\"o}der} C.,  {Reiners} A.,   {Schmitt} J.~H.~M.~M.,  2009, \mn@doi
  [\aap] {10.1051/0004-6361:200810377}, \href
  {http://adsabs.harvard.edu/abs/2009A%26A...493.1099S} {493, 1099}

\bibitem[\protect\citeauthoryear{{Smith}}{{Smith}}{2012}]{2012BASI...40..487S}
{Smith} G.~H.,  2012, Bulletin of the Astronomical Society of India, \href
  {http://adsabs.harvard.edu/abs/2012BASI...40..487S} {40, 487}

\bibitem[\protect\citeauthoryear{{Tabernero}, {Montes}  \& {Gonzalez
  Hernandez}}{{Tabernero} et~al.}{2012}]{2012yCat..35470013T}
{Tabernero} H.~M.,  {Montes} D.,   {Gonzalez Hernandez} J.~I.,  2012, VizieR
  Online Data Catalog, \href
  {http://adsabs.harvard.edu/abs/2012yCat..35470013T} {354}

\bibitem[\protect\citeauthoryear{{Tang}, {Worthey}  \& {Davis}}{{Tang}
  et~al.}{2014}]{2014MNRAS.445.1538T}
{Tang} B.,  {Worthey} G.,   {Davis} A.~B.,  2014, \mn@doi [\mnras]
  {10.1093/mnras/stu1867}, \href
  {http://adsabs.harvard.edu/abs/2014MNRAS.445.1538T} {445, 1538}

\bibitem[\protect\citeauthoryear{{Taylor}}{{Taylor}}{2005}]{2005ASPC..347...29T}
{Taylor} M.~B.,  2005, in {Shopbell} P.,  {Britton} M.,   {Ebert} R.,  eds,
  Astronomical Society of the Pacific Conference Series Vol. 347, Astronomical
  Data Analysis Software and Systems XIV. p.~29

\bibitem[\protect\citeauthoryear{{Turon} et~al.,}{{Turon}
  et~al.}{1993}]{1993BICDS..43....5T}
{Turon} C.,  et~al., 1993, Bulletin d'Information du Centre de Donnees
  Stellaires, \href {http://adsabs.harvard.edu/abs/1993BICDS..43....5T} {43}

\bibitem[\protect\citeauthoryear{{Upgren}, {Weis}  \& {Hanson}}{{Upgren}
  et~al.}{1985}]{1985AJ.....90.2039U}
{Upgren} A.~R.,  {Weis} E.~W.,   {Hanson} R.~B.,  1985, \mn@doi [\aj]
  {10.1086/113910}, \href {http://adsabs.harvard.edu/abs/1985AJ.....90.2039U}
  {90, 2039}

\bibitem[\protect\citeauthoryear{{Ventura}, {Zeppieri}, {Mazzitelli}  \&
  {D'Antona}}{{Ventura} et~al.}{1998}]{1998A&A...334..953V}
{Ventura} P.,  {Zeppieri} A.,  {Mazzitelli} I.,   {D'Antona} F.,  1998, \aap,
  \href {http://cdsads.u-strasbg.fr/abs/1998A%26A...334..953V} {334, 953}

\bibitem[\protect\citeauthoryear{{Ventura}, {D'Antona}  \&
  {Mazzitelli}}{{Ventura} et~al.}{2008}]{2008Ap&SS.316...93V}
{Ventura} P.,  {D'Antona} F.,   {Mazzitelli} I.,  2008, \mn@doi [\apss]
  {10.1007/s10509-007-9672-8}, \href
  {http://cdsads.u-strasbg.fr/abs/2008Ap%26SS.316...93V} {316, 93}

\bibitem[\protect\citeauthoryear{{White}, {Gabor}  \& {Hillenbrand}}{{White}
  et~al.}{2007}]{2007AJ....133.2524W}
{White} R.~J.,  {Gabor} J.~M.,   {Hillenbrand} L.~A.,  2007, \mn@doi [\aj]
  {10.1086/514336}, \href {http://adsabs.harvard.edu/abs/2007AJ....133.2524W}
  {133, 2524}

\bibitem[\protect\citeauthoryear{{Zinn}, {Huber}, {Pinsonneault}  \&
  {Stello}}{{Zinn} et~al.}{2017}]{2017ApJ...844..166Z}
{Zinn} J.~C.,  {Huber} D.,  {Pinsonneault} M.~H.,   {Stello} D.,  2017, \mn@doi
  [\apj] {10.3847/1538-4357/aa7c1c}, \href
  {http://adsabs.harvard.edu/abs/2017ApJ...844..166Z} {844, 166}

\bibitem[\protect\citeauthoryear{{de Bruijne}}{{de
  Bruijne}}{2014}]{2014arXiv1404.3896D}
{de Bruijne} J.~H.~J.,  2014, preprint, \href
  {http://adsabs.harvard.edu/abs/2014arXiv1404.3896D} {} (\mn@eprint {arXiv}
  {1404.3896})

\bibitem[\protect\citeauthoryear{{de Bruijne}, {Hoogerwerf}  \& {de Zeeuw}}{{de
  Bruijne} et~al.}{2000}]{2000ApJ...544L..65D}
{de Bruijne} J.~H.~J.,  {Hoogerwerf} R.,   {de Zeeuw} P.~T.,  2000, \mn@doi
  [\apjl] {10.1086/317296}, \href
  {http://adsabs.harvard.edu/abs/2000ApJ...544L..65D} {544, L65}

\bibitem[\protect\citeauthoryear{{de Bruijne}, {Hoogerwerf}  \& {de Zeeuw}}{{de
  Bruijne} et~al.}{2001}]{2001A&A...367..111D}
{de Bruijne} J.~H.~J.,  {Hoogerwerf} R.,   {de Zeeuw} P.~T.,  2001, \mn@doi
  [\aap] {10.1051/0004-6361:20000410}, \href
  {http://adsabs.harvard.edu/abs/2001A%26A...367..111D} {367, 111}

\bibitem[\protect\citeauthoryear{{van Bueren}}{{van
  Bueren}}{1952}]{1952BAN....11..385V}
{van Bueren} H.~G.,  1952, \bain, \href
  {http://adsabs.harvard.edu/abs/1952BAN....11..385V} {11, 385}

\bibitem[\protect\citeauthoryear{{van Leeuwen}}{{van
  Leeuwen}}{2007}]{2007ASSL..350.....V}
{van Leeuwen} F.,  ed. 2007, {Hipparcos, the New Reduction of the Raw Data}
  Astrophysics and Space Science Library Vol. 350,
  \mn@doi{10.1007/978-1-4020-6342-8.
}

\bibitem[\protect\citeauthoryear{{van Leeuwen}}{{van
  Leeuwen}}{2009}]{2009A&A...497..209V}
{van Leeuwen} F.,  2009, \mn@doi [\aap] {10.1051/0004-6361/200811382}, \href
  {http://adsabs.harvard.edu/abs/2009A%26A...497..209V} {497, 209}

\makeatother
\end{thebibliography}

\appendix

\section{Radial velocity data}\label{sec:RV}

We use radial velocity measurements from 12 selected literature sources. Details on these are provided in Appendix~\ref{subsec:RV_P98} through \ref{subsec:RV_GALAH}. Appendix~\ref{subsec:RV_epoch} describes how we treat epoch data. Appendix~\ref{subsec:RV_final} explains how we arrive at a final, merged list of radial velocities and associated standard errors for each object in our sample.

\subsection{\citet{1998A&A...331...81P}}\label{subsec:RV_P98}

\citet{1998A&A...331...81P} list 282 stars in their Table~2, 254 of which have radial velocities collected from 28 literature sources. We apply the same zero-point corrections as Perryman et al.\ (see their Sect.~3.2): Coravel radial velocities (column~r in Table~2 equals 24) are corrected by adding $+$0.4~km~s$^{-1}$ to the values listed in Table~2 and Griffin et al.\ radial velocities (column~r in Table~2 equals 1) are corrected by adding $-q(V) - 0.5~ {\rm km~s}^{-1}$ for stars fainter than $V = 6$~mag and $-0.5~ {\rm km~s}^{-1}$ for brighter stars, with $q(V) = 0.44 - 700 \times 10^{-0.4 V}$~km~s$^{-1}$.

\subsection{\citet{2012JApA...33...29G}}\label{subsec:RV_Griffin2012}

\citet{2012JApA...33...29G} presents orbital elements for 52 stars, 46 of which have {\it Hipparcos} or {\it Tycho-2} identifiers. We use the systematic ($\gamma$) radial velocities from Griffin's Table~1 and, after affirmative private communication with R. Griffin, apply the same zero-point corrections as detailed in Appendix~\ref{subsec:RV_P98}.

\subsection{\citet{2013Obs...133..144G}}\label{subsec:RV_Griffin2013}

Table~II in \citet{2013Obs...133..144G} lists mean radial velocities of 15 possible new Hyades members. Since the measurements do not have standard errors, we adopt an error of $1\ {\rm km~s}^{-1}$ for the first 13 stars and an error of $10\ {\rm km~s}^{-1}$ for the last 2 stars since their radial velocities are followed by a colon (:), indicative of an uncertain measurement. Following affirmative private communication with R. Griffin, we apply the same zero-point corrections as detailed in Appendix~\ref{subsec:RV_P98}.

\subsection{\citet{2009A&A...498..949M}}

\citet{2009A&A...498..949M} provides 776 radial velocity measurements for 157 Hyades stars, 148 of which have {\it Hipparcos} or {\it Tycho-2} identifiers.

\subsection{\citet{2007AJ....133.2524W}} 

\citet{2007AJ....133.2524W} list 474 radial velocities for 368 stars, 351 of which have {\it Hipparcos} or {\it Tycho-2} identifiers.

\subsection{\citet{2012yCat..35470013T}}

\citet{2012yCat..35470013T} list radial velocities for 61 stars in their Table~A.1 (including an added reference star, vB 153, in the online version), all of which have a {\it Hipparcos} identifier.

\subsection{\citet{2000A&A...354..881D}}

Table~2 in \citet{2000A&A...354..881D} lists 401 radial velocity measurements for 10 stars.

\subsection{HADES}

Table~A.2 in \citet{2017A&A...598A..27M} lists 71 stars with radial velocities, only four of which lie in the Hyades field (these are GJ 3186, TYC 1795-941-1, GJ 150.1B, and GJ 162). The authors state that the estimated uncertainties are in the range $0.3 - 0.6\ {\rm km~s}^{-1}$. We therefore conservatively assign a standard error of $1\ {\rm km~s}^{-1}$ to each radial velocity.

\subsection{XHIP}

The Extended {\it Hipparcos} Compilation \citep[][XHIP]{2012AstL...38..331A} contains 117\,955 entries, 46\,392 of which have radial velocity data collected from a large variety of literature sources (see their Table~4). We remove 5867 entries, namely those for which error bounds are not given (e\_RV = 999), those which have the worst radial velocity quality rating (q\_RV = D) indicative of unreliable measurements, and those originating from \citet[][r\_RV = 551]{2007AN....328..889K} since a subset of this catalogue contains astrometric instead of spectroscopic radial velocities.

\subsection{RAVE}

RAVE DR5 \citep{2017AJ....153...75K} contains radial velocities from 520\,629 spectra for 457\,588 unique stars, 264\,386 of which are TGAS or {\it Tycho-2} stars.

\subsection{APOGEE}

SDSS DR13 \citep{2016arXiv160802013S} includes data from the third release of APOGEE. Although there are 164\,562 targets in the allStar summary table, stars targeted in multiple fields appear as duplicates such that there are 157\,935 unique stars. Some 140\,486 of these survive the cross-match with TGAS.

\subsection{GALAH}\label{subsec:RV_GALAH}

GALAH's first data release \citep{2017MNRAS.465.3203M} gives radial velocities for 9860 stars derived from 10\,680 observations. Martell et al.\ report that 98\% of their radial velocities have a standard deviation less than 0.6~km~s$^{-1}$. We therefore conservatively adopt a radial velocity standard error of $1\ {\rm km~s}^{-1}$ for each measurement.

\subsection{Epoch radial velocities}\label{subsec:RV_epoch}

Epoch radial velocities have been published by \citet{2009A&A...498..949M}, \citet{2007AJ....133.2524W}, \citet{2000A&A...354..881D}, RAVE \citep{2017AJ....153...75K}, APOGEE \citep{2016arXiv160802013S}, and GALAH \citep{2017MNRAS.465.3203M}. In these cases, we adopt the median of the reported epoch data as radial velocity. To obtain an error estimate on this median, we first calculate the median absolute deviation around the median and quadratically combine it with the largest of the measurement uncertainties.

\subsection{Creating a final, merged table}\label{subsec:RV_final}

As a first step in the construction of a final, merged table, we exclude all radial velocities with standard errors larger than 10~km~s$^{-1}$ and we inflate all radial velocity standard errors smaller than $1\ {\rm km~s}^{-1}$ by quadratically adding a $1\ {\rm km~s}^{-1}$ noise floor in order to mask the inconsistency in and uncertainty on the zero-point of the radial velocity measurements across different literature sources and instruments / surveys. This leaves 908 stars that have at least one radial velocity measurement and 1388 stars without any radial velocity data.

As a consistency check, we perform a comparison between the radial velocities from different literature sources for objects that are present in more than one list. This only brings up six problem cases, i.e., inconsistencies larger than 3$\sigma$. Such cases can be explained by, for instance, unrecognised binarity and/or underestimated standard errors and/or differences in the zero-point used by different authors. As an example, one of the problem cases is HIP 21179, which is a known spectroscopic binary that has radial velocity measurements originating from three literature sources: $-12.527 \pm 4.1 \ {\rm km~s}^{-1}$ from RAVE \citep{2017AJ....153...75K}, $42.1 \pm 1 \ {\rm km~s}^{-1}$ from \citet{1998A&A...331...81P}, and $40.045 \pm 2.147 \ {\rm km~s}^{-1}$ from \citet{2012JApA...33...29G}. The latter radial velocity is the result of a full orbital solution based on 92 measurements. For the same star, SIMBAD contains two additional radial velocities: $40.4 \pm 0.6~ {\rm km~s}^{-1}$ from \citet{2013MNRAS.430.2154P}, which uses \citet{2006A&AT...25..143B} as a source, and $53.85 \pm 8.26~ {\rm km~s}^{-1}$ from \citet{2009A&A...498..949M}, which reports an RMS scatter of $47.43~ {\rm km~s}^{-1}$ in the velocity measurement based on 33 measurements spread over 1830 days. Clearly, in this particular case, the radial velocity reported in RAVE DR5 is an instantaneous, and hence non-representative, measurement of the (systemic) radial velocity of the spectroscopic binary.

If an object has radial velocity matches in several source catalogues, we use the following priority ranking: \citet{1998A&A...331...81P}, \citet{2012JApA...33...29G}, \citet{2013Obs...133..144G}, \citet{2009A&A...498..949M}, \citet{2007AJ....133.2524W}, \citet{2012yCat..35470013T}, \citet{2000A&A...354..881D}, HADES, XHIP, RAVE, APOGEE, and GALAH. This allows us to construct a final table in which each star is assigned a single radial velocity (or a null value in case no radial velocity measurement exists) plus an identifying number to track the associated literature source. In total, there are 228 radial velocities from \citet{1998A&A...331...81P}, 5 from \citet{2012JApA...33...29G}, 10 from \citet{2013Obs...133..144G}, 12 from \citet{2009A&A...498..949M}, 19 from \citet{2007AJ....133.2524W}, 3 from \citet{2012yCat..35470013T}, 0 from \citet{2000A&A...354..881D}, 4 from HADES, 591 from XHIP, 23 from RAVE, 12 from APOGEE, and 1 from GALAH.

\section{Johnson photometry}\label{sec:Johnson}

When available and as a default, Johnson $V$ and $(B-V)$ photometry is taken from the {\it Hipparcos} Catalogue \citep{1997ESASP1200.....E}. We assign a fixed 0.02-mag error to the errorless $V$ magnitudes reported in the {\it Hipparcos} Catalogue. For seven stars, {\it Hipparcos} photometry is available but suspicious: the $(B-V)$ error is zero for five objects (HIP 16529, HIP 20056, HIP 20441, HIP 20601, and HIP 21741) while the $(B-V)$ error is excessively large for two objects (HIP 20686 and HIP 21459). We hence use ASCC-2.5 Johnson photometry \citep{2001KFNT...17..409K} for these seven stars.

For 63 members, (photometric) {\it Hipparcos} data are not available. For 59 of these, we can resort to ASCC-2.5 Johnson photometry. For the remaining four stars, photometry is taken from \citet{1985AJ.....90.2039U} (TYC 1265-1118-1) or from SIMBAD (TYC 1821-255-1, TYC 1289-1656-1, and TYC 1786-572-1). Since TYC 632-608-1 has faulty photometry in the {\it Tycho-2} catalogue, and therefore also in ASCC-2.5, we take its Johnson photometry from SIMBAD and assume a standard error of 0.1~mag on $(B-V)$.

As a check of the homogeneity of our photometry, we compare our data, as described above, with the data of \citet{1967BOTT....4..149M}. For the sample of 115 (presumably non-double) stars in Figure~\ref{fig:FST}, 43 are in common. The differences in the $V$ magnitudes and $(B-V)$ colour indices in this sample show a random scatter around $\pm0.01$~mag (compared to single-observation, unity-airmass probable errors of $\sigma_V=0.014$/0.020~mag in $V$ and $\sigma_{B-V}=0.010$/0.014~mag in $(B-V)$ for $V$ brighter/fainter than 9~mag quoted by \citealt{1967BOTT....4..149M}), comparable to the symbol sizes in Figure~\ref{fig:FST}.

We finally recall that extinction is negligible for the Hyades such that there is no need to perform corrections to the absolute magnitude and/or colour indices.

\section{Double stars}\label{sec:double_stars}

Inspired by \citet{1999A&A...348.1040D} and \citet{2000A&A...356.1119L}, we take the following definition to identify known or suspected double stars:
\begin{enumerate}
\item Objects flagged in columns s and/or u in Table 2 of \citet{1998A&A...331...81P}. Column s labels spectroscopic binaries and stars with variable radial velocities. Column u contains the {\it Hipparcos} double-star information from Catalogue field H59 (see point~iv below):
\item Visual binaries with magnitude difference $\Delta m < 4$~mag and a separation $\rho < 20$~arcsec according to the Hipparcos Input Catalogue \citep[][HIC]{1993BICDS..43....5T}.
\item Spectroscopic binaries from the ninth catalogue of spectroscopic binary orbits \citep[][SB9]{2004A&A...424..727P}.
\item {\it Hipparcos} double stars, i.e., objects identified in the {\it Hipparcos} Catalogue \citep{1997ESASP1200.....E} as a solution of type component ({\it Hipparcos} field H59 equals C), acceleration (H59 equals G), orbital (H59 equals O), variability-induced mover (H59 equals V), or stochastic (H59 equals X).
\item Objects present in the {\it Tycho} Double Star Catalogue \citep[][TDSC]{2002A&A...384..180F}.
\end{enumerate}
These criteria, combined through non-exclusive or's, result in 105 of our 251 members being labelled as double star.

\section{Isochrones}\label{sec:isochrones}

To explore the role played by the treatment of the convective instability on the shape of evolutionary tracks and isochrones, we calculate and show in Figure~\ref{fig:FST} isochrones (for 700~Myr) based on the classical mixing length theory \citep[MLT;][]{1958ZA.....46..108B} and on the full spectrum of turbulence (FST) convection model \citep{1991ApJ...370..295C,1996ApJ...473..550C}. The evolutionary sequences for stars of different mass and metallicity have been calculated by means of the ATON code \citep{1998A&A...334..953V,2008Ap&SS.316...93V}. 

MLT modelling involves an assumption on the parameter $\alpha$ (the mixing length in terms of the pressure scale height $H_{\rm p}$), which is usually calibrated on the fit of the Solar model and then extended to other masses, compositions, and ages.

The FST model is characterised by convective fluxes in which the full distribution of eddies is accounted for, and by a convective scale length defined as the harmonic mean between the distance of the layer from the top and the bottom of the convective zone. In addition, the distance to the top (and bottom) is increased by a fraction $\beta H_{\rm p}$. The parameter $\beta$ allows finetuning of the Solar radius at the Solar age and can hence be considered as an efficiency parameter. The FST choice for the scale of mixing allows a more detailed description of the layers close to the formal border of convection because this scale length tends to vanish as the stability region is approached. This is particularly useful when focusing on phenomena of which the scale is smaller than $H_{\rm p}$, for instance when describing convection at effective temperatures close to the transition between fully radiative and partially convective envelopes.

We use a non-instantaneous formulation to describe convective mixing. Core overshooting in both the MLT and FST cases is modelled by assuming an exponential decay of the convective velocity from the border, with scale length $\zeta$ times the thickness of the convective region \citep[see][]{1998A&A...334..953V}. We use the opacity tables from the OPAL website \citep{1996ApJ...464..943I} and we follow \citet{2005ApJ...623..585F} for low-temperature opacities. We adopt the conductive opacities from \citet{1999A&A...346..345P}. We use grey atmospheric conditions for temperatures above 5000~K. For cooler effective temperatures, we adopt NextGen atmosphere models \citet{1999ApJ...512..377H,1999ApJ...525..871H}. We match the atmosphere models with the interior models at an optical depth $\tau = 3$.

We adopt an Iron content $[{\rm Fe}/{\rm H}] = +0.15$ for the Hyades as advocated by \citet{2013AJ....146..143M}. We assume the isotopic initial distribution of each mode to be Solar-scaled according to the Solar distribution presented by \citet{2003ApJ...591.1220L}. In both cases, we fix the core overshooting to $\zeta = 0.03$ in order to reproduce the luminosity of the red-clump stars. The degeneracy between the convection parameters ($\alpha$ in the MLT and $\beta$ in the FST), metallicity, and Helium content ($Y$) implies $\alpha=2.2$ and $Y=0.27$ for the MLT models and $\beta=0.3$ and $Y=0.30$ for the FST models. These discrepancies are deemed acceptable in the context of this work in which the focus is purely on showing the comparison between the results obtained by changing the convection modelling.

\section{Expansion and rotation}\label{sec:expansion_rotation}

An estimate of the expansion (or contraction) and rotation velocity of the cluster can be obtained, respectively, by the (member-averaged) vectorial dot and cross products of the position and velocity vectors of the members (denoted by index $i$) with respect to the cluster centre (after subtracting the common space motion of the cluster):
\begin{eqnarray}
v_{{\rm exp},i} &=& \frac{\vec{v_i} \cdot \vec{r_i}}{|\vec{r_i}|};\\
v_{{\rm rot},i} &=& \frac{|\vec{v_i} \times \vec{r_i}|}{|\vec{r_i}|}.
\end{eqnarray}
The inner vector product projects the velocity vector of each star onto the radial direction that connects the cluster centre and the star and therefore represents the radial velocity component as seen from the cluster centre (i.e., expansion if positive and contraction if negative). The cross vector product represents the remaining velocity, i.e., the rotation velocity around the cluster centre. After averaging over the 200 candidate members with known radial velocities, this gives $v_{\rm exp} = -0.05 \pm 0.19$~km~s$^{-1}$ and $v_{\rm rot} =  0.55 \pm 0.27$~km~s$^{-1}$; for the 149 members within 10~pc of the centre, the values are $v_{\rm exp} =  0.08 \pm 0.12$~km~s$^{-1}$ and $v_{\rm rot} =  0.46 \pm 0.28$~km~s$^{-1}$ (the standard errors reported here represent the sample variance; for $v_{\rm rot}$, we take a mean separately in each of the three dimensions before calculating the magnitude of the vector). In short, the data do not support a significant expansion (or contraction) or rotation of the cluster.

\onecolumn

\section{Extension of \citet{2000A&A...356.1119L}}\label{sec:ML}

The maximum-likelihood method developed by \citet{2000A&A...356.1119L} works on astrometric data, i.e., it excludes radial velocity as observable. The likelihood function ($L$, or, equivalently, the log-likelihood function $\mathcal{L} \equiv \ln L$) considers three-dimensional observables (the vector $\vec{\theta}$), namely trigonometric parallax and the two proper-motion components for each star. The probability density function for the observables is modelled as a three-dimensional normal distribution and, since the observed errors of separate stars are assumed to be independent \citep[see also Sect.~5.4 in][]{2000A&A...356.1119L}, the full probability density function of all observables is the product over the stars of these individual distributions. The gradient and the (approximation of the) Hessian of the likelihood function are given by Eqs~(A.3) and (A.7) in \citet{2000A&A...356.1119L}, respectively.

We extend the method to also include radial velocity as fourth observable. Since radial velocities are available only for a subset of the stars in our sample, the full probability density function of all observables is a mix of individual distributions with three- and four-dimensional observable vectors. Both the gradient and the Hessian need to be derived for this new case. Starting from Eq.~(A3) in \citet{2000A&A...356.1119L}:
\begin{equation}
f_j = \frac{d\mathcal{L}}{d\theta_j} = \sum_{i =1}^N \left[ \sum_{\alpha = 1}^4 \sum_{\beta = 1}^4 D_{i\alpha\beta}^{-1} \frac{\partial c_{i\alpha}}{\partial \theta_j} \left(a_{i \beta} - c_{i \beta} \right) - \frac{1}{2} \frac{\partial \ln \mathrm{det} \vec{D}_i}{\partial \theta_j} -
\frac{1}{2} \sum_{\alpha = 1}^4 \sum_{\beta = 1}^4 \frac{\partial D_{i\alpha\beta}^{-1}}{\partial \theta_j} \left(a_{i\alpha}-c_{i\alpha}\right) \left(a_{i\beta}-c_{i\beta}\right) \right].
\end{equation}
Since:
\begin{equation}
\frac{\partial \ln \mathrm{det} \vec{D}_i}{\partial \theta_j}  = \mathrm{Tr}\, \left(\vec{D}_i^{-1} \frac{\partial \vec{D}_i}{\partial \theta_j}\right)
\end{equation}
and, following Jacobi's formula,
\begin{equation}
 \frac{\partial \vec{D}_i^{-1}}{\partial \theta_j}  = - \vec{D}_i^{-1} \, \frac{\partial \vec{D}_i}{\partial \theta_j} \vec{D}_i^{-1},
\end{equation}
the gradient of the four-dimensional likelihood is:
\begin{equation}
f_j = \sum_{i =1}^N \left[ \sum_{\alpha = 1}^4 \sum_{\beta = 1}^4 D_{i\alpha\beta}^{-1} \frac{\partial c_{i\alpha}}{\partial \theta_j} \left(a_{i \beta} - c_{i \beta} \right) - \frac{1}{2} \mathrm{Tr}\, \left(\vec{D}_i^{-1} \frac{\partial \vec{D}_i}{\partial \theta_j}\right) -
\frac{1}{2} \sum_{\alpha = 1}^4 \sum_{\beta = 1}^4  \left[- \vec{D}_i^{-1} \, \frac{\partial \vec{D}_i}{\partial \theta_j} \vec{D}_i^{-1} \right]_{\alpha \beta}  \left(a_{i\alpha}-c_{i\alpha}\right) \left(a_{i\beta}-c_{i\beta}\right) \right].
\end{equation}

Next, we need to derive the Hessian of the likelihood function,
\begin{equation}
H = \frac{\partial^2 \mathcal{L}}{\partial \theta_j \partial \theta_k},
\end{equation}
and the Hessian for $U(\vec{\theta})$, with $U(\vec{\theta}) = -2\ \ln L(\vec{\theta}) \equiv -2\ \mathcal{L}(\vec{\theta})$, which is therefore:
\begin{equation}
\frac{\partial^2 U}{\partial \theta_j \, \partial \theta_k} = -2 \frac{\partial^2 \mathcal{L}}{\partial \theta_j \,  \partial \theta_k}.
\end{equation}
We calculate the approximate Hessian as the matrix $N_{jk} = -{\rm E}(H)$, which is also required to estimate the covariance matrix, which is defined as:
\begin{equation}
{\rm Cov}(\hat{\vec{\theta}}) \ge - \left[E \left(\frac{\partial^2 \mathcal{L}(\vec{\theta})}{\partial \theta_j \, \partial \theta_k} \right) \right]_{\vec{\theta} = \hat{\vec{\theta}}} = -E(H) = N_{jk}.
\end{equation}
The Hessian of $U(\vec{\theta})$ is then $\sim 2N_{jk}$.

Starting from Eq.~(A4) in \citet{2000A&A...356.1119L}:
\begin{equation}
N_{jk}= \sum_{i =1}^N \left[ \sum_{\alpha = 1}^4 \sum_{\beta = 1}^4 D_{i\alpha\beta}^{-1} \frac{\partial c_{i\alpha}}{\partial \theta_j} \frac{\partial c_{i\beta}}{\partial \theta_k} + \frac{1}{2} \frac{\partial^2 \ln \mathrm{det} \vec{D}_i}{\partial \theta_j \partial \theta_k} +
\frac{1}{2} \sum_{\alpha = 1}^4 \sum_{\beta = 1}^4 D_{i\alpha\beta}^{-1}\frac{\partial^2 D_{i\alpha\beta}^{-1}}{\partial \theta_j \partial \theta_k}  \right].
\end{equation}

After simplification, we obtain:
\begin{eqnarray}
N_{jk} &=& \sum_{i =1}^N \left[ \sum_{\alpha = 1}^4 \sum_{\beta = 1}^4 D_{i\alpha\beta}^{-1} \frac{\partial c_{i\alpha}}{\partial \theta_j} \frac{\partial c_{i\beta}}{\partial \theta_k} + 
\frac{1}{2} \mathrm{Tr} \left[ -\vec{D}_i^{-1} \frac{\partial \vec{D}_i}{\partial \theta_j} \vec{D}_i^{-1}\frac{\partial \vec{D}_i}{\partial \theta_k} + \vec{D}_i^{-1} \frac{\partial^2 \vec{D}_i}{\partial \theta_j \partial \theta_k}\right]\right.\nonumber\\
&+&\left.
\frac{1}{2} \sum_{\alpha = 1}^4 \sum_{\beta = 1}^4 D_{i\alpha\beta}\left[ 2 \left(\vec{D}_i^{-1} \frac{\partial \vec{D}_i}{\partial \theta_k} \vec{D}_i^{-1}  \right)\frac{\partial \vec{D}_i}{\partial \theta_j} \vec{D}_i^{-1} - \vec{D}_i^{-1} \frac{\partial^2 \vec{D}_i}{\partial \theta_j \partial \theta_k} \vec{D}_i^{-1}\right]_{\alpha\beta}    \right].
\end{eqnarray}

\label{lastpage}
\end{document}